\documentclass[prd,amsmath,amssymb,floatfix,superscriptaddress,notitlepage,nofootinbib,preprintnumbers]{revtex4-1}
\usepackage{amsfonts,amssymb,amsmath,graphicx,color,bm,enumitem,ulem}
\definecolor{ultramarine}{rgb}{0.07, 0.04, 0.56}
\definecolor{cadmiumgreen}{rgb}{0.0, 0.42, 0.24}
\definecolor{indigo(dye)}{rgb}{0.0, 0.25, 0.42}
\usepackage[linktocpage=true]{hyperref}
\hypersetup{
colorlinks=true,
citecolor=red,
linkcolor=blue,
urlcolor=indigo(dye),
}

\newcommand{\be}{\begin{equation}}  
\newcommand{\ee}{\end{equation}}
\newcommand{\bem}{\begin{pmatrix}}
\newcommand{\eem}{\end{pmatrix}}

\newcommand{\tr}{\tilde{r}}

\begin{document}

\title{Spontaneous scalarization of black holes in the Horndeski theory}

\author{Masato Minamitsuji and Taishi Ikeda}
\affiliation{Center for Astrophysics and Gravitation (CENTRA), Instituto Superior T\'ecnico, University of Lisbon,
Lisbon 1049-001, Portugal.}

\begin{abstract}
We investigate the possibility of spontaneous scalarization of static, spherically symmetric, and asymptotically flat black holes (BHs) in the Horndeski theory. Spontaneous scalarization of BHs is a phenomenon that the scalar field spontaneously obtains a nontrivial profile in the vicinity of the event horizon via the nonminimal couplings and eventually the BH possesses a scalar charge. In the theory in which spontaneous scalarization takes place, the Schwarzschild solution with a trivial profile of the scalar field exhibits a tachyonic instability in the vicinity of the event horizon, and evolves into a hairy BH solution. Our analysis will extend the previous studies about the Einstein-scalar-Gauss-Bonnet (GB) theory to other classes of the Horndeski theory. First, we clarify the conditions for the existence of the vanishing scalar field solution $\phi=0$ on top of the Schwarzschild spacetime, and we apply them to each individual generalized galileon coupling. For each coupling, we choose the coupling function with minimal power of $\phi$ and $X:=-(1/2)g^{\mu\nu}\partial_\mu\phi\partial_\nu\phi$ that satisfies the above condition, which leaves nonzero and finite imprints in the radial perturbation of the scalar field. Second, we investigate the radial perturbation of the scalar field about the $\phi=0$ solution on top of the Schwarzschild spacetime. While each individual generalized galileon coupling except for a generalized quartic coupling does not satisfy the hyperbolicity condition or realize a tachyonic instability of the Schwarzschild spacetime by itself, a generalized quartic coupling can realize it in the intermediate length scales outside the event horizon. Finally, we investigate a model with generalized quartic and quintic galileon couplings, which includes the Einstein-scalar-GB theory as the special case, and show that as one increases the relative contribution of the generalized quartic galileon term the effective potential for the radial perturbation develops a negative region in the vicinity of the event horizon without violation of hyperbolicity, leading to a pure imaginary mode(s) and hence a tachyonic instability of the Schwarzschild solution.
\end{abstract}
\maketitle

\section{Introduction}
\label{sec1}

Although scalar-tensor theories have been popular for a long time,
in the recent years,
these theories have attracted renewed interest
from different aspects.
The first direction of recent studies
is the extension of the known scalar-tensor theories to a more general framework.
For a long time,
there has been the belief that 
scalar-tensor theories with higher order derivative interactions  
suffer instabilities due to the presence of Ostrogradsky ghosts.
Although generically this remains true,
recent studies have revealed several new classes
of scalar-tensor theories with higher derivative interactions 
that still possess only $2+1$ degrees of freedom,
namely two gravitational wave (GW) polarizations and one scalar mode,
and hence do not suffer an Ostrogradsky instability.
In the first class of such theories,
the equations of motion remain of the second order 
by the antisymmetrization of higher derivatives interactions in the Lagrangian.
While this class of the scalar-tensor theory was constructed by Horndeski a long time ago \cite{Horndeski:1974wa},
the same theory was rediscovered very recently 
with the different representation in Refs. \cite{Deffayet:2011gz,Kobayashi:2011nu}.
We call this class of the scalar-tensor theory the Horndeski theory,
which contains four free functions of the scalar field $\phi$
and the canonical kinetic term $X=-(1/2)g^{\mu\nu}\partial_\mu\phi\partial_\nu\phi$.
After the rediscovery of the Horndeski theory,
it has been noticed that 
even if the equations of motion contain higher order time derivatives,
it is still possible to keep the $2+1$ degrees of freedom
by imposing the certain degeneracy conditions.
The existence of scalar-tensor theories beyond Horndeski
was initially recognized via the disformal transformation of the Horndeski theory \cite{Zumalacarregui:2013pma}.
Such theories are currently known as 
Gleyzes-Langlois-Piazza-Vernizzi (GLPV)~\cite{Gleyzes:2014dya}
and
degenerate higher-order scalar-tensor (DHOST)
theories 
\cite{Langlois:2015cwa,Crisostomi:2016czh,Kimura:2016rzw,BenAchour:2016fzp}.
Studies in this direction still leave some room for further extension.
Since some of these theories are directly related to the Horndeski theory
via disformal transformation \cite{Achour:2016rkg},
studies of gravitational aspects of the Horndeski theory
will provide us direct implications for 
more general higher derivative scalar-tensor theories.

The other important direction of studies of scalar-tensor theories
is their application to issues in cosmology and black hole (BH) physics.
While applications to cosmology have been argued in most studies \cite{Koyama:2015vza},
BHs or relativistic stars
are also very important and intriguing subjects 
for testing the new classes of scalar-tensor theories 
in strong field regimes 
in light of the forthcoming GW astronomy \cite{Berti:2015itd}.
In this paper, 
we will study the possibility of spontaneous scalarization
of BHs in the context of the Horndeski theory.

Spontaneous scalarization
is a phenomenon
which is caused by
a tachyonic instability of a metric solution in general relativity (GR) with a constant profile of the scalar field,
via couplings of the scalar field to the scalar invariants composed of the metric, its derivatives, and matter fields
$f(\phi) I(g_{\mu\nu};\Psi)$,
where $g_{\mu\nu}$ is the metric tensor and $\Psi$ represents matter fields.
As the consequence of a tachyonic instability, 
the scalar field obtains the nontrivial profile, $\phi=\phi (x^\mu)$,
and
the BH possesses a scalar charge.
Although spontaneous scalarization may be potentially relevant for any metric solution in GR, 
it would be exhibited most efficiently in/around BHs and relativistic compact stars.
The most famous example of spontaneous scalarization
is that of compact stars induced by the coupling to matter fields
with extremely high density and pressure \cite{Damour:1993hw,Damour:1996ke,Harada:1997mr,Harada:1998ge}.
Spontaneous scalarization 
can also be caused for BHs
via the coupling to the Gauss-Bonnet (GB) term $f(\phi)(R^2-4R^{\alpha\beta}R_{\alpha\beta}+R^2)$ 
\cite{Doneva:2017bvd,Silva:2017uqg,Antoniou:2017acq,Antoniou:2017hxj,Blazquez-Salcedo:2018jnn,Minamitsuji:2018xde,Silva:2018qhn}
and to the electromagnetic terms \cite{Herdeiro:2018wub,Doneva:2010ke,Stefanov:2007eq}.
Since relativistic stars and BHs of the theory in which spontaneous scalarization takes place can be different from those in GR, these theories can be distinguished from GR by several astrophysical observations.

Recalling that the Einstein-scalar-GB theory is a class of the Horndeski theory,
a natural and interesting question
is
whether
spontaneous scalarization of a BH
can be caused in another class of the Horndeski theory.
While it would be difficult to construct fully backreacted scalarized BH solutions,
we will study whether 
the constant scalar solution $\phi=0$ on top of the Schwarzschild spacetime exists
in another class of the Horndeski theory,
and if it exists,
whether it exhibits a tachyonic instability against the radial perturbation.
For simplicity, for the radial stability analysis,
we will focus on the scalar field perturbation, by neglecting the metric perturbations.
Our analysis will reveal
that 
in contrast to the naive expectation
the existence of the $\phi=0$ solution on top of the Schwarzschild spacetime
and
its radial stability 
crucially depend on the class of the Horndeski theory.

This paper is constructed as follows.
In Sec. \ref{sec2}, we will review the Horndeski theory
and the properties of the Schwarzschild solution with a constant scalar field $\phi=0$.
In Sec. \ref{sec3},
for each individual generalized galileon coupling in the Horndeski theory,
we will clarify the conditions
for the existence of the constant scalar field $\phi=0$ on top of the Schwarzschild spacetime.
We then choose the coupling functions with the minimal powers of $\phi$ and $X$ 
that satisfy the above conditions,
which leave nonzero and finite imprints in the linear perturbations. 
In Sec. \ref{sec4},
focusing on each individual generalized galileon coupling specified in Sec. \ref{sec3},
we will check
the existence of the $\phi=0$ solution,
and the possibility of a tachyonic instability without violation of hyperbolicity of the radial perturbation.
In Sec. \ref{sec5},
we will closely
investigate the model composed of $G_4$ and $G_5$ which includes
the Einstein-scalar-GB theory as the special limit.
In Sec. \ref{sec6},
we will close the paper after giving the brief summary and conclusion.

\section{The Horndeski theory and the Schwarzschild solution}
\label{sec2}

\subsection{The Horndeski theory}
\label{sec21}

In this paper, 
we consider the Horndeski theory
\begin{eqnarray}
\label{action}
S=\frac{M_p^2}{2}
\int d^4x {\cal L},
\end{eqnarray}
with the Lagrangian density
\begin{eqnarray}
\label{lagrangian}
\frac{{\cal L}}{\sqrt{-g}}
&=&
R+
\alpha X
+G_2(\phi,X)
-G_3(\phi,X)\Box\phi
+G_4(\phi,X) R
+G_{4X}
\left[
(\Box\phi)^2-(\nabla_\mu \nabla_\nu\phi)^2 
\right]
\nonumber\\
&+&G_5(\phi,X)G_{\mu\nu}\nabla^\mu\nabla^\nu\phi
-\frac{1}{6}G_{5X}
\left[
(\Box\phi)^3
-3\Box\phi (\nabla_\mu \nabla_\nu\phi)^2 
+2(\nabla_\mu \nabla_\nu\phi)^3 
\right],
\end{eqnarray}
where
the indices $\mu,\nu,\cdots$ run the four-dimensional spacetime;
$g_{\mu\nu}$ is the metric tensor;
$g:={\rm det}(g_{\mu\nu})$ is its determinant;
$R$ and $G_{\mu\nu}$ are the Ricci scalar and Einstein tensor 
associated with the metric $g_{\mu\nu}$;
$\phi$ is the scalar field,
$X:= -(1/2)g^{\mu\nu} \partial_{\mu}\phi \partial_{\nu}\phi$
is the canonical kinetic term;
$(\nabla_\mu \nabla_\nu\phi)^2:= (\nabla_\mu \nabla_\nu\phi)(\nabla^\mu \nabla^\nu \phi)$;
$(\nabla_\mu \nabla_\nu\phi)^3:= (\nabla_\mu \nabla_\nu\phi)(\nabla^\nu\nabla^\alpha \phi )(\nabla_\alpha\nabla^\mu \phi)$;
$G_i (\phi,X)$'s ($i=2,3,4,5$) are free functions of $\phi$ and $X$;
and $G_{iX}:=\partial G_i/\partial X$
($i=2,3,4,5$).
The Horndeski theory 
known as the most general scalar-tensor theory with the second order equations of motion
was originally formulated in Ref. \cite{Horndeski:1974wa}
and more recently
reorganized into the form of Eq. \eqref{lagrangian} in Refs. \cite{Deffayet:2011gz,Kobayashi:2011nu}
mainly for cosmological applications.

We will further assume that each generalized galileon coupling function in Eq. \eqref{lagrangian} is given by  
\begin{eqnarray}
\label{g2}
G_2(\phi,X) &=&  f_2(\phi) g_2(X)-V(\phi) ,
\\
\label{g3}
G_3(\phi,X) &=&f_3(\phi) g_3(X),
\\ 
\label{g4}
G_4(\phi,X) &=& f_4(\phi) g_4(X),
\\
\label{g5}
G_5(\phi,X) &=& f_5(\phi) g_5(X),
\end{eqnarray}
with 
$f_i(\phi)$ and $g_i(X)$ ($i=2,3,4,5$)
being functions of $\phi$ and $X$, respectively,
and $V(\phi)$ being the potential.
Throughout this paper, 
we will assume that $\alpha>0$,
so that the ordinary kinetic term takes the correct sign.

\subsection{Scalar field on top of the Schwarzschild spacetime}
\label{sec22}

We consider the general form of the metric for a static and spherically symmetric spacetime
\begin{eqnarray}
\label{metric}
ds^2=g_{\mu\nu}dx^\mu dx^\nu
=
-A(r)dt^2
+\frac{dr^2}{B(r)}
+ r^2
\left(
 d\theta^2
+\sin^2\theta d\varphi^2
\right),
\end{eqnarray}
where $r$ is the radial coordinate, $t$ is the time coordinate,
and $(\theta,\varphi)$ are coordinates for the unit two-sphere.
We also assume that the scalar field depends on $r$ and $t$, $\phi=\phi(r,t)$.
We substitute them 
into the action \eqref{action} with the Lagrangian density \eqref{lagrangian}.

Varying it with respect to 
the variables $\psi(=A, B)$ and $\phi$, respectively,
we obtain the equations of motion:
\begin{eqnarray}
\label{eoms}
0=
{\cal E}_\psi
&=&
\frac{\partial {\cal L}}{\partial \psi}
-
\frac{\partial}{\partial r}
\left(
\frac{\partial {\cal L}}{\partial \psi'}
\right)
+
\frac{\partial^2}{\partial r^2}
\left(
\frac{\partial {\cal L}}{\partial \psi''}
\right),
\\
\label{eomt}
0=
{\cal E}_\phi
&=&
\frac{\partial {\cal L}}{\partial \phi}
-
\left[
\frac{\partial}{\partial r}
\left(
\frac{\partial {\cal L}}{\partial \phi'}
\right)
+\frac{\partial}{\partial t}
\left(
\frac{\partial {\cal L}}{\partial \dot{\phi}}
\right)
\right]
+
\left[
\frac{\partial^2}{\partial r^2}
\left(
\frac{\partial {\cal L}}{\partial \phi''}
\right)
+\frac{\partial^2}{\partial r\partial t}
\left(
\frac{\partial {\cal L}}{\partial \dot{\phi}'}
\right)
+
\frac{\partial^2}{\partial t^2}
\left(
\frac{\partial {\cal L}}{\partial \ddot{\phi}}
\right)
\right],
\end{eqnarray}
where a prime and a dot
denote the derivatives with respect to $r$ and $t$, 
respectively.
We decompose the scalar field $\phi$ into the background part $\phi_0(r)$ 
and the test field part $\phi_1$ on top of it:
\begin{eqnarray}
\phi&=&\phi_0(r)+ \epsilon \phi_1(r,t),
\end{eqnarray}
where $\epsilon\ll 1$ is the bookkeeping parameter for the perturbation around a static solution.
Since we focus on the radial perturbation with the multipole index $\ell=0$,
$\phi_1$ only depends on $r$ and $t$.

By expanding ${\cal E}_\psi$ and ${\cal E}_\phi$
in terms of $\epsilon$,
the background solution for $A$, $B$, and $\phi_0(r)$ can be obtained from 
the ${\cal O} (\epsilon^0)$ part of ${\cal E}_A=0$, ${\cal E}_B=0$, and ${\cal E}_\phi=0$.
On top of the background solution, the test scalar field solution $\phi_1$ can be found
from the ${\cal O} (\epsilon^1)$ part of ${\cal E}_\phi=0$.
We would like to emphasize 
that starting with the metric \eqref{metric} does not lose the generality of our analysis at all.
If the scalar field is time dependent,  
in general
the variation of the action \eqref{action} with respect to the $(t,r)$ component of the metric
gives rise to an independent component of the equations of motion \cite{Motohashi:2016prk}.
However, 
since we are only interested in the background solution of the static metric
and then at ${\cal O}(\epsilon^0)$ the scalar field is static $\phi=\phi_0(r)$,
the equation of motion obtained from the variation of the action
with respect to the $(t,r)$ component of the metric
is trivially satisfied at ${\cal O}(\epsilon^0)$.
Thus, setting the $(t,r)$ component of the metric to zero before 
varying the action does not lose the generality of our analysis at all.
Similarly, 
without loss of generality,
we can also fix the angular component of the metric as in Eq. (\ref{metric}) 
from the beginning,
since at ${\cal O} (\epsilon^0)$
the equation of motion obtained from the variation of the action \eqref{action} with respect to this component 
is not independent from the other components
of the background equations, i.e., ${\cal E}_A=0$, ${\cal E}_B=0$, and ${\cal E}_\phi=0$ at ${\cal O} (\epsilon^0)$.
Thus, 
at ${\cal O} (\epsilon^0)$
Eqs. \eqref{eoms} and \eqref{eomt} provide the independent and complete set of the equations to determine the background solutions.

We assume the background of the Schwarzschild spacetime:
\begin{eqnarray}
\label{sch}
A(r)=B(r)=1-\frac{2M}{r},
\end{eqnarray}
where $M>0$ is mass,
while for the moment the background scalar field $\phi=\phi_0(r)$ can take a nontrivial profile.
In our analysis,
we will neglect the metric perturbations.
In the ordinary scalar-tensor and Einstein-scalar-GB theories,
even if we take the metric perturbations into consideration,
on the background of the Schwarzschild spacetime and the constant scalar field
the master equation for the radial perturbation reduces to the same equation 
as the scalar field equation of motion in the test field analysis \cite{Minamitsuji:2018xde}. Thus, we expect that in the generic Horndeski theories on the background of the constant scalar field $\phi_0={\rm const}$
the scalar field and metric perturbations would be decoupled, and 
neglecting the metric perturbations would not modify the causal properties and the effective potential of the scalar field perturbation
at least qualitatively.
We will also mention this in more detail  in Sec. \ref{sec6}.

We expand the gravitational equations of motion ${\cal E}_A=0$ and ${\cal E}_B=0$
up to ${\cal O}(\epsilon^0)$, 
and the scalar field equation of motion ${\cal E}_\phi=0$ up to ${\cal O}(\epsilon^1)$.
At ${\cal O}(\epsilon^0)$,
we will check 
whether the constant scalar field $\phi_0(r)=0$ can satisfy all the equations of motion;
\footnote{
The value of the scalar field may not be $\phi_0=0$,
but any constant field value can be made to $\phi_0=0$
via the redefinition of the field with a constant shift.}
more precisely,
the sufficient condition for the existence of the $\phi_0=0$ solution on top of the Schwarzschild solution spacetime is
whether
the ${\cal O} (\epsilon^0)$ part of the equations of motion 
can be satisfied 
when the limits of 
$\phi_0\to 0$, $\phi_{,\tr}\to 0$, and $\phi_{0,\tr\tr}\to 0$
are taken simultaneously and independently,
where we have defined the proper length in the radial direction
\begin{eqnarray}
\label{propr}
{\tilde r}:=\int \frac{dr}{\sqrt{1-\frac{2M}{r}}},
\end{eqnarray}
and $\phi_{0,\tr}:=\partial \phi_0/\partial{\tilde r}$.
We note that
$\phi_0'(r)$ and $\phi_0''(r)$ are not coordinate invariant
and for the proper measure of the derivatives those with respect to $\tr$ are employed.
We also note that 
the ${\cal O}(\epsilon^0)$ part of the kinetic term $X$ (denoted by $X_0$)
is given by 
\begin{eqnarray}
\label{x0}
X_0 (r):= -\frac{B(r)}{2} \phi'_0(r)^2
= -\frac{r-2M}{2r}\phi'_0(r)^2
= -\frac{1}{2}\phi_{0,{\tilde r}}^2.
\end{eqnarray}

On top of the Schwarzschild spacetime,
the perturbation $\phi_1(r,t)$ satisfies the ${\cal O}(\epsilon)$ part of
the scalar field equation of motion ${\cal E}_\phi=0$,
which is given in the form of 
\begin{eqnarray}
\label{pert_eq}
-\rho_1
\ddot{\phi}_1
+\rho_2\phi_1''
+\rho_3 \phi_1'
+\rho_4\phi_1
=0,
\end{eqnarray}
where the coefficients  $\rho_i$ ($i=1,2,3,4$)  are 
determined by the background Schwarzschild metric, 
the scalar field $\phi_0(r)$, 
and its derivatives,
$\phi_0'(r)$ and $\phi_0''(r)$.
\footnote{
We note that since Eq. \eqref{pert_eq} is the linear differential equation for $\phi_1$,
there is an ambiguity about the rescalings of $\rho_i$'s ($i=1,2,3,4$) by a common factor $Q(r)$, such as $\rho_i(r)\to Q(r) \rho_i(r)$.
This ambiguity is not relevant for $U_{\rm eff} (r)$ and $\rho_5(r)$ defined below,
since 
they are fixed
only by the ratios of two different $\rho_i$'s.}
Assuming the separable ansatz
\begin{align}
\phi_1(r,t)=e^{-i\omega t} {\tilde C}(r) \Psi(r),
\end{align}
where the constant $\omega$ denotes the frequency
and the function $\tilde C$ 
is given by
\begin{align}
\ln \tilde C
:=\int dr
\left(
\frac{M}{r(r-2M)}
-\frac{\rho_3}{2\rho_2}
\right),
\end{align}
the radial part of Eq. \eqref{pert_eq}
can be rewritten in the form 
of the Schr\"{o}dinger-type equation \cite{Minamitsuji:2018xde}
\begin{eqnarray}
\label{sch2}
&&
\left[
-\frac{d^2}{dr_\ast^2}
+U_{\rm eff}(r)
\right]
\Psi(r)
=\omega^2
\rho_5(r)
 \Psi(r),
\end{eqnarray}
where we have introduced the tortoise coordinate $dr_\ast:=dr/(1-2M/r)$,
with the effective potential  
\begin{align}
\label{effpot_full}
U_{\rm eff}(r)
:=
\left(1-\frac{2M}{r}\right)^2
\left\{
-\frac{M(2r-3M)}{r^2(r-2M)^2}
-\frac{1}{2}
\left(\frac{\rho_3}{\rho_2}\right)'
-\frac{\rho_3^2}{4\rho_2^2}
+\frac{\rho_4}{\rho_2}
\right\},
\end{align}
and 
\begin{eqnarray}
\rho_5(r)
:= 
\left(1-\frac{2M}{r}\right)^2
\frac{\rho_1}{\rho_2}.
\end{eqnarray}
As we will see later,
in the coefficients $\rho_i$'s in Eq. \eqref{pert_eq},
especially in $\rho_1$,
the ratios such as $\phi_0''(r)/\phi_0'(r)$
and $\phi_0(r)/\phi_0'(r)$
appear,
which become ambiguous
when the limits of $\phi_0(r)\to 0$, $\phi_{0,\tr}(r)\to 0$, and $\phi_{0,\tr\tr}(r)\to 0$ are taken
simultaneously.

To circumvent this issue,
first, we will solve the ${\cal O}(\epsilon^0)$ part of ${\cal E}_\phi=0$.
In the models which will be finally obtained in Sec. \ref{sec3} and discussed in Sec. \ref{sec4},
the ${\cal O}(\epsilon^0)$ part of ${\cal E}_\phi=0$ will reduce to
a linear differential equation for $\phi_0(r)$.
We will solve it under the regularity boundary conditions at the event horizon, i.e.,
$|\phi_0(2M)|<\infty$ and $|\phi_{0,\tr}(2M)|<\infty$.
The solution to this equation can be schematically written as 
\begin{align}
\label{sca_phi0}
\phi_0(r)=C_0 \zeta_0 (r),
\end{align}
where $C_0$ is an integration constant and $\zeta_0(r)$ is a function of $r$.
The $\phi_0(r)=0$ solution can be obtained after taking the limit of $C_0\to 0$.
But 
if there is the case that $\zeta_0(r)$ or their derivatives
with respect to the proper length $\tr$
diverge at some $r>2M$, it is subtle whether the $\phi_0=0$ solution exists,
since $\phi_0$ may remain nonzero at this point in the limit of $C_0\to 0$.
Thus, in order
to ensure the existence of the  $\phi_0=0$ solution in the limit of $C_0\to 0$,
furthermore,
we have to impose 
that 
$\zeta_0$ and its derivatives
are regular and finite
everywhere outside the event horizon $r>2M$,
namely
$|\zeta_0(r)|<\infty$, $|\zeta_{0,{\tilde r}}|<\infty$, $|\zeta_{0,\tr\tr}|<\infty$,
and 
$\zeta_0'(r)$ never crosses $0$ 
outside the event horizon $r>2M$.
The ratios $\phi_0(r)/\phi_0'(r)$ and $\phi_0''(r)/\phi_{0}'(r)$
in the limit of the $\phi_0(r)=0$ solution
can be calculated 
by taking the limit of $C_0\to 0$ after calculating $\rho_i$'s ($i=1,2,3,4,5$).

Hyperbolicity of Eq. \eqref{sch2} is then ensured if
\begin{eqnarray}
\label{rho5}
\rho_5>0,
\end{eqnarray}
everywhere outside the event horizon $r>2M$.
In most cases,
the ratios as $\phi_0''(r)/\phi_0'(r)$ and $\phi_0(r)/\phi_0'(r)$
appear only in $\rho_1$,
which does not affect $U_{\rm eff} (r)$ given in Eq. \eqref{effpot_full}. 
On the other hand, 
 $\rho_5$ given in Eq. \eqref{rho5}
contains the above ratios,
and hence
the hyperbolicity depends on the background solution $\phi_0(r)$.

As the amplitude of the scalar field perturbation grows,
at some moment 
the test scalar field analysis would break down
and the backreaction to the spacetime geometry would no longer be negligible.
In the models with the canonical ordinary term $\alpha X$ in Eq.~\eqref{lagrangian}, 
nonlinearities of the perturbations cannot be neglected,
when the effective energy-momentum tensor of the backreaction 
$\delta T_{\mu\nu}\sim   \partial_\mu \phi_1\partial_\nu \phi_1$
(by setting $\epsilon=\alpha=1$)
becomes comparable to the background curvature $M/r^3$,
namely, when $\phi_1\sim \sqrt{M/r}$,
which becomes of ${\cal O}(1)$ in the vicinity of the event horizon $r=2M$.

In addition,
since we focus on the radial perturbation with the multipole index $\ell=0$,
from our analysis
it will not be clear 
whether the perturbation modes with higher multipole indices $\ell\geq 1$ also become unstable or not.
Since the effective potential for a mode with a higher multipole index $\ell\geq 1$
is given by adding the positive contribution
to the one for the radial perturbation \eqref{effpot_full},
the instability may not be caused
for the perturbation modes with higher multipole indices $\ell\geq 1$,
even if it is caused for the radial perturbation.
For the modes with higher multipole indices $\ell\geq 2$,
the scalar field perturbation is also coupled to the gravitational wave perturbations
\cite{Kobayashi:2012kh,Kobayashi:2014wsa,Kase:2014baa,Franciolini:2018uyq}.
If the radial perturbation is unstable 
while the ones with higher multipole indices are stable, 
no gravitational wave polarizations would be excited. 
On the other hand, 
if the higher modes with higher multipole indices also become unstable,
the gravitational wave polarizations would also be excited.

\subsection{Special cases}
\label{sec23}

Before going to the main analysis, we review the $\phi_0=0$ solution
on top of the Schwarzschild spacetime
in a few special classes of the Horndeski theory.

\subsubsection{The Einstein-scalar theory with a potential}
\label{sec231}

The simplest example is the case of the Einstein-scalar theory with a potential:
\begin{align}
\label{st}
\frac{{\cal L}}{\sqrt{-g}}
=
R
+\alpha X
-V(\phi),
\end{align}
which is equivalent to the Horndeski theory Eq. \eqref{lagrangian}
\cite{Kobayashi:2011nu} with  
\begin{align}
\label{gb}
G_2&=-V(\phi),
\quad
G_3=
G_4=
G_5=0.
\end{align}
The theory \eqref{st}
admits 
the Schwarzschild metric \eqref{sch} and the constant scalar field $\phi_0=0$ 
as a solution,
if $V(0)=0$ and $V_\phi (0)=0$,
where $V_\phi:= \partial V/\partial \phi$ and so on.
The general model satisfying this requirement is given by
\begin{align}
 V(\phi)
=\frac{1}{2} \mu^2\phi^2
 + \sum_{n=3}^\infty v_n \phi^n,
\end{align}
where $\mu$ and $v_{n}$ ($n=3,4,5,\cdots$) are constants.
The coefficients $\rho_i$s in Eq. \eqref{pert_eq} are given by 
\begin{align}
\rho_1=\frac{\alpha r^3}{r-2M},
\quad
\rho_2=\alpha r(r-2M),
\quad
\rho_3=2\alpha (r-M),
\quad
\rho_4=-r^2V_{\phi\phi} (0),
\quad
\rho_5=1,
\end{align}
where $\rho_1>0$ and $\rho_2>0$ for $\alpha>0$.
The effective potential for the radial perturbation \eqref{effpot_full}
is given by 
\begin{align}
U_{\rm eff}(r)
=\frac{2M}{r^4}(r-2M)
\left[
1+\frac{r^3 V_{\phi\phi} (0)}{2M \alpha}
\right].
\end{align}
For the positive effective mass $V_{\phi\phi} (0)>0$,
$U_{\rm eff}(r)$ is non-negative
and hence the Schwarzschild solution is linearly stable against the radial perturbation.
On the other hand,
for $V_{\phi\phi} (0)<0$,
$U_{\rm eff}(r)$ becomes negative in the region far away from the event horizon $r\gg 2M$,
suggesting a tachyonic instability of the global Minkowski vacuum. 
We emphasize
that such an instability
has nothing to do with spontaneous scalarization,
since it should be caused by 
an $\omega^2<0$ mode which is trapped mostly in the vicinity of the event horizon
and affects only the BH and its vicinity.

\subsubsection{The Einstein-scalar-GB theory}
\label{sec232}

The well-studied example of spontaneous scalarization in BH spacetime
is the case of the Einstein-scalar-GB theory: 
\begin{align}
\label{sgb}
\frac{{\cal L}}{\sqrt{-g}}
=
R
+\alpha X
+ f(\phi)
\left(
R^2
-4R^{\rho\sigma}R_{\rho\sigma}
+R^{\alpha\beta\rho\sigma}R_{\alpha\beta\rho\sigma}
\right),
\end{align}
which is equivalent to the Horndeski theory Eq. \eqref{lagrangian}
\cite{Kobayashi:2011nu} with  
\begin{align}
\label{gb}
G_2&=8 f_{\phi\phi\phi\phi}(\phi) X^2 (3-\ln X),
\nonumber\\
G_3&=4 f_{\phi\phi\phi}(\phi) X (7-3\ln X),
\nonumber\\
G_4&=4 f_{\phi\phi}(\phi) X (2-\ln X),
\nonumber\\
G_5&=-4 f_{\phi}(\phi) \ln X,
\end{align}
where $f_{\phi}:=\partial f/\partial\phi$ and so on.
This theory admits the Schwarzschild metric \eqref{sch} and $\phi_0=0$
as a solution, if $f_{\phi}(0)=0$~\cite{Doneva:2017bvd,Silva:2017uqg,Antoniou:2017acq,Antoniou:2017hxj,Blazquez-Salcedo:2018jnn,Minamitsuji:2018xde,Silva:2018qhn}.
The general model satisfying this requirement is given by 
\begin{align}
f(\phi)
=  \frac{\eta}{8}\phi^2
 + \sum_{n=3}^{\infty}
   f_{n} \phi^n,
\end{align}
where $\eta$ and $f_n$ ($n=3,4,5,\cdots$) are constants.
On this solution, 
the coefficients $\rho_i$s in Eq. \eqref{pert_eq} are given by 
\begin{align}
\rho_1=\frac{\alpha r^3}{r-2M},
\quad
\rho_2=\alpha r(r-2M),
\quad
\rho_3=2\alpha (r-M),
\quad
\rho_4=\frac{48M^2 f_{\phi\phi} (0)}{r^4},
\quad
\rho_5=1,
\end{align}
where $\rho_1>0$ and $\rho_2>0$ for $\alpha>0$.
The effective potential for the radial perturbation is given by
\begin{align}
U_{\rm eff}(r)
=\frac{2M}{r^4}(r-2M)
\left[
1-\frac{24M f_{\phi\phi} (0)}{\alpha r^3}
\right].
\end{align}
For $f_{\phi\phi} (0)>0$,
the negative region of $U_{\rm eff}(r)$
appears only in the vicinity of the BH event horizon.
Hence,
the modes with the pure imaginary frequencies
are effectively trapped in the vicinity of the event horizon
and do not affect the asymptotic Minkowski vacuum.
The first mode with $\omega^2<0$ was shown to appear for $\eta/(\alpha M^2)>2.903$,
where the leading behavior in the vicinity of $\phi=0$ is given by $f(\phi)=\eta \phi^2/8+{\cal O}(\phi^3)$
\cite{Doneva:2017bvd,Silva:2017uqg,Blazquez-Salcedo:2018jnn,Minamitsuji:2018xde,Silva:2018qhn}.

\subsubsection{The shift-symmetric Horndeski theory with regular coupling functions}
\label{sec233}

The shift-symmetric Horndeski theory 
corresponds to the case of $f_i(\phi)=1$ ($i=2,3,4,5$)
and $V(\phi)=0$ in Eqs. \eqref{g2}-\eqref{g5}.
For the models where $g_i(X)$'s ($i=2,3,4,5$) in Eqs. \eqref{g2}-\eqref{g5}
and their derivatives are analytic at $X=0$,
a BH no-hair theorem was shown
for static, spherically symmetric, and asymptotically flat BH solutions \cite{Hui:2012qt},
which holds
unless one assumes a linearly time-dependent scalar field ansatz
$\phi=qt+\chi (r)$ \cite{Babichev:2013cya},
where $q$ is a constant.
The Schwarzschild metric and  $\phi_0={\rm const}$
is a solution if $g_2(0)=0$.
The radial perturbation $\phi_1$ about it
obeys Eq. \eqref{pert_eq} with
\begin{eqnarray}
\label{mleq}
\rho_1=
 \frac{\left(\alpha+g_{2X}(0)\right)r^3}{r-2M},
\quad
\rho_2= 
\left(\alpha+g_{2X}(0)\right)
r(r-2M),
\quad
\rho_3
=2
\left(\alpha+g_{2X}(0)\right)
(r-M),
\quad
\rho_4
=0,
\quad
\rho_5=1,
\end{eqnarray}
where we have assumed
$\alpha+g_{2X}(0)>0$ for the correct sign of the effective kinetic term of the perturbation,
i.e., $\rho_1>0$ and $\rho_2>0$ outside the event horizon.
The effective potential $U_{\rm eff}(r)$, Eq. \eqref{effpot_full},
is then given by  
\begin{eqnarray}
\label{u_standard}
U_{\rm eff} (r)
=\frac{2M}{r^4} \left(r-2M\right),
\end{eqnarray}
which is always non-negative.
Thus, 
the Schwarzschild solution in this example
is linearly stable against the radial perturbation,
which confirms the no-hair theorem
for the shift-symmetric Horndeski theory \cite{Hui:2012qt}.

\subsubsection{The Horndeski theory without the shift symmetry}
\label{sec234}

In order to realize a tachyonic instability of the Schwarzschild BH in the Horndeski theory,
one has to break at least one of the assumptions in Sec. \ref{sec233}.
The first is to break the shift symmetry,
and
the second is to consider the functions $g_i(X)$ ($i=2,3,4,5$),
such that $g_i(X)$ and/or its derivatives are singular at $X=0$.
Here, we focus on the case without the shift symmetry,
while $g_i(X)$'s are assumed to be regular at $X=0$.

We consider 
the Horndeski theory breaking the shift symmetry 
with regular $f_i(\phi)$'s ($i=2,3,4,5$), $V(\phi)$,
and  $g_i(X)$'s ($i=2,3,4,5$) in Eqs. \eqref{g2}-\eqref{g5}.
The Schwarzschild metric and $\phi_0=0$ is a solution, if 
\begin{eqnarray}
\label{cond20}
f_2(0)g_2(0)-V(0)=0,
\qquad
f_{2 \phi}(0)g_2(0)-V_\phi(0)=0.
\end{eqnarray}
The radial perturbation $\phi_1$ about it
obeys Eq. \eqref{pert_eq} with
\begin{eqnarray}
\rho_1&=&
 \frac{\left(\alpha+f_2(0)g_{2X}(0)-2 f_{3\phi} (0)g_3(0) \right)r^3}{r-2M},
\quad
\rho_2= 
\left(\alpha+f_2(0)g_{2X}(0)-2 f_{3\phi} (0)g_3(0) \right)
r(r-2M),
\nonumber\\
\rho_3
&=&2
\left(\alpha+f_2(0)g_{2X}(0)-2 f_{3\phi} (0)g_3(0) \right)
(r-M),
\quad
\rho_4
=
r^2
\left(f_{2\phi\phi} (0) g_2(0)- V_{\phi\phi} (0)\right),
\quad
\rho_5=1,
\end{eqnarray}
where $g_{iX}:=\partial g_i/\partial X$ and so on, and
we have assumed $\alpha+f_2(0)g_{2X}(0)-2\alpha f_{3\phi} (0)g_3(0) >0$ for the correct sign of the effective kinetic term of the perturbation, 
i.e., $\rho_1>0$ and $\rho_2>0$ outside the event horizon.
The effective potential $U_{\rm eff}(r)$, Eq. \eqref{effpot_full},
is then given by  
\begin{eqnarray}
\label{eff_regular}
U_{\rm eff} (r)
=\frac{2M}{r^4} \left(r-2M\right)
\left[
1
+\frac{r^3 (V_{\phi\phi}(0)- f_{2\phi\phi}(0)g_2(0))}
       {2M \left(\alpha+f_2(0)g_{2X}(0)-2 f_{3\phi} (0)g_3(0) \right)}
\right].
\end{eqnarray}
For $V_{\phi\phi}(0) -f_{2\phi\phi}(0)g_2(0)>0$,
$U_{\rm eff} (r)\geq 0$ outside the event horizon $r>2M$
and hence the Schwarzschild solution is linearly stable
against the radial perturbation. 
On the other hand,
for $V_{\phi\phi}(0) -f_{2\phi\phi}(0)g_2(0)<0$,
$U_{\rm eff}(r)$ becomes negative in the region far away from the event horizon $r\gg 2M$.
We emphasize again
that such an instability
has nothing to do with spontaneous scalarization.
We note that the effective potential Eq. \eqref{eff_regular} slightly differs from that for the scalar field perturbation obtained in Ref. \cite{Tattersall:2018nve} [in (24) and (25) with $\ell=0$] by the contribution of $G_{4\phi}$. 

\section{The existence of the $\phi_0=0$ solution for singular galileon couplings}
\label{sec3}

From the analyses in Sec. \ref{sec2},
we found that 
for a successful tachyonic instability of the Schwarzschild solution in the vicinity of the BH event horizon, 
$g_i(X)$'s ($i=2,3,4,5$) which are nonanalytic at $X=0$
will be necessary,
so that their contributions to the background equations of motion and the effective potential
become important in the limit of $X\to 0$.
On the other hand, 
too singular choices of $g_i(X)$'s give rise to the divergent contributions at $X=0$,
which do not admit the Schwarzschild and $\phi_0=0$ solutions.
In this section, 
we will investigate 
the conditions for the existence of the $\phi_0=0$ solution
for singular galileon coupling functions.
We note that the classification of the shift-symmetric Horndeski theories in terms of the existence of the constant scalar field solution was given in Ref.
\cite{Saravani:2019xwx}.

In this section, we will focus on each individual generalized galileon coupling in the Horndeski theory:

\begin{enumerate}

\item{Model with $G_2$ given by Eq. \eqref{g2} and $G_3=G_4=G_5=0$.}
\label{model1}

\item{Model with $G_3$ given by Eq. \eqref{g3} and $G_2=G_4=G_5=0$.}
\label{model2}

\item{Model with $G_4$ given by Eq. \eqref{g4} and $G_2=G_3=G_5=0$.}
\label{model3}

\item{Model with $G_5$ given by Eq. \eqref{g5} and $G_2=G_3=G_4=0$.}
\label{model4}

\end{enumerate}

In each of models \ref{model1}-\ref{model4},
in the case that $f_i(\phi)$ and $g_i(X)$
are expressed in terms of Laurant series expansion
with respect to $\phi=0$ and $X=0$,
respectively,
the ${\cal O}(\epsilon^0)$ part of the equations of motion is schematically given by 

\begin{eqnarray}
\label{expansion}
\sum_{j}
C_j(r)
(\phi_0)^{a_j} (\phi_{0,\tr})^{b_j} (\phi_{0,\tr\tr})^{c_j}=0,
\end{eqnarray}
where the index $j=1,2,3,\cdots$ labels each contribution,
$C_j(r)$ denotes the $r$-dependent coefficient for the $j$th term,
$a_j$ and $b_j$ are constants,
and 
$c_j$ is either $0$ or $1$,
because of the quasilinearity of the equations of motion in the Horndeski theory.
As one of the conditions for the existence of the $\phi_0=0$ solution,
for any $j$,
we impose
\begin{eqnarray}
\label{suff_cond}
a_j\geq 0,
\quad
b_j\geq 0.
\end{eqnarray}
We note that for $c_j=0$ the case of $a_j=b_j=0$ is excluded,
since otherwise this term does not vanish
in the limit of $\phi_0\to 0$ and $\phi_{0,\tr} \to 0$.
We also note that 
even if the equations of motion contain terms which do not satisfy 
the condition Eq. \eqref{suff_cond},
we still admit them if they cancel each other.

A more rigorous approach would be first to solve the nonlinear equations 
for $\phi_0(r)$ in a given class of the theory.
If the $\phi_0=0$ solution exists in a certain limit of the integration constant,
$\phi_0(r)$, $\phi_{0,\tr}$, and $\phi_{0,\tr\tr}$ would approach $0$
with different powers of the parameter.
However,
even if one could still find the $\phi_0=0$ solution,
how $\phi_0$ and its derivatives approach $0$
would highly depend on the model,
and in general it would be difficult to estimate their scalings {\it a priori}
for a given model.
Hence,
for a model-independent analysis,
instead of solving for $\phi_0$ in a given model,
we assume that all $\phi_0$, $\phi_{0,\tr}$, and $\phi_{0,\tr\tr}$
approach $0$,
but for the moment do not specify their relative ratios.
The $\phi_0=0$ solution then exists
if each term of Eq. \eqref{expansion} vanishes
when the limits of 
$\phi_0\to 0$, $\phi_{0,\tr}\to 0$, and $\phi_{0,\tr\tr}\to 0$
are taken independently and separately.
In this regards,
the condition Eq. \eqref{suff_cond} can be viewed as a sufficient condition.

Besides the terms as in Eq. \eqref{expansion},
there will be also the case
that the functions $g_i(X)$'s contain some power of $\ln X$.
Then,
the ${\cal O}(\epsilon^0)$ part of the equations of motion contains the terms as
\begin{eqnarray}
\label{subsidiary1}
(\phi_0)^{a_j} 
\times
\left[(\phi_{0,\tr})^{b_j'} \left(\ln (\phi_{0,\tr}) \right)^{b_j''}\right]
\times
 (\phi_{0,\tr\tr})^{c_j}.
\end{eqnarray}
In such a case, 
in addition to $a_j\geq 0$,
we have to impose $b_j'>0$ for the existence of the $\phi_0=0$ solution,
since
$(\phi_{0,\tr})^{b_j'} (\ln (\phi_{0,\tr}))^{b_j''}\to 0$
for an arbitrary $b_j''$
in the limit of $\phi_{0,\tr} \to 0$.

Similarly,
if $f_i(\phi)$'s contain some power of $\ln \phi$,
its contribution to the ${\cal O}(\epsilon^0)$ part of the equations of motion
is given by
\begin{eqnarray}
\label{subsidiary2}
\left[(\phi_0)^{a_j'} \left(\ln (\phi_0) \right)^{a_j''} \right]
\times
(\phi_{0,\tr})^{b_j}
\times
 (\phi_{0,\tr\tr})^{c_j}.
\end{eqnarray}
Then,  we also have to impose $a_j'>0$
as well as $b_j\geq 0$
for an arbitrary $a_j''$.

For each model,
among the coupling functions satisfying the above conditions, 
we will select the coupling functions with the minimal power of $\phi$ and $X$ in $f_i(\phi)$ and $g_i(X)$,
respectively,
such that 
the given generalized galileon coupling leaves the finite and nonvanishing contributions to the radial perturbation,
which is necessary for a tachyonic instability and spontaneous scalarization of a Schwarzschild BH.

\subsection{Model \ref{model1}}
\label{sec31}

Let us start with generalized quadratic galileon coupling
model \ref{model1}.
Substituting the Schwarzschild metric \eqref{sch} into the ${\cal O} (\epsilon^0)$ part of the equations of motion,
the contributions of the generalized quadratic galileon coupling, in which the limits of $\phi_{0}\to 0$, $\phi_{0,\bar{r}}\to 0$, and $\phi_{0,\bar{r}\bar{r}}\to 0$ can be nontrivial, are given by  
\begin{eqnarray}
\label{eomg2}
{\cal E}_A,
\,\,\,
{\cal E}_B
&\supset&
f_2 g_2,
\,\,\,
f_2 X_0 g_{2X},
\,\,\,
V,
\nonumber\\
{\cal E}_\phi
&\supset&
f_2 g_{2X} \phi_{0,\tr\tr},
\,\,
f_2 g_{2XX} X_0\phi_{0,\tr\tr},
\,\,
f_2g_{2X}\sqrt{-X_0}, 
\,\,
f_2g_{2XX} (-X_0)^{3/2}, 
\,\,
f_{2\phi}g_2,
\,\,
f_{2\phi}g_{2X} X_0,
\,\,
V_{\phi},
\end{eqnarray}
where 
$f_2$, $V$, and $g_2$ (and their derivatives) are evaluated at $\phi=\phi_0(r)$ and $X=X_0(r)$ given by Eq. \eqref{x0}.
The $\phi_0(r)=0$ solution is allowed,
if each term in Eq. \eqref{eomg2} independently goes to $0$.
Here, we do not need to consider the contribution of the ordinary kinetic term $\alpha X$,
since they trivially vanish in the limit of $\phi_{0,\tr}\to 0$ and $\phi_{0,\tr\tr}\to 0$.

We assume that the lowest order contributions to $f_2$, $V$, and $g_2$ 
in the vicinity of $\phi_0=0$ and $X_0\propto \phi_{0,\tr}^2=0$, respectively, 
are given by 
\begin{eqnarray}
\label{fg2}
f_2 \sim \phi^{\alpha_2},
\qquad
g_2\sim (-X)^{\beta_2/2},
\qquad
V\sim  \phi^{\gamma_2},
\end{eqnarray}
where 
$\alpha_2$, $\beta_2$, and $\gamma_2$ are constants.
The contributions in the ${\cal O}(\epsilon^0)$ part of ${\cal E}_A=0$ and ${\cal E}_B=0$ are given by 
\begin{eqnarray}
f_2 g_2,
\,\,\,
f_2 X_0 g_{2X}
&\sim &
\phi_0^{\alpha_2} (\phi_0')^{\beta_2},
\nonumber\\
V&\sim& \phi_0^{\gamma_2}.
\end{eqnarray}
The contributions in the ${\cal O}(\epsilon^0)$ part of ${\cal E}_\phi=0$ are given by 
\begin{eqnarray}
f_2 g_{2X} \phi_{0,\tr\tr},
\,\,
f_2 g_{2XX} X_0\phi_{0,\tr\tr}
&\sim &
\phi_0^{\alpha_2} (\phi_{0,\tr})^{\beta_2-2} \phi_{0,\tr\tr},
\nonumber\\
f_2g_{2X}\sqrt{-X_0}, 
\,\,
f_2g_{2XX} (-X_0)^{3/2}
&\sim &
\phi_0^{\alpha_2} (\phi_{0,\tr})^{\beta_2-1},
\nonumber\\
f_{2\phi}g_2,
\,\,
f_{2\phi}g_{2X} X_0
&\sim &
\phi_0^{\alpha_2-1} (\phi_{0,\tr})^{\beta_2},
\nonumber\\
V_{\phi}
&\sim& 
\phi_0^{\gamma_2-1}.
\end{eqnarray}
The minimal choice which satisfies the condition Eq. \eqref{suff_cond}
is then given by $\alpha_2=\beta_2=1$ and $\gamma_2=2$,
i.e., $f_2\sim \phi$, $g_2\sim \sqrt{-X}$, and $V\sim \phi^2$,

for which 
the terms proportional to $f_2 g_{2X} \phi_{0,\tr\tr}$ and $f_2 g_{2XX} X_0\phi_{0,\tr\tr}$ cancel each other,
since their summation is proportional to $g_{2X}+2X_0 g_{2XX}$.
We note that 
for $g_2\sim \sqrt{-X} (-X)^{\beta_2'}$ with $\beta_2'\neq 0$
they do not cancel each other and no $\phi_0=0$ solution exists.

We note that
by taking the corrections from the logarithmic factors into consideration
the model with
$f_2 \sim \phi (\ln \phi )^{\alpha_2'}$
and 
$V\sim \phi^2 (\ln \phi)^{\gamma_2'}$
with $\alpha_2'$ and $\gamma_2'$ being constants
still allows the $\phi_0=0$ solution,
since the $(\ln\phi )^{\alpha_2'}$ terms
obtained from $f_{2\phi}g_2$ and $f_{2\phi}g_{2X}X_0 $ cancel each other 
in the ${\cal O}(\epsilon^0)$ part of ${\cal E}_\phi=0$.
For simplicity, we consider the case of $\alpha_2'=\gamma_2'=0$.
Then,
the ${\cal O}(\epsilon^0)$ part of ${\cal E}_\phi=0$ reduces to a linear differential equation of $\phi_0$
at the leading order,
and then
$\phi_0$, $\phi_{0,\tr}$, and $\phi_{0,\tr\tr}$
approach $0$ with the same speed
toward the $\phi_0=0$ solution.
Adding the higher order terms in $f_2$, $g_2$, and $V$ to Eq. \eqref{fg2} 
gives the nonlinear corrections to the linear equation at the leading order,
and hence 
does not alter the existence of the $\phi_0=0$ solution.
Thus, the model with the generalized quadratic galileon coupling 
alone which satisfies the condition Eq. \eqref{suff_cond}
is given by  
\begin{eqnarray}
\label{g2_conc}
f_2 &=&f_{20} \phi+ \sum_{n=2}^\infty f_{2n} \phi^n,
\qquad
g_2 = (-X)^{\frac{1}{2}} + \sum_{n=2}^\infty d_{2n} (-X)^{\frac{n}{2}},
\qquad 
V=\frac{\mu^2}{2} \phi^2  + \sum_{n=3}^\infty v_{n} \phi^n,
\end{eqnarray}
where 
$\mu$, $f_{2n}$ $d_{2n}$, and $v_{n}$ are constants.

\subsection{Model \ref{model2}}
\label{sec32}

Next, we consider the 
generalized cubic galileon coupling model 
\ref{model2}.
Substituting the Schwarzschild metric \eqref{sch} into the ${\cal O} (\epsilon^0)$ part of the equations of motion,
the contributions of the generalized cubic galileon coupling which can be nontrivial in the limits of $\phi_{0}\to 0$, $\phi_{0,\bar{r}}\to 0$, and  $\phi_{0,\bar{r}\bar{r}}\to 0$ are given by
\begin{eqnarray}
{\cal E}_A,
\,\,\,
{\cal E}_B
&\supset&
f_3 g_{3X} (-X_0)^{3/2},
\,\,\,
f_3 g_{3X} X_0 \phi_{0,\tr\tr},
\,\,\,
f_{3\phi} g_{3} X_0,
\nonumber\\
{\cal E}_\phi
&\supset&
f_3 g_{3X} X_0,
\,\,\,
f_3 g_{3XX} X_0^2,
\,\,\,
f_3 g_{3X} \sqrt{-X_0} \phi_{0,\tr\tr},
\,\,\,
f_3 g_{3XX}  (-X_0)^{3/2}\phi_{0,\tr\tr},
\,\,\,
f_{3\phi} g_3 \phi_{0,\tr\tr},
\nonumber\\
&&
f_{3\phi} g_{3X} X_0 \phi_{0,\tr\tr},
\,\,\,
f_{3\phi} g_{3} \sqrt{-X_0},
\,\,\,
f_{3\phi} g_{3X} (-X_0)^{3/2},
\,\,\,
f_{3\phi\phi}g_3 X_0,
\end{eqnarray}
where 
$f_3$ and $g_3$ (and their derivatives) are evaluated at $\phi(r)=\phi_0(r)$ and $X=X_0(r)$,
respectively.
Here, we do not need to consider the contribution of the ordinary kinetic term $\alpha X$,
since they trivially vanish in the limit of $\phi_{0,\tr}\to 0$ and $\phi_{0,\tr\tr}\to 0$.

We assume that the lowest order contributions to $f_3$ and $g_3$
in the vicinity of $\phi_0=0$ and $X_0\propto \phi_{0,\tr}^2=0$, respectively, 
are given by
\begin{eqnarray}
\label{fg3}
f_3 \sim  \phi^{\alpha_3},
\qquad
g_3\sim  (-X)^{\beta_3/2},
\end{eqnarray}
where 
$\alpha_3$ and $\beta_3$ are constants. 
The contributions in the ${\cal O}(\epsilon^0)$ part of ${\cal E}_A=0$ and ${\cal E}_B=0$ are given by 
\begin{eqnarray}
f_3 g_{3X} (-X_0)^{3/2}
&\sim& \phi_0^{\alpha_3} (\phi_{0,\tr})^{\beta_3+1},
\nonumber\\
f_3 g_{3X} X_0 \phi_{0,\tr\tr}
&\sim &
\phi_0^{\alpha_3}
(\phi_0')^{\beta_3}
\phi_{0,\tr\tr},
\nonumber\\
f_{3\phi} g_{3} X_0
&\sim& 
\phi_0^{\alpha_3-1}
(\phi_{0,\tr})^{\beta_3+2}.
\end{eqnarray}
The contributions in the ${\cal O}(\epsilon^0)$ part of ${\cal E}_\phi=0$ are given by 
\begin{eqnarray}
f_3 g_{3X} X_0,
\,\,\,
f_3 g_{3XX} X_0^2
&\sim&
\phi_0^{\alpha_3}
(\phi_{0,\tr})^{\beta_3},
\nonumber\\
f_3 g_{3X} \sqrt{-X_0} \phi_{0,\tr\tr},
\,\,\,
f_3 g_{3XX}  (-X_0)^{3/2}
\phi_{0,\tr\tr}
&\sim&
\phi_0^{\alpha_3}
(\phi_{0,\tr})^{\beta_3-1}
\phi_{0,\tr\tr},
\nonumber\\
f_{3\phi} g_3 \phi_{0,\tr\tr},
\,\,\,
f_{3\phi} g_{3X} X_0 \phi_{0,\tr\tr}
&\sim&
\phi_0^{\alpha_3-1}
(\phi_{0,\tr})^{\beta_3}
\phi_{0,\tr\tr},
\nonumber\\
f_{3\phi} g_{3} \sqrt{-X_0},
\,\,\,
f_{3\phi} g_{3X} (-X_0)^{3/2}
&\sim&
\phi_0^{\alpha_3-1}
(\phi_{0,\tr})^{\beta_3+1},
\nonumber\\
f_{3\phi\phi}g_3 X_0
&\sim& 
\phi_0^{\alpha_3-2}
(\phi_{0,\tr})^{\beta_3+2}.
\end{eqnarray}
The minimal choice which satisfies the condition Eq. \eqref{suff_cond}
is then given by $\alpha_3=1$ and $\beta_3=0$,
where $f_{3\phi\phi}g_3 X_0$ vanishes.
The model with $f_3\propto \phi$ and $g_3=1$ 
can be absorbed into the ordinary kinetic of the scalar field after the partial integration,
as discussed in Sec. \ref{sec231} with $V(\phi)=0$.

By taking the corrections from the logarithmic factor into consideration,
only for $g_3\sim \ln (-X)$
the terms proportional to $f_3 g_{3X} \sqrt{-X_0} \phi_{0,\tr\tr}$ and $f_3 g_{3XX}  (-X_0)^{3/2}\phi_{0,\tr\tr}$
cancel each other,
since their summation is proportional to $g_{3X}+X_0 g_{3XX}$.
However,
since
$f_{3\phi} g_3 \phi_{0,\tr\tr}\propto  \phi_{0,\tr\tr}  \ln \phi_{0,\tr}$
is divergent in the limit of $\phi_{0,\tr}\to 0$,
this model does not allow the $\phi_0(r)=0$ solution. 
We have explicitly confirmed this by solving the ${\cal O} (\epsilon^0)$ part of
the equations of motion.
Thus, we will not consider model \eqref{model2} in the rest of the paper.

\subsection{Model \ref{model3}}
\label{sec33}

We then consider the
generalized quartic galileon coupling model \ref{model3}.
Substituting the Schwarzschild metric \eqref{sch} into the ${\cal O} (\epsilon^0)$ part of the equations of motion,
the contributions of the generalized quartic galileon coupling which can be nontrivial in the limits of $\phi_{0}\to 0$, $\phi_{0,\bar{r}}\to 0$, and  $\phi_{0,\bar{r}\bar{r}}\to 0$ are given by
\begin{eqnarray}
{\cal E}_A,
\,\,\,
{\cal E}_B
&\supset&
f_4 g_{4X} X_0,
\,\,\,
f_4 g_{4XX} X_0^2,
\,\,\,
f_{4}g_{4X} (-X_0)^{\frac{1}{2}} \phi_{0,\tr\tr},
\,\,\,
f_{4}g_{4XX} (-X_0)^{\frac{3}{2}} \phi_{0,\tr\tr},
\nonumber\\
&&
f_{4\phi} g_4 \phi_{0,\tr\tr},
\,\,\,
f_{4\phi} g_{4X}  X_0 \phi_{0,\tr\tr},
\,\,\,
f_{4\phi} g_{4X} (-X_0)^{\frac{3}{2}},
\,\,\,
f_{4\phi} g_{4}  (-X_0)^{\frac{1}{2}},
\,\,\,
f_{4\phi\phi} g_4 X_0
\nonumber\\
{\cal E}_\phi
&\supset&
f_4 g_{4XX} X_0 \phi_{0,\tr\tr},
\,\,\,
f_4 g_{4XXX} X_0^2\phi_{0,\tr\tr},
\,\,\,
f_4 g_{4XX} (-X_0)^{3/2},
\,\,\,
f_4 g_{4XXX} (-X_0)^{5/2},
\nonumber\\
&&
f_{4\phi} g_{4X} (-X_0)^{1/2}\phi_{0,\tr\tr},
\,\,\,
f_{4\phi} g_{4X} X_0,
\,\,\,
f_{4\phi} g_{4XX} (-X_0)^{3/2}\phi_{0,\tr\tr},
\,\,\,
f_{4\phi} g_{4XX} X_0^2,
\,\,\,
f_{4\phi\phi}
g_{4X}
(-X_0)^{3/2},
\end{eqnarray}
where 
$f_4$ and $g_4$ (and their derivatives) are evaluated at $\phi(r)=\phi_0(r)$ and $X=X_0(r)$, respectively.
Here, we do not need to consider the contribution of the ordinary kinetic term $\alpha X$,
since they trivially vanish in the limits of $\phi_{0,\tr}\to 0$ and $\phi_{0,\tr\tr}\to 0$.

We assume that the lowest order contributions of each function
in the vicinity of $\phi_0=0$ and $X_0\propto \phi_{0,\tr}^2=0$, respectively,
are given by 
\begin{eqnarray}
f_4\sim \phi^{\alpha_4},
\qquad
g_4\sim (-X)^{\beta_4/2},
\end{eqnarray}
where 
$\alpha_4$ and $\beta_4$ are constants. 
The contributions in the ${\cal O}(\epsilon^0)$ part of ${\cal E}_A=0$ and ${\cal E}_B=0$ are given by 
\begin{eqnarray}
f_4 g_{4X} X_0,
\,\,\,
f_4 g_{4XX} X_0^2
&\sim&
\phi_0^{\alpha_4}
(\phi_{0,\tr})^{\beta_4},
\nonumber\\
f_{4}g_{4X} (-X_0)^{\frac{1}{2}} \phi_{0,\tr\tr},
\,\,\,
f_{4}g_{4XX} (-X_0)^{\frac{3}{2}} \phi_{0,\tr\tr}
&\sim&
\phi_0^{\alpha_4}
(\phi_{0,\tr})^{\beta_4-1}
\phi_{0,\tr\tr},
\nonumber\\
f_{4\phi} g_4 \phi_{0,\tr\tr},
\,\,\,
f_{4\phi} g_{4X}  X_0 \phi_{0,\tr\tr}
&\sim&
\phi_0^{\alpha_4-1}
(\phi_{0,\tr})^{\beta_4}
\phi_{0,\tr\tr},
\nonumber\\
f_{4\phi} g_{4X} (-X_0)^{\frac{3}{2}},
\,\,\,
f_{4\phi} g_{4}  (-X_0)^{\frac{1}{2}}
&\sim&
\phi_0^{\alpha_4-1}
(\phi_{0,\tr})^{\beta_4+1},
\nonumber\\
f_{4\phi\phi} g_4 X_0
&\sim&
\phi_0^{\alpha_4-2}
(\phi_{0,\tr})^{\beta_4+2}.
\end{eqnarray}
The contributions in the ${\cal O}(\epsilon^0)$ part of ${\cal E}_\phi=0$ are given by 
\begin{eqnarray}
f_4 g_{4XX} X_0\phi_{0,\tr\tr},
\,\,\,
f_4 g_{4XXX} X_0^2\phi_{0,\tr\tr}
&\sim&
\phi_0^{\alpha_4}
(\phi_{0,\tr})^{\beta_4-2}
\phi_{0,\tr\tr},
\nonumber\\
f_4 g_{4XX} (-X_0)^{3/2},
\,\,\,
f_4 g_{4XXX} (-X_0)^{5/2}
&\sim&
\phi_0^{\alpha_4}
(\phi_{0,\tr})^{\beta_4-1},
\nonumber\\
f_{4\phi} g_{4X} (-X_0)^{1/2}\phi_{0,\tr\tr},
\,\,\,
f_{4\phi} g_{4XX} (-X_0)^{3/2}\phi_{0,\tr\tr}
&\sim&
\phi_0^{\alpha_4-1}
(\phi_{0,\tr})^{\beta_4-1}
\phi_{0,\tr\tr},
\nonumber\\
f_{4\phi} g_{4X} X_0,
\,\,\,
f_{4\phi} g_{4XX} X_0^2
&\sim&
\phi_0^{\alpha_4-1}
(\phi_{0,\tr})^{\beta_4},
\nonumber\\
f_{4\phi\phi}
g_{4X}
(-X_0)^{3/2}
&\sim&
\phi_0^{\alpha_4-2}
(\phi_{0,\tr})^{\beta_4+1}.
\end{eqnarray}
The minimal choice which satisfies the condition Eq. \eqref{suff_cond}
is then given by $\alpha_4=\beta_4=1$,
i.e., $f_4\sim \phi$ and $g_4\sim \sqrt{-X}$,
for which 
$f_{4\phi\phi} g_4 X_0$ and $f_{4\phi\phi} g_{4X} (-X_0)^{3/2}$ vanish
and 
the terms proportional to 
$f_4 g_{4XX}\phi_{0,\tr\tr} X_0$ and $f_4 g_{4XXX} X_0^2\phi_{0,\tr\tr}$
cancel each other,
since their summation is proportional to $3g_{4XX}+2X_0 g_{4XXX}$.

Moreover, 
$f_4 \sim \phi (\ln \phi )^{\alpha_4'}$ with $\alpha_4'\neq 0$
is not allowed,
since 
the terms 
$f_{4\phi} g_4 \phi_{0,\tr\tr}$,
$f_{4\phi} g_{4X}  X_0 \phi_{0,\tr\tr}$,
$f_{4\phi} g_{4X} (-X_0)^{\frac{3}{2}}$,
$f_{4\phi} g_{4}  (-X_0)^{\frac{1}{2}}$,
$f_{4\phi} g_{4X} (-X_0)^{1/2}\phi_{0,\tr\tr}$,
$f_{4\phi} g_{4XX} (-X_0)^{3/2}\phi_{0,\tr\tr}$,
$f_{4\phi} g_{4X} X_0$,
and 
$f_{4\phi} g_{4XX} X_0^2$
in the ${\cal O}(\epsilon^0)$ part of the equations of motion
give rise to the contributions as $(\ln \phi_0)^{\alpha_4'}$,
for which the $\phi_0\to 0$ limit
becomes singular.
Thus, 
$f_4\sim\phi$ is the unique minimal model which allows the $\phi_0=0$ solution.
Then, the ${\cal O}(\epsilon^0)$ part of ${\cal E}_\phi=0$ reduces to a linear differential equation of $\phi_0$
at the leading order,
and hence $\phi_0$, $\phi_{0,\tr}$, and $\phi_{0,\tr\tr}$
approach to $0$ with the same speed in the vicinity of the $\phi_0=0$ solution.
The model with the generalized quartic galileon coupling alone which could potentially realize spontaneous scalarization
is given by  
\begin{eqnarray}
\label{g4_conc}
f_4 &=&f_{40} \phi  + \sum_{n=2}^\infty f_{4n} \phi^n,
\qquad
g_4 = \sqrt{-X}+ \sum_{n=2}^\infty d_{4n} (-X)^{\frac{n}{2}}.
\end{eqnarray}
where $f_{40}$, $f_{4n}$, and $d_{4n}$ are constants.

\subsection{Model \ref{model4}}
\label{sec34}

Finally, 
we consider the generalized quintic galileon coupling model \ref{model4}.
Substituting the Schwarzschild metric \eqref{sch} into the ${\cal O} (\epsilon^0)$ part of the equations of motion,
the contributions of generalized quintic galileon coupling which can be nontrivial in the limits of $\phi_{0}\to 0$, $\phi_{0,\bar{r}}\to 0$, and $\phi_{0,\bar{r}\bar{r}}\to 0$ are given by 
\begin{eqnarray}
{\cal E}_A,
\,\,\,
{\cal E}_B
&\supset&
f_5 g_{5X} (-X_0)^{3/2},
\,\,\,
f_5 g_{5XX} (-X_0)^{5/2},
\,\,\,
f_5 g_{5X} X_0 \phi_{0,\tr\tr},
\,\,\,
f_5 g_{5XX} X_0^2 \phi_{0,\tr\tr}
\nonumber\\
&&
f_{5\phi} g_{5X}X_0^2,
\,\,\,
f_{5\phi} g_{5}X_0,
\,\,\,
f_{5\phi} g_{5}\sqrt{-X_0}\phi_{0,\tr\tr},
\,\,\,
f_{5\phi} g_{5X} (-X_0)^{3/2}\phi_{0,\tr\tr},
\,\,\,
f_{5\phi\phi} g_{5} (-X_0)^{3/2},
\nonumber\\
{\cal E}_\phi
&\supset&
f_{5} g_{5X}X_0,
\,\,\,
f_5 g_{5XX}  X_0^{2},
\,\,\,
f_5 g_{5XXX}  X_0^{3},
\,\,\,
f_5 g_{5X} \sqrt{-X_0}\phi_{0,\tr\tr},
\,\,\,
f_5 g_{5XX} (-X_0)^{3/2}\phi_{0,\tr\tr},
\,\,\,
f_5 g_{5XXX}  (-X_0)^{5/2}\phi_{0,\tr\tr},
\nonumber\\
&&
f_{5\phi} g_{5X} X_0 \phi_{0,\tr\tr},
\,\,\,
f_{5\phi} g_{5XX} X_0^{2}\phi_{0,\tr\tr},
\,\,\,
f_{5\phi} g_{5X} (-X_0)^{3/2},
\,\,\,
\,\,\,
f_{5\phi} g_{5XX} (-X_0)^{5/2},
\,\,\,
f_{5\phi\phi} g_{5X} X_0^{2},
\end{eqnarray}
where 
$f_5$ and $g_5$ (and their derivatives) are evaluated at $\phi(r)=\phi_0(r)$ and $X=X_0(r)$, respectively.
Here, we do not need to consider the contribution of the ordinary kinetic term $\alpha X$,
since they trivially vanish in the limit of $\phi_{0,\tr}\to 0$ and $\phi_{0,\tr\tr}\to 0$.

We assume that the lowest order contributions of each function
in the vicinity of $\phi_0=0$ and $X_0\propto \phi_{0,\tr}^2=0$, respectively,
are given by 
\begin{eqnarray}
f_5\sim \phi^{\alpha_5},
\qquad
g_5\sim (-X)^{\beta_5/2},
\end{eqnarray}
where $\alpha_5$ and $\beta_5=0$ are constants.
The contributions in the ${\cal O}(\epsilon^0)$ part of ${\cal E}_A=0$ and ${\cal E}_B=0$ are given by 
\begin{eqnarray}
f_5 g_{5X} (-X_0)^{3/2},
\,\,\,
f_5 g_{5XX} (-X_0)^{5/2}
&\sim&
\phi_0^{\alpha_5}
(\phi_{0,\tr})^{\beta_5+1},
\nonumber\\
f_5 g_{5X} X_0 \phi_{0,\tr\tr},
\,\,\,
f_5 g_{5XX} X_0^2 \phi_{0,\tr\tr}
&\sim&
\phi_0^{\alpha_5}
(\phi_{0,\tr})^{\beta_5}
\phi_{0,\tr\tr},
\nonumber\\
f_{5\phi} g_{5X}X_0^2,
\,\,\,
f_{5\phi} g_{5}X_0
&\sim&
\phi_0^{\alpha_5-1}
(\phi_{0,\tr})^{\beta_5+2},
\nonumber\\
f_{5\phi} g_{5}\sqrt{-X_0}\phi_{0,\tr\tr},
\,\,\,
f_{5\phi} g_{5X} (-X_0)^{3/2}\phi_{0,\tr\tr}
&\sim&
\phi_0^{\alpha_5-1}
(\phi_{0,\tr})^{\beta_5+1}
\phi_{0,\tr\tr},
\nonumber\\
f_{5\phi\phi} g_{5} (-X_0)^{3/2}
&\sim&
\phi_0^{\alpha_5-2}
(\phi_{0,\tr})^{\beta_5+3}.
\end{eqnarray}
The contributions in the ${\cal O}(\epsilon^0)$ part of ${\cal E}_\phi=0$ are given by
\begin{eqnarray}
f_{5} g_{5X}X_0,
\,\,\,
f_5 g_{5XX}  X_0^{2},
\,\,\,
f_5 g_{5XXX}  X_0^{3}
&\sim&
\phi_0^{\alpha_5}
(\phi_{0,\tr})^{\beta_5},
\nonumber\\
f_5 g_{5X} \sqrt{-X_0}\phi_{0,\tr\tr},
\,\,\,
f_5 g_{5XX} (-X_0)^{3/2}\phi_{0,\tr\tr},
\,\,\,
f_5 g_{5XXX}  (-X_0)^{5/2}\phi_{0,\tr\tr}
&\sim&
\phi_0^{\alpha_5}
(\phi_{0,\tr})^{\beta_5-1}
\phi_{0,\tr\tr},
\nonumber\\
f_{5\phi} g_{5X} X_0 \phi_{0,\tr\tr},
\,\,\,
f_{5\phi} g_{5XX} X_0^{2}\phi_{0,\tr\tr}
&\sim&
\phi_0^{\alpha_5-1}
(\phi_{0,\tr})^{\beta_5}
\phi_{0,\tr\tr},
\nonumber\\
f_{5\phi} g_{5X} (-X_0)^{3/2},
\,\,\,
\,\,\,
f_{5\phi} g_{5XX} (-X_0)^{5/2}
&\sim&
\phi_0^{\alpha_5-1}
(\phi_{0,\tr})^{\beta_5+1},
\nonumber\\
f_{5\phi\phi} g_{5X} X_0^{2}
&\sim&
\phi_0^{\alpha_5-2}
(\phi_{0,\tr})^{\beta_5+2}.
\end{eqnarray}
The minimal choice which satisfies the condition Eq. \eqref{suff_cond}
is then given by $\alpha_5=1$ and $\beta_5=0$,
i.e., $f_5\sim \phi$ and $g_5=1$,
which can be absorbed into a shift-symmetric theory after the partial integration.

By taking the corrections from the logarithmic factor into consideration,
only for $g_5\sim \ln (-X)$, 
the terms proportional to 
$f_5 g_{5X} \sqrt{-X_0}\phi_{0,\tr\tr}$, $f_5 g_{5XX} (-X_0)^{3/2}\phi_{0,\tr\tr}$, and $f_5 g_{5XXX}  (-X_0)^{5/2}\phi_{0,\tr\tr}$
cancel each other,
since their summation is proportional to 
$(r-3M)(g_{5X}+X_0 g_{5XX}) +(r-2M) (2g_{5XX}+X_0  g_{5XXX})$.
$f_{5\phi\phi} g_{5} (-X_0)^{3/2}$ and $f_{5\phi\phi} g_{5X} X_0^{2}$ vanish.

Moreover, 
$f_5 \sim \phi (\ln \phi )^{\alpha_5'}$ with $\alpha_5'\neq 0$
is not allowed,
since 
the terms
$f_{5\phi} g_{5X}X_0^2$,
$f_{5\phi} g_{5}X_0$,
$f_{5\phi} g_{5}\sqrt{-X_0}\phi_{0,\tr\tr}$,
$f_{5\phi} g_{5X} (-X_0)^{3/2}\phi_{0,\tr\tr}$,
$f_{5\phi} g_{5X} X_0 \phi_{0,\tr\tr}$,
$f_{5\phi} g_{5XX} X_0^{2}\phi_{0,\tr\tr}$,
$f_{5\phi} g_{5X} (-X_0)^{3/2}$,
and 
$f_{5\phi} g_{5XX} (-X_0)^{5/2}$
in the ${\cal O}(\epsilon^0)$ part of the equations of motion
give rise to the terms as $(\ln \phi_0)^{\alpha_5'}$,
for which the $\phi_0\to 0$ becomes singular.
Thus, 
$f_5\sim\phi$ is the unique minimal model which allows the $\phi_0=0$ solution.
Then the ${\cal O}(\epsilon^0)$ part of ${\cal E}_\phi=0$ reduces to a linear differential equation of $\phi_0$
at the leading order, 
$\phi_0$, $\phi_{0,\tr}$, and $\phi_{0,\tr\tr}$ approach $0$ with the same speed
in the vicinity of the $\phi_0=0$ solution.
The model with the generalized quintic galileon coupling alone which could potentially realize spontaneous scalarization
is given by  
\begin{eqnarray}
\label{g5_conc}
f_5 (\phi) &=&-f_{50} \phi  + \sum_{n=2}^\infty f_{5n} \phi^n,
\qquad
g_5(X)= \ln (-X)+ \sum_{n=1}^\infty d_{5n} (-X)^{\frac{n}{2}}.
\end{eqnarray}
where $f_{50}$, $f_{5n}$, and $d_{5n}$ are constants.

\subsection{Discussions}

The no-hair theorem
for the static, spherically symmetric, and asymptotically flat BH solutions
in the shift-symmetric Horndeski theories \cite{Hui:2012qt}
was established
on the basis of the properties of 
the Noether current $J^\mu$ associated with the shift symmetry.
In a generic static, spherically symmetric, and asymptotically flat BH spacetime,
integrating $\partial_r (\sqrt{-g}J^r)=0$
and
imposing the regularity at the horizon $J_\mu J^\mu < \infty$
on the event horizon yield $J^r=0$
everywhere outside the event horizon.
On the other hand, 
in a generic class of the Horndeski theory
the contributions to $J^r$ are given by 
\begin{eqnarray}
\label{jr}
J^r
\supset
\sqrt{-X} G_{2X},
\,\,\,
X G_{3X},
\,\,\,
\sqrt{-X}G_{4X},
\,\,\,
(-X)^{3/2}G_{4XX},
\,\,\, 
X G_{5X},
\,\,\,
X^2 G_{5XX}.
\end{eqnarray}
In order to obtain a nontrivial solution $\phi_{,\tr} \propto \sqrt{-X}\neq 0$,
each contribution in Eq. \eqref{jr} should be nonzero and finite,
which requires the following leading order behavior of $G_i(X)$'s 
\begin{eqnarray}
G_{2}\sim \sqrt{-X},
\quad
G_{3}\sim \ln (-X),
\quad
G_{4}\sim \sqrt{-X},
\quad
G_{5}\sim \ln (-X).
\end{eqnarray}
They agree with the leading order part of $g_i(X)$'s 
in Eqs. \eqref{g2_conc},  \eqref{g4_conc}, and \eqref{g5_conc},
respectively (see also \cite{Babichev:2017guv}).
Since in the models of Eqs. \eqref{g2_conc}, \eqref{g4_conc}, and \eqref{g5_conc},
$f_i(\phi)\sim \phi$ ($i=2,3,4,5$), 
the shift symmetry is minimally broken by multiplying an additional power of $\phi$
to the above coupling functions.

The situation is very similar to the case of the Einstein-scalar-GB theory \eqref{sgb}
where both hairy BH solutions and spontaneous scalarization have been studied.
In the shift-symmetric Einstein-scalar-GB theory with $f(\phi)\propto \phi$ in Eq. \eqref{sgb},
a hairy BH solution was explicitly constructed in Refs. \cite{Sotiriou:2014pfa}.
On the other hand,
spontaneous scalarization of the Schwarzschild solution 
was shown to take place for a quadratic coupling $f(\phi)\propto \phi^2$
\cite{Doneva:2017bvd,Silva:2017uqg,Blazquez-Salcedo:2018jnn,Minamitsuji:2018xde,Silva:2018qhn}.
Thus, 
in order to discuss spontaneous scalarization, 
an additional power of $\phi$ is multiplied to the linear coupling in the shift symmetric theory by which
the shift symmetry is minimally broken.

\section{The $\phi_0=0$ solution and linear stability against the radial perturbation}\label{sec4}

In this section, 
we will closely look at the solution of the scalar field on top of the Schwarzschild spacetime,
and investigate the existence of the $\phi_0=0$ solution and the linear stability against the radial perturbation.

\subsection{Model \ref{model1}}
\label{sec41}

We focus on the leading order part of \eqref{g2_conc}:
\begin{eqnarray}
g_2(X)= (-X)^{1/2},
\qquad
f_2(\phi)= f_{20} \phi,
\qquad
V(\phi)= \frac{1}{2}\mu^2 \phi^2,
\end{eqnarray}
since the higher order terms would not affect the existence of the $\phi_0=0$ solution and the linear stability.
The ${\cal O} (\epsilon^1)$ part of ${\cal E}_\phi=0$ is given by Eq. \eqref{pert_eq} with
\begin{eqnarray}
\label{rhos2}
\rho_1
&=&
\frac{r^3\alpha}{r-2M}
\mp
\frac{f_{20} r^{7/2} \phi_0 }
        {\sqrt{2} (r-2M)^{3/2}\phi_0'},
\quad
\rho_2
=
r(r-2M)\alpha,
\quad
\rho_3
=
2(r-M)\alpha,
\nonumber\\
\rho_4
&=&
-
\frac{1}{2}
\sqrt{\frac{r}{r-2M}}
\left[
\pm \sqrt{2} f_{20} (2r-3M)
+2\sqrt{r^3(r-2M)}  \mu^2
\right],
\nonumber\\
\rho_5 
&=&1
\mp
\frac{f_{20} \phi_0}
        {\sqrt{2(1-2M/r)} \alpha \phi_0'},
\end{eqnarray}
where the upper and lower branches correspond to the cases of $\phi_0'>0$ and $\phi_0'<0$, respectively.
For $\alpha>0$,
$\rho_2>0$ outside the event horizon.
We assume that $\phi_0'$ never cross $0$, where $\rho_1$ blows up and hyperbolicity is broken.
$U_{\rm eff} (r)$ in Eq. \eqref{sch2} is given by 
\begin{eqnarray}
\label{u2}
U_{\rm eff}(r)
=\frac{2M }{r^4} (r-2M)
+\frac{ 
\pm \sqrt{2}f_{20}\sqrt{r-2M}(2r-3M) 
+2r^{3/2} (r-2M)\mu^2}{2\alpha r^{5/2}}.
\end{eqnarray}
In order to evaluate $\phi_0/\phi_0'$ in the limit of the $\phi_0=0$ solution,
we investigate the ${\cal O} (\epsilon^0)$ part of ${\cal E}_\phi=0$
on the Schwarzschild background,
which is explicitly given by 
\begin{eqnarray}
\label{g2_eq}
- \left(
\pm \sqrt{2} f_{20} (2r-3M)
+2\sqrt{r-2M} r^{3/2} \mu^2
 \right)
\phi_0
+2\sqrt{1-\frac{2M}{r}}
\alpha
\left(
2(r-M)\phi_0'
+r(r-2M)\phi_0''
\right)
=0.
\end{eqnarray}
Solving Eq. \eqref{g2_eq} under the regularity boundary condition at the event horizon,
$|\phi_0(2M)|<\infty$ and $|\phi_{0,\tr}(2M)|<\infty$,
the general scalar field solution $\phi_0(r)$ near the event horizon $r\gtrsim 2M$ is given by
\begin{eqnarray}
\phi_0(r)
&=&
C_0
\left[
1
\pm
\frac{2f_{20} \sqrt{M (r-2M)}}
       {\alpha}
+\frac{M(f_{20}^2+2\alpha \mu^2) (r-2M)}
        {\alpha^2} 
+
{\cal O}
\left(
 r-2M
\right)^{3/2}
\right].
\label{horizon_vicinity}
\end{eqnarray}
If we choose $C_0>0$,
both the branches of $\phi_0'>0$ and $\phi_0'<0$ at the event horizon require
\begin{eqnarray}
\label{f20p}
f_{20}>0.
\end{eqnarray}
We note that 
$\phi_{0,\tr}$ and $\phi_{0,\tr\tr}$ are regular at the event horizon
and satisfy the condition discussed in Sec. \ref{sec22}. We also note that for both the branches the other solution of Eq. \eqref{g2_eq}
contains a term proportional to $\ln (r-2M)\sim \ln \tilde{r}$,
which is singular at $r=2M$,
and hence
does not satisfy the condition discussed in Sec. \ref{sec22}.

After calculating $\phi_{0}'(r)$ and $\phi_0''(r)$,
the simultaneous limit $\phi_0\to 0$, $\phi_0'\to 0$, and $\phi_0''\to 0$
can be taken by $C_0\to 0+$,
and then
\begin{eqnarray}
\rho_5
=
\pm
\frac{\sqrt{M} (2\mu^2 \alpha -f_{20}^2)}
       {f_{20}\alpha}
\sqrt{r-2M}
+{\cal O}
\left(
 r-2M
\right).
\end{eqnarray}
For the upper and lower branches, 
in order to ensure hyperbolicity near the event horizon,
we have to impose
\begin{eqnarray}
\label{f201}
0<f_{20}<\sqrt{2\alpha}\mu,
\qquad
f_{20}>\sqrt{2\alpha}\mu,
\end{eqnarray}
respectively.
For the branch of $\phi_0'>0$,
Eq. \eqref{u2} is always non-negative for $r>2M$.
Thus, the Schwarzschild solution is linearly stable against the radial perturbation 
and no tachyonic instability takes place.

On the other hand,
for the branch of $\phi_0'<0$,
the general solution $\phi_0(r)$ for $\mu^2>0$ at $r\gg 2M$
is given by $\phi_0(r) \sim  -  e^{r\mu/\sqrt{\alpha}}/r$,
which satisfies $\phi_{0}'<0$.
However, 
a runaway growth of $\phi_0(r)$ 
does not satisfy the condition
of the regularity at the spatial infinity, $|\zeta_0(\infty)|<\infty$.

For $\mu^2= 0$,
the general solution $\phi_0(r)$ at $r\gg 2M$
is given by 
a linear combination of the approximated solutions written in terms of Bessel and Neumann functions
$J_1
\left(
\frac{2^{5/4}\sqrt{f_{20} r}}{\sqrt{\alpha}}
\right)/\sqrt{r}$
and 
$Y_1
\left(
\frac{2^{5/4}\sqrt{f_{20}r}}{\sqrt{\alpha}}
\right)/\sqrt{r}$, respectively.
Thus, $\phi_0(r)$ is oscillating, and $\phi_{0}'$ crosses $0$ many times,
where $\rho_5$ blows up
and hyperbolicity is broken.

We note that for $C_0<0$
the role of $\phi_0'>0$ and $\phi_0'<0$ (as well as $f_{20}$ and $-f_{20}$)
is simply exchanged,
and the conclusion remains unchanged.
Therefore, in model \ref{model1}
spontaneous scalarization does not take place.

\subsection{Model \ref{model3}}
\label{sec43}

We then focus on the leading order part of \eqref{g4_conc}:
\begin{eqnarray}
\label{G4}
g_4(X)= (-X)^{1/2},
\qquad
f(\phi)= f_{40} \phi,
\end{eqnarray}
for which the ${\cal O} (\epsilon^0)$ part of ${\cal E}_\phi=0$
is given by 
\begin{eqnarray}
\label{scaeq_g4}
&\mp&
\sqrt{2}f_{40} M\phi_0
+2
\left(
\pm \sqrt{2}f_{40} (3M^2+2Mr-2r^2)
+\alpha r^{3/2}\sqrt{r-2M} (r-M)
\right)
\phi_0'
\nonumber\\
&+&r(r-2M)
\left(
\pm 2\sqrt{2}f_{40}(3M-2r)
+\alpha r^{3/2} \sqrt{r-2M}
\right)
\phi_0''
=0,
\end{eqnarray}
where the upper and lower branches correspond to the cases of $\phi_0'>0$ and for $\phi_0'<0$, respectively.
The ${\cal O} (\epsilon^1)$ part of ${\cal E}_\phi=0$
is given by Eq. \eqref{pert_eq} with 
\begin{eqnarray}
\rho_1
&=&
\frac{\sqrt{r}}{ (r-2M)\phi_0'}
\left[
\pm
\sqrt{2}f_{40} (4M-3r)\phi_0
+r
\left(
(\pm 2\sqrt{2} f_{40}(7M-4r) +r^{3/2}\sqrt{r-2M}\alpha)\phi_0'
\pm
2\sqrt{2}f_{40}(2M-r)r\phi_0''
\right)
\right],
\nonumber\\
\rho_2
&=&
\frac{r-2M}{\sqrt{r}}
\left(
 \pm2\sqrt{2}f_{40}(3M-2r)
+r^{3/2} \sqrt{r-2M}\alpha
\right),
\nonumber\\
\rho_3
&=&
\frac{2}{r^{3/2}}
\left(
\pm 
 \sqrt{2} f_{40}(3M^2+2Mr-2r^2)
+r^{3/2}\sqrt{r-2M} (r-M)\alpha
\right),
\nonumber\\
\rho_4
&=&
\mp
\frac{\sqrt{2} f_{40}M}{r^{3/2}},
\end{eqnarray}
and 
\begin{eqnarray}
\rho_5=
1
\mp
 \frac{\sqrt{2} f_{40} 
         \left((3r-4M)\phi_0+2r(r-2M) (r\phi_0''+2\phi_0')\right)
          }
      {r\left(\pm2\sqrt{2}f_{40}(3M-2r)
           +r^{3/2} \sqrt{r-2M}\alpha\right)\phi_0'}.
\end{eqnarray}
$U_{\rm eff} (r)$ in Eq. \eqref{sch2} is given by 
\begin{eqnarray}
&&U_{\rm eff}(r)
=\frac{2M }{r^4} (r-2M)
\pm
\frac{f_{40}}
      {r^4 \left(\pm 2\sqrt{2} f_{40}(3M-2r)+r^{3/2}\alpha \sqrt{r-2M}\right)^2}
\nonumber\\
&\times&
\left[
\pm 2f_{40} (81M^4-132M^3 r+54M^2 r^2+4M r^3-4r^4)
+\sqrt{2(r-2M)}M r^{3/2} 
\left(
7r^2-28M r+27M^2
\right)
\alpha
\right].
\nonumber\\
\end{eqnarray}
There are two general scalar field solutions to Eq. \eqref{scaeq_g4}
which satisfy the regularity boundary conditions at the event horizon $r=2M$,
\begin{eqnarray}
\label{g4_sol1}
\phi_0(r)
=
C_1
\left[
1
-\frac{r-2M}{2M}
\mp
\frac{\alpha (r-2M)^{3/2}}{3f_{40}\sqrt{M}}
+
{\cal O}
\left(
 (r-2M)^{2}
\right)
\right],
\end{eqnarray}
and 
\begin{eqnarray}
\label{g4_sol2}
\phi_0(r)
=
C_2
\left[
\sqrt{r-2M}
\pm \frac{\sqrt{M}\alpha}{2f_{40}} (r-2M)
+
\left(
-\frac{3}{4M}
+\frac{M\alpha^2}{3f_{40}^2}
\right)
(r-2M)^{3/2}
+
{\cal O}
\left(
 (r-2M)^{2}
\right)
\right],
\end{eqnarray}
where $C_1$ and $C_2$ are integration constants.
We note that 
$\phi_{0,\tr}$ and $\phi_{0,\tr\tr}$ are regular at the event horizon
and satisfy the condition discussed in Sec. \ref{sec22}.

\subsubsection{The case of the solution \eqref{g4_sol1}}
\label{sec421}

For the solution \eqref{g4_sol1},
$C_1<0$ for the $\phi_0'>0$ branch and $C_1>0$ for the $\phi_0'<0$ one. 
The $C_1 \to 0$ limit is taken after calculating $\phi_{0}'(r)$ and $\phi_0''(r)$,
and then 
\begin{eqnarray}
\rho_5
=
\pm
\frac{\sqrt{M} \alpha}{f_{40}}
 (r-2M)^{1/2}
+{\cal O}
\left(
 r-2M
\right).
\end{eqnarray}
For the upper and lower branches,
hyperbolicity in the vicinity of the event horizon 
imposes, respectively,
\begin{eqnarray}
\label{f40}
f_{40}>0,
\qquad
f_{40}<0,
\end{eqnarray}
for which 
there is always a point $r_s(>2M)$ where $\pm 2\sqrt{2}f_{40}(3M-2r_s) +\alpha r_s^{3/2} \sqrt{r_s-2M}=0$ in Eq. \eqref{scaeq_g4}.
At $r=r_s$,
$\rho_5$ diverges 
and changes the sign,
and 
hence the hyperbolicity is broken.

\subsubsection{The case of the solution \eqref{g4_sol2}}
\label{sec422}

On the other hand, 
for the solution \eqref{g4_sol2},
in the vicinity of the event horizon $r\gtrsim 2M$ we obtain 
\begin{eqnarray}
\rho_5
=\frac{M\alpha^2 (r-2M)}{f_{40}^2} 
\pm
\frac{(-5f_{40}^2\alpha+2M^2\alpha^3)}{f_{40}^3\sqrt{M}}
(r-2M)^{3/2}
+
{\cal O}
\left(
(r-2M)^2
\right),
\end{eqnarray}
which allows both $f_{40}>0$ and $f_{40}<0$.
Thus, 
if we choose $f_{40}<0$ and $f_{40}>0$ 
for the $\phi_0'>0$ and $\phi_0'<0$ branches, respectively,
Eq. \eqref{scaeq_g4} can be integrated without crossing the singularity,
and $\rho_1>0$ and $\rho_2>0$
outside the event horizon.

Because of the symmetry in the background and perturbation equations,
the $\phi_0'(r)>0$ branch with $f_{40}<0$
is physically equivalent to 
the $\phi_0'(r)<0$ branch with $f_{40}>0$.
Hence 
in the rest of this subsection we focus on the $\phi_0'>0$ branch.
For numerical analysis, 
without loss of generality, we may set $M=1$ and $\alpha=1$ by appropriate rescaling of $f_{40}$.

\begin{figure}[h]
\unitlength=1.1mm
\begin{center}
  \includegraphics[height=4.5cm,angle=0]{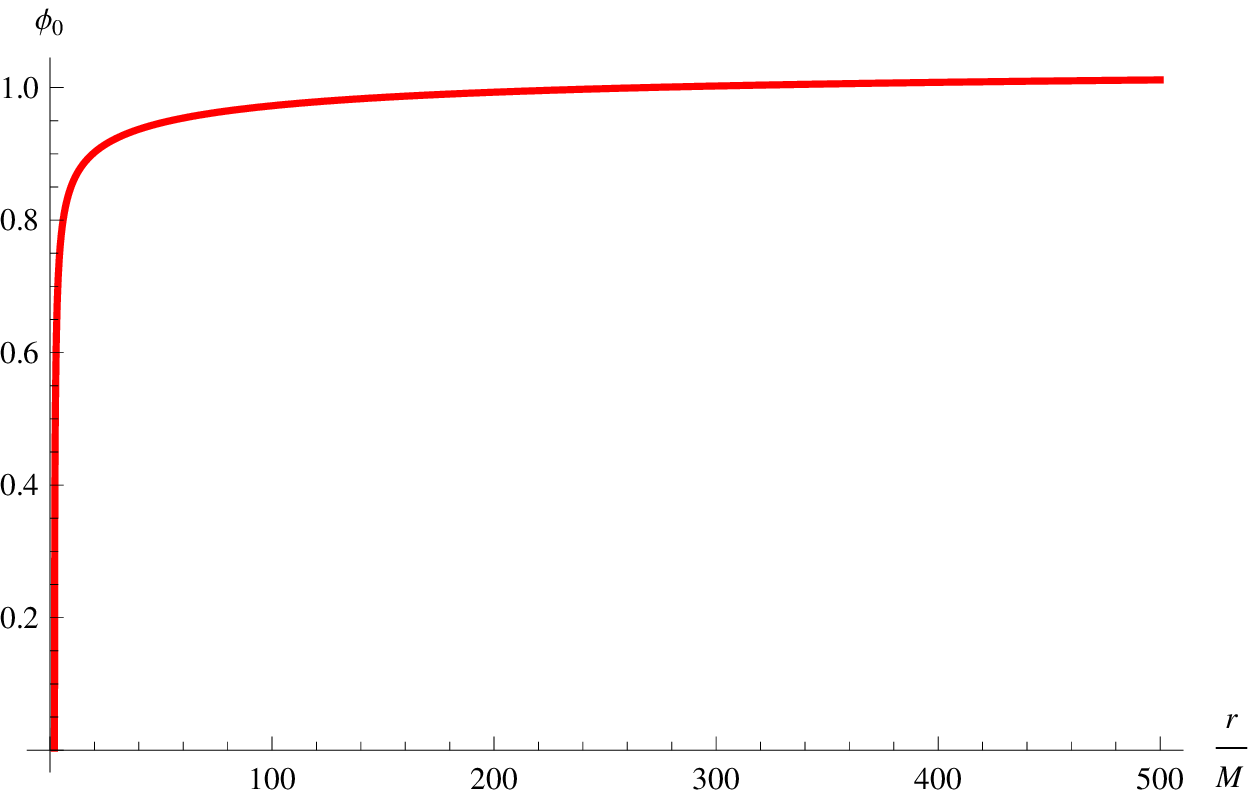}
  \includegraphics[height=4.5cm,angle=0]{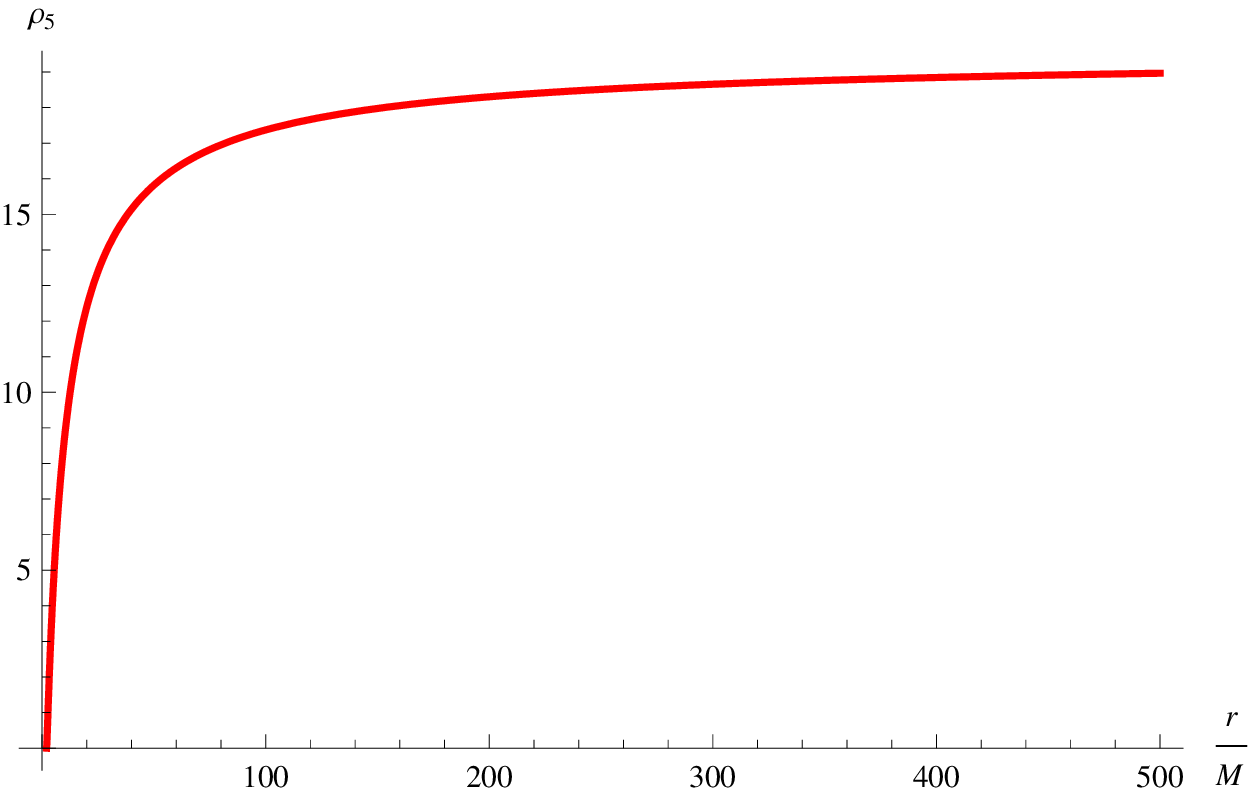}
  \includegraphics[height=4.5cm,angle=0]{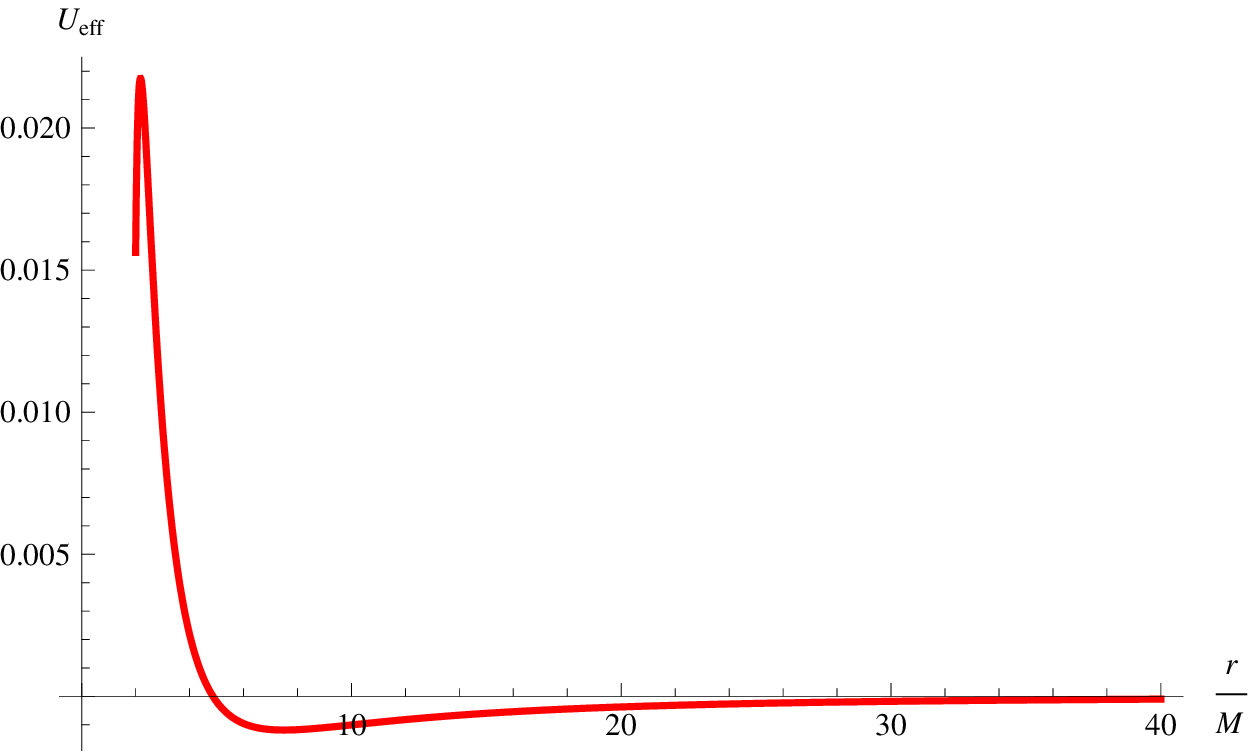}
\caption{
The solution to Eq \eqref{scaeq_g4} for $f_{40}=-50$ is shown
for the $\phi_0'>0$ branch.
The plots for 
$\rho_5$
and
$U_{\rm eff}$ are also shown.
}
\label{figg4}
\end{center}
\end{figure}

In Fig. \ref{figg4},
the solution to Eq. \eqref{scaeq_g4} for $f_{40}=-50$ is shown
for the $\phi_0'(r)>0$ branch.
The plots for 
$\rho_5$
and
$U_{\rm eff}$ are also shown.
We find that 
$\phi_0(r)$ is monotonically increasing and 
approaching a positive finite value,
and hence
there is no violation of hyperbolicity everywhere outside the event horizon.
We also find 
$U_{\rm eff}$ possesses a negative region
in the intermediate length scales
and gradually approaches $0$ at the asymptotic infinity.
Thus, 
the $\phi_0=0$ solution in this model is expected to be unstable,
but
this tachyonic instability is different from 
the case of $\mu^2<0$ in the ordinary scalar-tensor theory \eqref{st}
and
the case of the Einstein-scalar-GB theory \eqref{sgb}.
The negative region of $U_{\rm eff}$ appears for $f_{40}/(\alpha M) \gtrsim -3.78$,
and the depth of it is saturated for a sufficiently negative value of $|f_{40}|/(\alpha M) \gg {\cal O} (100)$.
The end point of the instability may result in a new hairy BH solution in the given model. Since $U_{\rm eff}$ approaches $0$ at the spatial infinity, such an instability would not affect the Minkowski vacuum at the asymptotic infinity. 
The existence of the resultant asymptotically flat hairy BH solutions and their stability will be left for future studies.

\subsubsection{The case of a linear combination of \eqref{g4_sol1} and \eqref{g4_sol2}}
\label{sec432}

Finally, we discuss the case of the linear combination of 
the two solutions \eqref{g4_sol1} and \eqref{g4_sol2}.
Setting $C_1=-C_2 g$,
where $g$ is the constant,
in the vicinity of the event horizon,
we obtain 
\begin{eqnarray}
\phi_0'(r)= C_2
\left(
\frac{1}{2\sqrt{r-2M}}
+ 
\frac{1}{2M}
\left(
g
\pm \frac{M^{3/2}\alpha}{f_{40}}
\right)
+ {\cal O} 
\left( (r-2M)^{1/2}  \right)
\right),
\end{eqnarray}
and 
\begin{eqnarray}
\rho_5
&=&
\frac{\alpha}{f_{40}^2}
\left(
\pm 
 \frac{f_{40} g}{\sqrt{M}}
+M\alpha
\right)
(r-2M)
+
\frac{1}{f_{40}^2}
\left(
-\frac{3f_{40}^2 g}{M^2}
\mp
\frac{f_{40} (g^2+5M)\alpha}{M^{3/2}}
+g \alpha^2
\pm
\frac{2 M^{3/2}\alpha^3}{f_{40}}
\right)
(r-2M)^{3/2}
\nonumber\\
&+&{\cal O}
\left(
(r-2M)^{2}
\right),
\end{eqnarray}
where we require $C_2>0$ and $C_2<0$ for $\phi_0'(r)>0$ and $\phi_0'(r)<0$ branches.
In order for $\phi_0'(r)$ not to change the sign at the intermediate $r>2M$, we require $g>0$.
Hyperbolicity in the vicinity of the event horizon then requires $\rho_5>0$,
which leads to 
\begin{eqnarray}
f_{40}>-\frac{M^{3/2}\alpha}{g},
\qquad 
f_{40}<\frac{M^{3/2}\alpha}{g}.
\end{eqnarray}
In order for Eq. \eqref{scaeq_g4} to be integrated, we require $f_{40}<0$ and $f_{40}>0$
for the $\phi_0'(r)>0$ and $\phi_0'(r)<0$ for the corresponding branches, respectively.
Thus, we have to impose
\begin{eqnarray}
\label{bound}
0>f_{40}>-\frac{M^{3/2}\alpha}{g},
\qquad 
0<f_{40}<\frac{M^{3/2}\alpha}{g}.
\end{eqnarray}
For $g\to \infty$, there is no allowed region for $f_{40}$, which is consistent with our analysis in Sec. \ref{sec421},
while 
for $g\to 0$ we recover the results in Sec. \ref{sec422}.
As long as the bound \eqref{bound} is satisfied, 
the basic properties of the solutions remain the same as those in Sec. \ref{sec422},
and hence we omit to show the numerical results here.

\subsection{Model \ref{model4}}
\label{sec44}

Finally, 
we focus on the leading order part of \eqref{g5_conc}
\begin{eqnarray}
\label{G5}
g_5(X)=\ln (-X),
\quad
f_5(\phi)=
-f_{50}\phi,
\end{eqnarray}
where the ${\cal O} (\epsilon^1)$ part of ${\cal E}_\phi=0$ is given by Eq. \eqref{pert_eq}
with 
\begin{eqnarray}
\rho_1
&=&
\frac{
(-8f_{50}M+12 f_{50}r-r^3\alpha)\phi_0'-8f_{50} r(2M-r)\phi_0''}
      {(2M-r)\phi_0'},
\qquad
\rho_2
=
\frac{(r-2M) (-4f_{50}+ r^2\alpha)}{r},
\nonumber\\
\rho_3
&=&
2
\left(
- \frac{4f_{50} M}{r^2}
+\alpha (r-M)
\right),
\qquad
\rho_4
=
\frac{12f_{50} M^2}{r^4},
\end{eqnarray}
and 
\begin{eqnarray}
\label{r5g5}
\rho_5
&=&
\frac{ \left(r^3\alpha-4f_{50} (3r-2M)\right) \phi_0'
      -8 f_{50}(r-2M) r\phi_0''} 
      {r (-4f_{50}+ r^2\alpha) \phi_0'}.
\end{eqnarray}
In Eq. \eqref{effpot_full},
\begin{eqnarray}
U_{\rm eff}(r)
&=&
(r-2M)
\left[
\frac{2M }{r^4} 
-
\frac{4f_{50}\left(-4f_{50} (3M-2r)+r^2 (3M^2-4Mr+r^2)\alpha\right)}
       {r^5 (-4f_{50}+r^2\alpha)^2}
\right].
\end{eqnarray}
$U_{\rm eff}$ is regular outside the event horizon for
\begin{eqnarray}
\label{f50}
f_{50}<\alpha M^2.
\end{eqnarray} 
Then, since $r^2-4f_{50}/\alpha>r^2-4M^2>0$ for $\alpha>0$, $\rho_2>0$ outside the event horizon.
In order to estimate $\phi_0''/\phi_0'$,
we then investigate the ${\cal O} (\epsilon^0)$ part of ${\cal E}_\phi=0$
\begin{eqnarray}
\label{scaeq_g5}
12 f_{50}M^2 \phi_0
+r^2
\left[
  2 (-4f_{50} M+r^2(r-M)\alpha) \phi_0'
+r(r-2M)
 (-4f_{50}+r^2\alpha)
\phi_0''
\right]
=0.
\end{eqnarray}
The general scalar field solution near the event horizon $r\gtrsim 2M$ is
\begin{eqnarray}
\phi_0(r)
&=&
C_0
\left[
1
+
\frac{3f_{50}}
       {8f_{50} M-8\alpha M^3}
(r-2M)
+
{\cal O}
\left(
 (r-2M)^2
\right)
\right].
\end{eqnarray}
The $\phi_0=0$ solution can be obtained by taking the limit of $C_0\to 0$.
After calculating the derivatives $\phi_0'(r)$ and $\phi_0''(r)$
and then taking the limit of $C_0 \to 0$,
we obtain 
\begin{eqnarray}
\rho_5
=
\frac{-2f_{50}+ M^2\alpha}
       {-f_{50}+M^2\alpha}
+{\cal O}
\left(
r-2M
\right).
\end{eqnarray}
Combined with Eq. \eqref{f50}, hyperbolicity in the vicinity of the event horizon is ensured for
\begin{eqnarray}
\label{hypg5}
f_{50}<\frac{\alpha M^2}{2}.
\end{eqnarray}

\begin{figure}[h]
\unitlength=1.1mm
\begin{center}
  \includegraphics[height=4.5cm,angle=0]{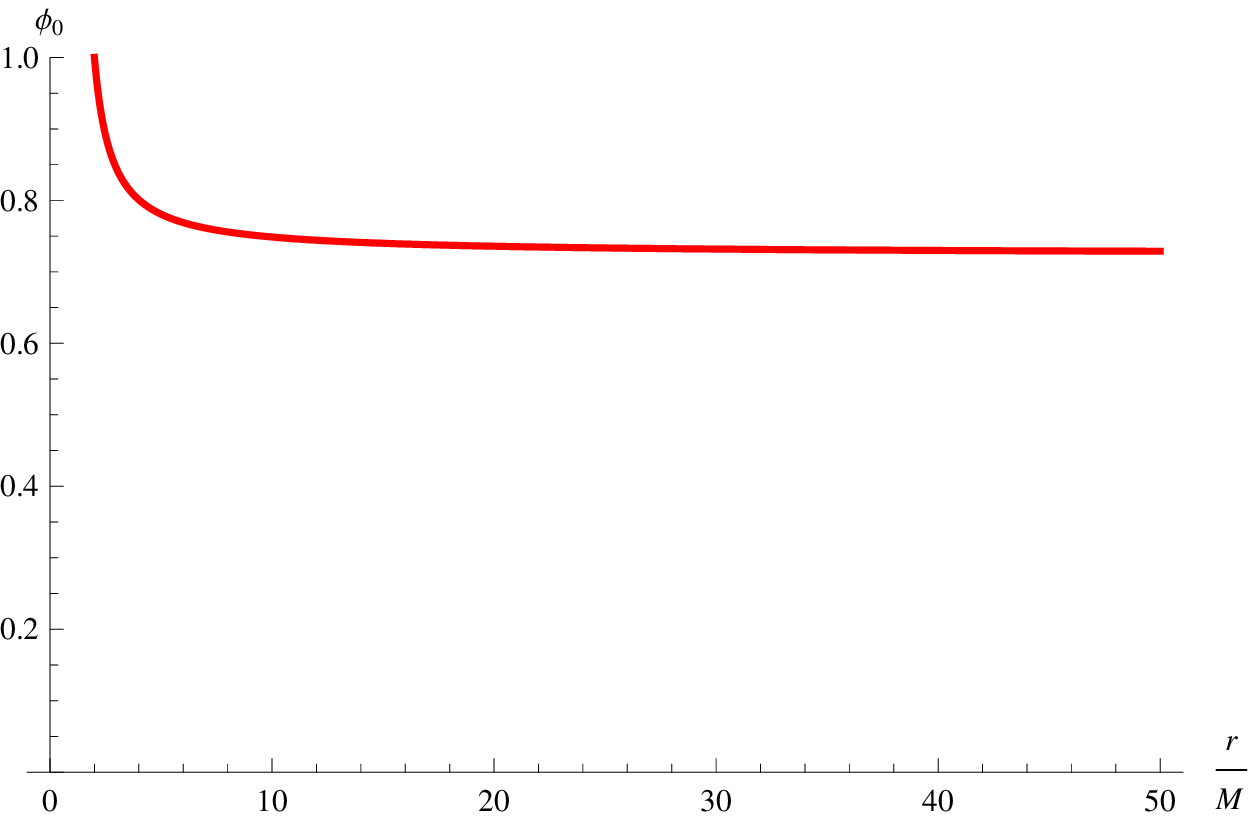}
  \includegraphics[height=4.5cm,angle=0]{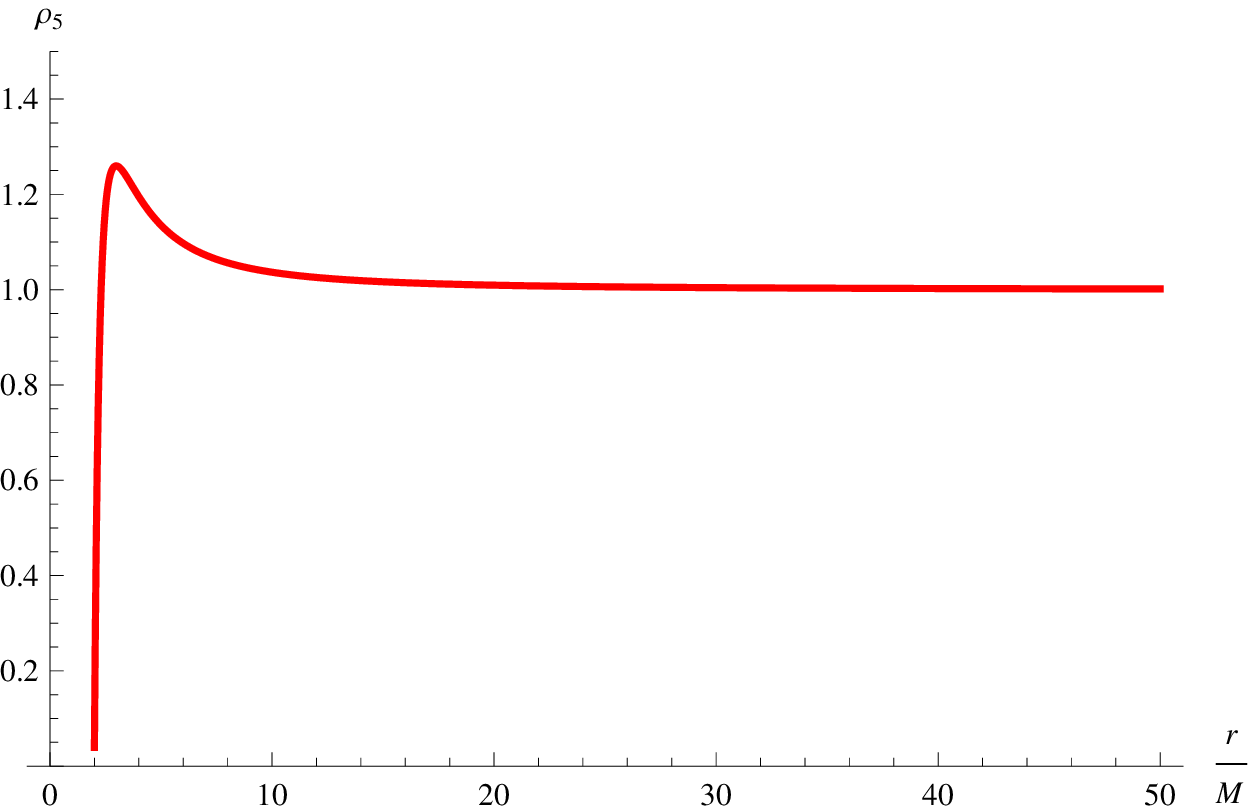}
  \includegraphics[height=4.5cm,angle=0]{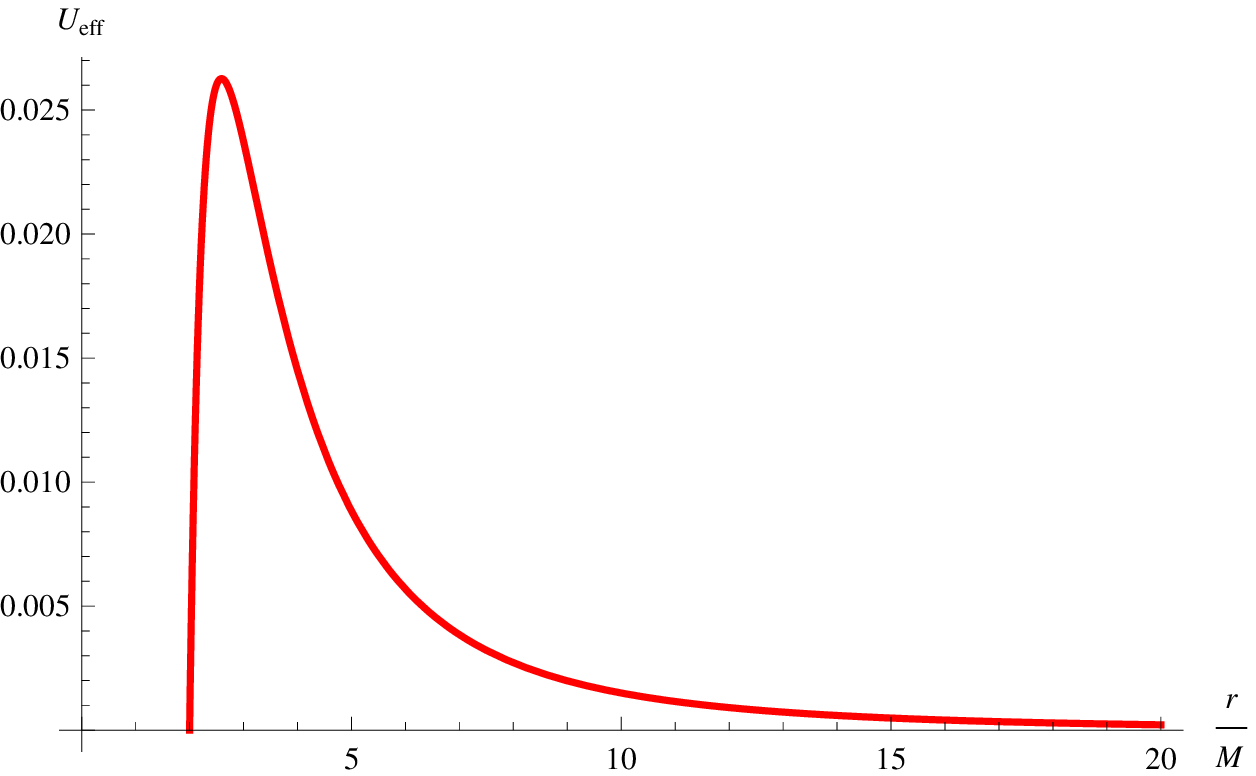}
\caption{
The solution to Eq \eqref{scaeq_g5} for $f_{50}=0.49$ is shown,
which satisfies the bound Eq. \eqref{hypg5}. 
The plots for 
$\rho_5$
and
$U_{\rm eff}$ are also shown.
}
\label{figg5}
\end{center}
\end{figure}

The theory with $\alpha \neq 1$
can be rewritten
in terms of the theory with $\alpha=1$
by the redefinition of $f_{50} \to f_{50}/\alpha$.
Furthermore,  the $M$ dependence can be absorbed 
by the rescalings of $r\to r/M$ and $f_{50}\to f_{50}/M^2$.
Hence, without loss of generality,
for the numerical analysis
we may set $\alpha=1$ and $M=1$.
In Fig. \ref{figg5},
the solution to Eq \eqref{scaeq_g5} for $f_{50}=0.49$
with $C_0=1.0$ 
is shown,
which satisfies Eq. \eqref{hypg5}. 
The plots for 
$\rho_5$
and
$U_{\rm eff}$ are also shown.
We find that 
whenever the bound Eq. \eqref{hypg5} is satisfied,
$U_{\rm eff}$ is always non-negative,
leading to no tachyonic instability in model \ref{model4}. 

\subsection{Discussions}

In this section,
we have clarified 
whether
each individual generalized galileon coupling in the Horndeski theory specified in Sec. \ref{sec3}
could have the $\phi_0=0$ solution
on top of the Schwarzschild spacetime,
and 
whether 
their solution is linearly unstable against the radial perturbation.
We have found that 
no individual class of the Horndeski theory,
except for the generalized quartic coupling model \ref{model3}, 
can realize a tachyonic instability 
without violation of the hyperbolicity,
although the reason is different between models.

Model \ref{model2}
was already excluded from our analysis in Sec. \ref{sec3}.
In model \ref{model1},
the solution to the ${\cal O}(\epsilon^0)$ part of the scalar field equation of motion
does not satisfy
the condition for the existence of the $\phi_0=0$ solution
discussed in Sec. \ref{sec22}.
In model \ref{model4}, 
although the $\phi_0=0$ solution exists on top of the Schwarzschild spacetime,
the effective potential for the radial perturbation 
was always non-negative for parameters without violation of hyperbolicity. 
However, 
we have also found  
that
the behaviors of the background solution and the radial perturbation 
in model \ref{model4} with $G_5\neq 0$ alone
are very similar to the requested one.
In fact, 
the coupling \eqref{G5}
can be regarded 
as the pure $G_5$ part of the Einstein-scalar-GB theory Eq. \eqref{gb}
with $f(\phi)=(f_{50}/8)\phi^2$.

In the next section,
we will consider a model composed of generalized quartic and quintic galileon couplings
which includes the Einstein-scalar-GB theory with the quadratic GB coupling
as a special limit,
and 
we investigate how large deviation from the Einstein-scalar-GB theory 
is allowed
in this class of the model
for a successful realization of a tachyonic instability of the Schwarzschild BH.

On the other hand, 
in model \ref{model3},
the $\phi_0=0$ solution exists on top of the Schwarzschild background,
and 
the effective potential for the perturbation about it  
possesses a negative region 
in the intermediate length scales outside the event horizon
without violation of the hyperbolicity.
This may suggest the existence of a new hairy BH solution
which would not modify the global Minkowski spacetime,
whose construction will be left for future work.

\section{A Model with generalized quartic and quintic galileon couplings}
\label{sec5}

\subsection{Model}

In this section, we consider the model with $G_2=G_3=0$,
and 
\begin{eqnarray}
\label{g4g5}
&&
f_4(\phi)=\beta \eta,
\qquad
g_4(X)=X (2-\ln X),
\nonumber\\
&&
f_5(\phi)=-\eta \phi,
\qquad
g_5=  \ln X,
\end{eqnarray}
where $0<\beta<1$ is a parameter.
$\beta=0$ corresponds to model \ref{model4} discussed
in Sec. \ref{sec34} with the replacement of $f_{50}\to \eta$,
and  
$\beta=1$ is equivalent to the Einstein-scalar-GB theory \eqref{gb} with $f(\phi)=\eta\phi^2/8$.

The ${\cal O} (\epsilon^1)$ part of ${\cal E}_\phi=0$ is then given by Eq. \eqref{pert_eq}
with 
\begin{eqnarray}
\rho_1
&=&
\frac{ \left(-r^3\alpha+4\eta (3r-2M)(1-\beta)\right) \phi_0'
      +8 \eta(r-2M) r(1-\beta)\phi_0''} 
      {(2M-r)\phi_0'},
\quad
\rho_2
=
\frac{(r-2M) (-4\eta(1-\beta)+ r^2\alpha)}{r},
\nonumber\\
\rho_3
&=&
2
\left(
 -\frac{4\eta (1-\beta) M}{r^2}
+\alpha (r-M)
\right),
\quad
\rho_4
=
\frac{12\eta M^2}{r^4},
\end{eqnarray}
and 
\begin{eqnarray}
\label{r5g4g5}
\rho_5
&=&
\frac{ \left(r^3\alpha-4\eta (3r-2M)(1-\beta)\right) \phi_0'
      -8 \eta(r-2M) r(1-\beta)\phi_0''} 
      {r (-4\eta(1-\beta)+ r^2\alpha) \phi_0'},
\end{eqnarray}
and 
the effective potential \eqref{effpot_full}  
\begin{eqnarray}
U_{\rm eff}(r)
&=&
(r-2M)
\left[
\frac{2M }{r^4} 
-
\frac{
4\eta 
\left[
r^2\alpha 
(3M^2
-4M r(1-\beta)
+r^2(1-\beta))
-
4\eta M 
(3M-2r(1-\beta))
(1-\beta)
\right]}
{r^5(r^2\alpha-4\eta(1-\beta))^2}
\right].
\label{ug4g5}
\end{eqnarray}
$U_{\rm eff}$
is regular outside the event horizon for
\begin{eqnarray}
\label{f0}
\eta<\frac{\alpha M^2}{1-\beta}.
\end{eqnarray} 
Then, since $r^2-4\eta(1-\beta)/\alpha>r^2-4M^2>0$ for $\alpha>0$, $\rho_2>0$ outside the event horizon.

We then investigate the ${\cal O} (\epsilon^0)$ part of ${\cal E}_\phi=0$:
\begin{eqnarray}
\label{scaeq_g}
 12 \eta M^2\phi_0
+r^2
\left(
  2 \left(r^2(r-M)\alpha+4 \eta M (-1+\beta)\right)\phi_0'
+r (r-2M) (r^2\alpha+4\eta (-1+\beta))\phi_0''
\right) 
=0.
\end{eqnarray}
The solution $\phi_0(r)$
satisfying the regularity boundary conditions at the event horizon $r=2M$,
discussed in Sec. \ref{sec22},
is given by
\begin{eqnarray}
\phi_0(r)
&=&
C_0
\left[
1
+
\frac{3\eta}
       {8(\eta M-M^3\alpha -\eta M\beta)}
(r-2M)
+
{\cal O}
\left(
 (r-2M)^2
\right)
\right].
\end{eqnarray}
The $\phi_0=0$ solution can be obtained by taking the $C_0 \to 0$ limit. 
After calculating $\phi_0'(r)$ and $\phi_0''(r)$
and taking the $C_0\to 0$ limit,
we find in the vicinity of the event horizon
\begin{eqnarray}
\rho_5
=
\left(
\frac{-2\eta(1-\beta) +M^2\alpha}
      {-\eta(1-\beta)+M^2\alpha}
\right)
+{\cal O}
\left(
 r-2M
\right).
\end{eqnarray}
Hyperbolicity in the vicinity of the event horizon is ensured for
\begin{eqnarray}
\label{hypg}
\eta<\frac{\alpha M^2}{2(1-\beta)}.
\end{eqnarray}

\begin{figure}[h]
\unitlength=1.1mm
\begin{center}
  \includegraphics[height=5.0cm,angle=0]{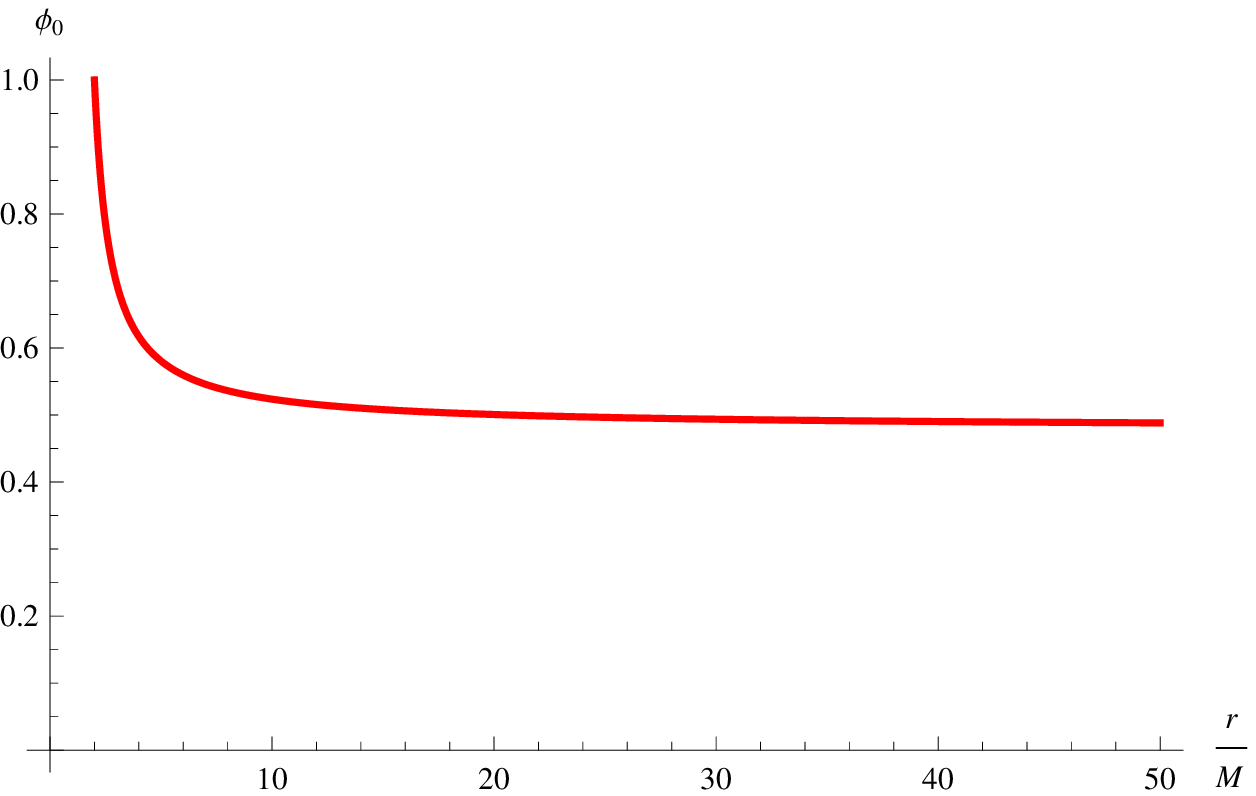}
  \includegraphics[height=5.0cm,angle=0]{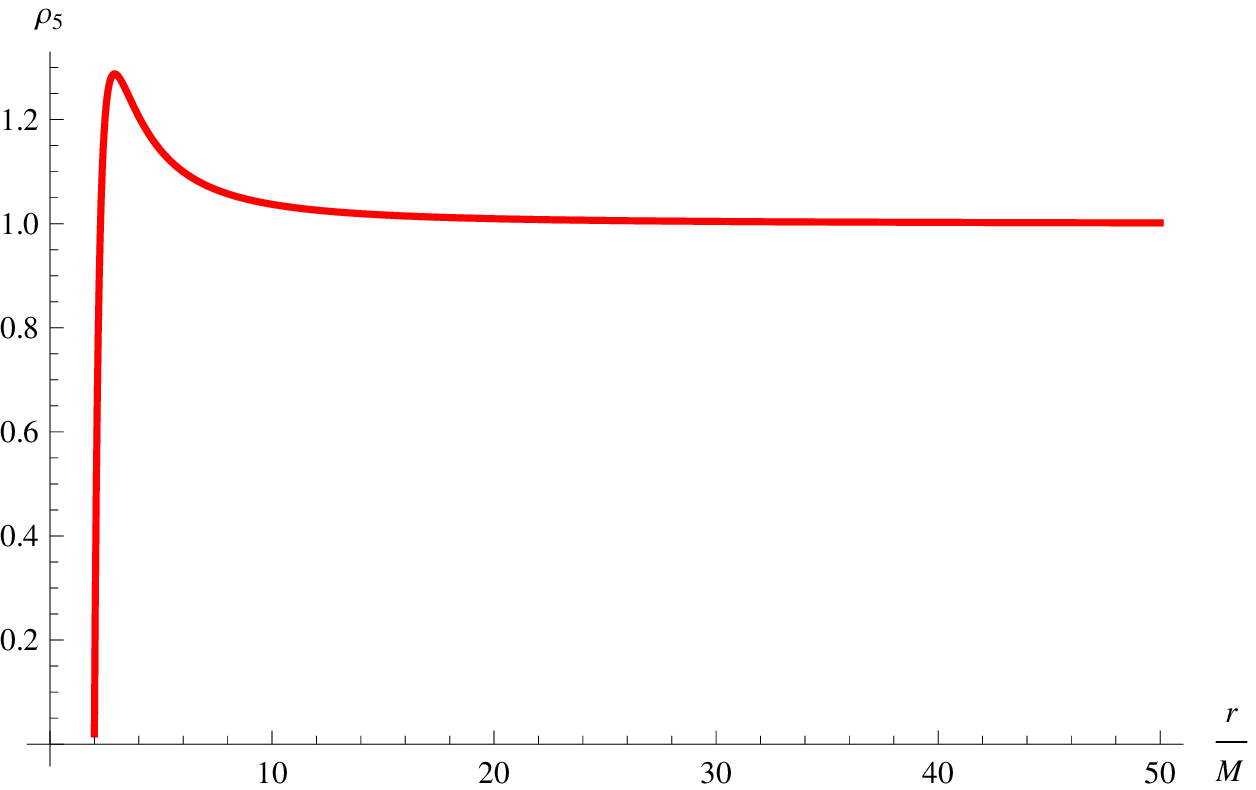}
  \includegraphics[height=5.0cm,angle=0]{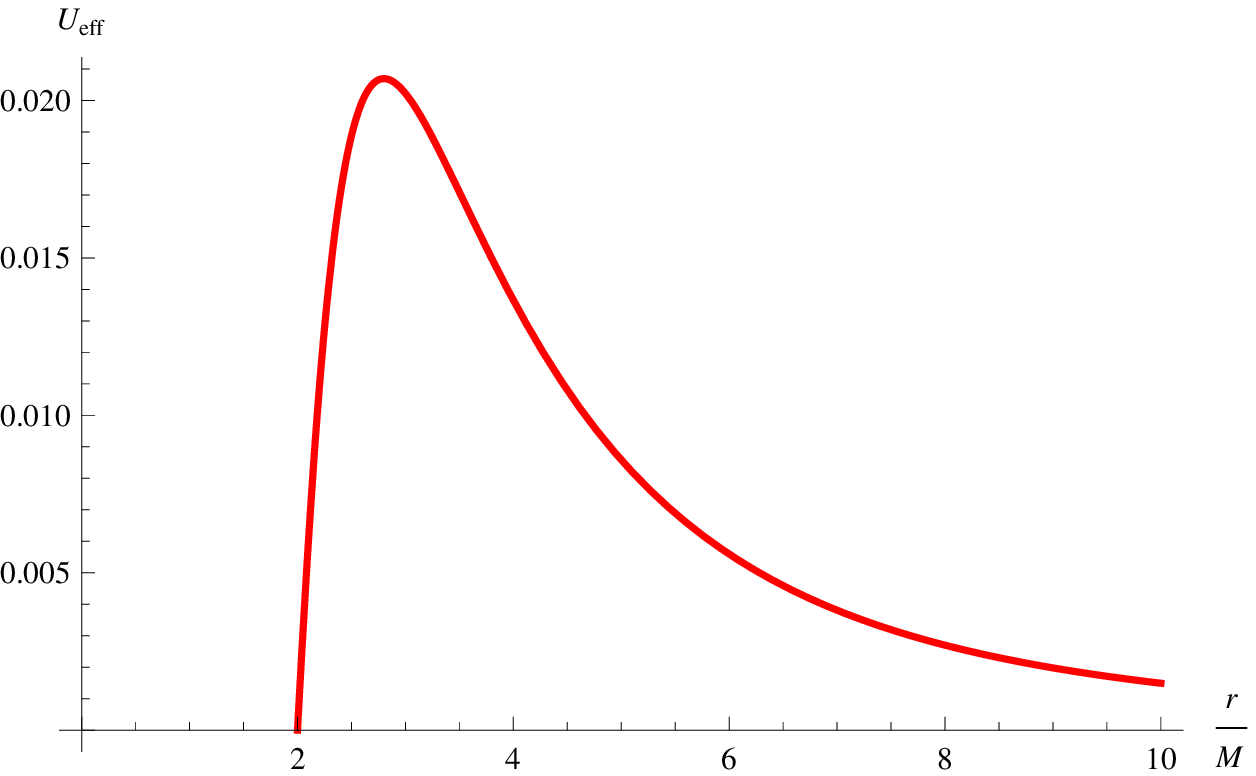}
\caption{
The solution to Eq \eqref{scaeq_g} for $\beta=0.50$ 
and $\eta=0.99/[2(1-\beta)]$
is shown,
which satisfies the bound Eq. \eqref{hypg}. 
The plots for 
$\rho_5$ and
$U_{\rm eff}$ are also shown
for the same parameters.
}
\label{figbeta2}
\end{center}
\end{figure}

\begin{figure}[h]
\unitlength=1.1mm
\begin{center}
  \includegraphics[height=4.5cm,angle=0]{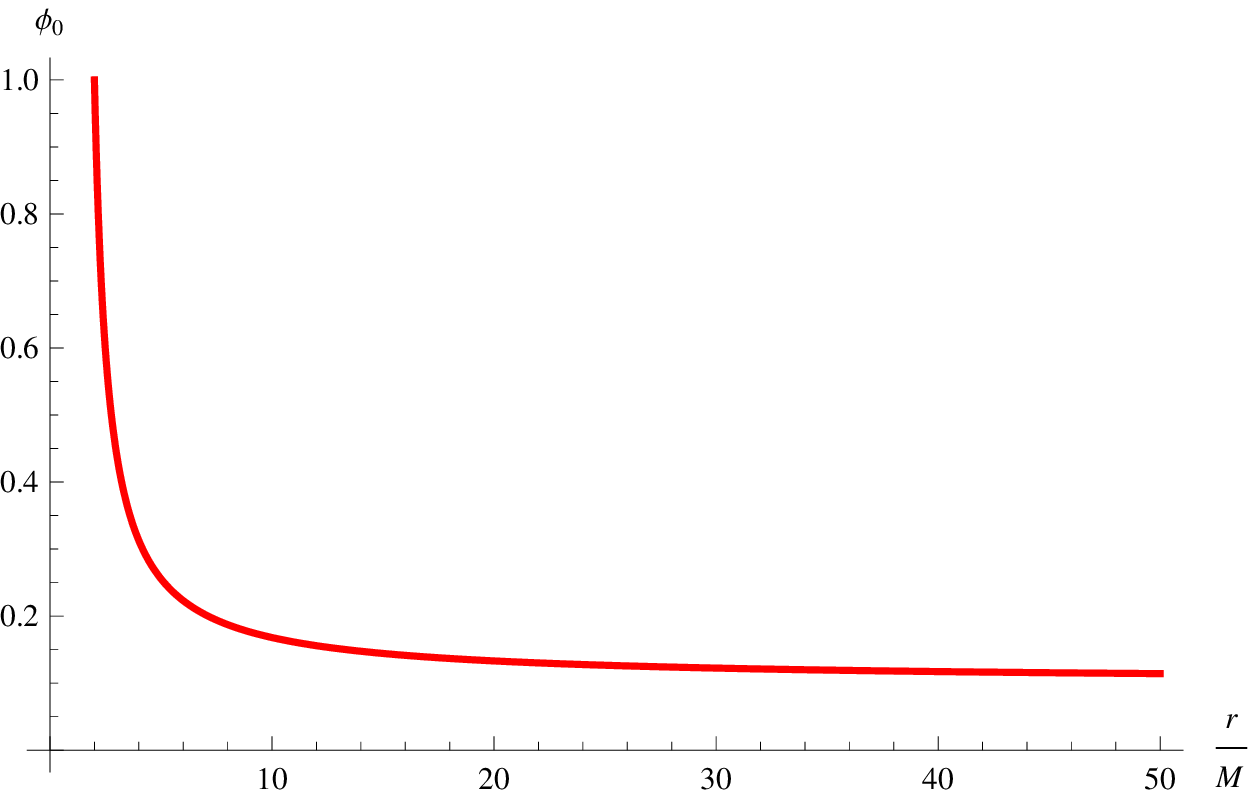}
  \includegraphics[height=4.5cm,angle=0]{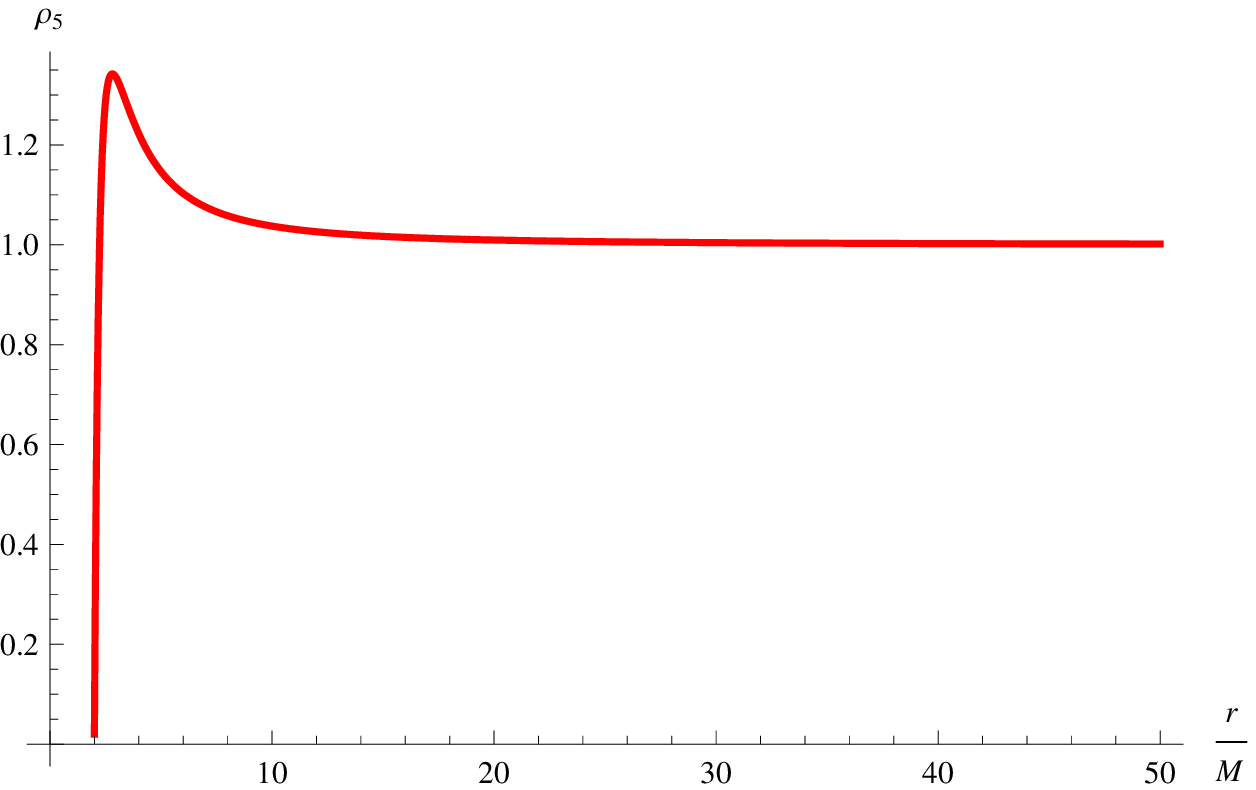}
  \includegraphics[height=4.5cm,angle=0]{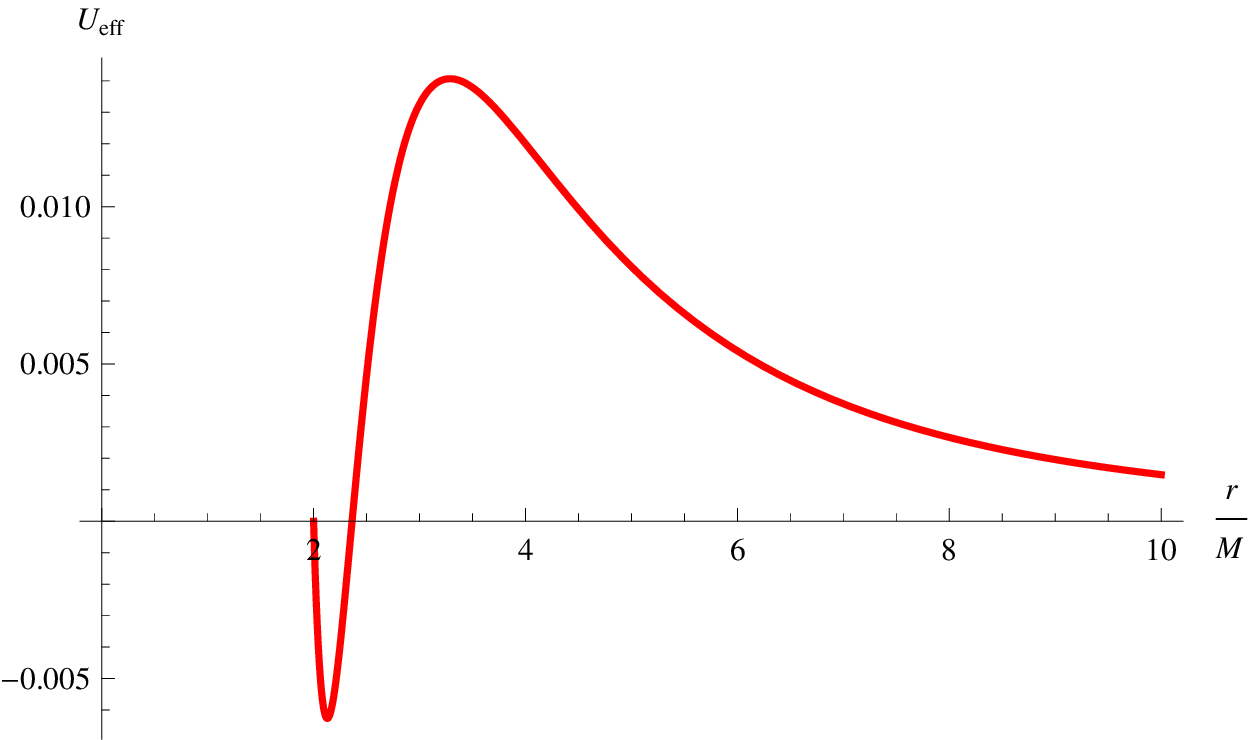}
\caption{
The same plots for $\beta=0.75$ and $\eta=0.99/[2(1-\beta)]$.
}
\label{figbeta3}
\end{center}
\end{figure}

\begin{figure}[h]
\unitlength=1.1mm
\begin{center}
  \includegraphics[height=4.5cm,angle=0]{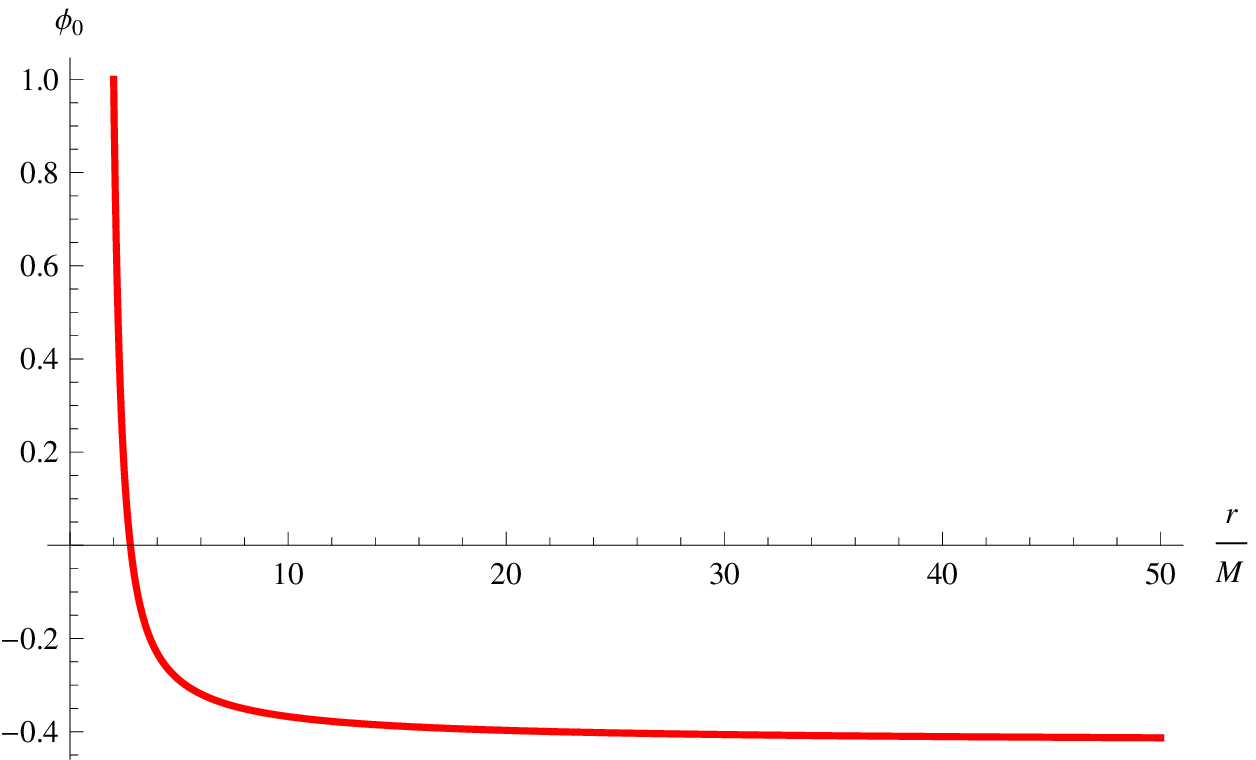}
  \includegraphics[height=4.5cm,angle=0]{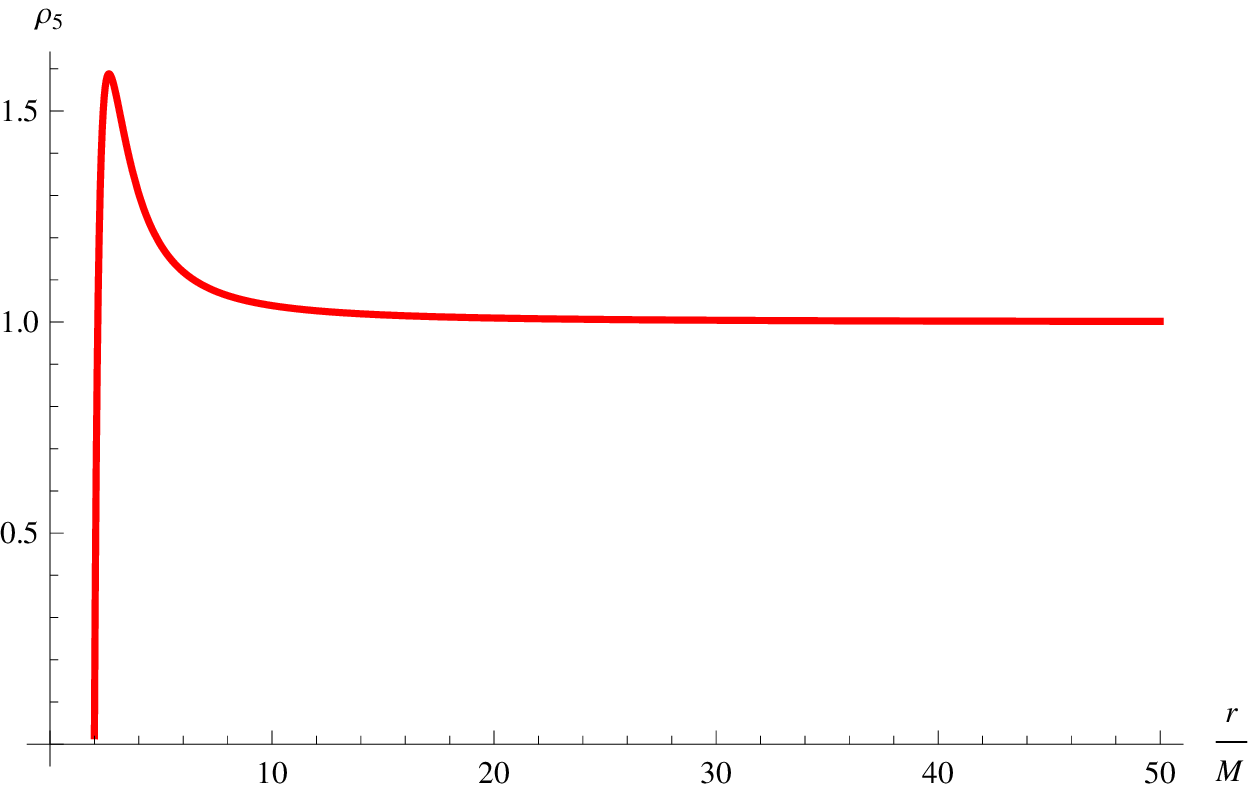}
  \includegraphics[height=4.5cm,angle=0]{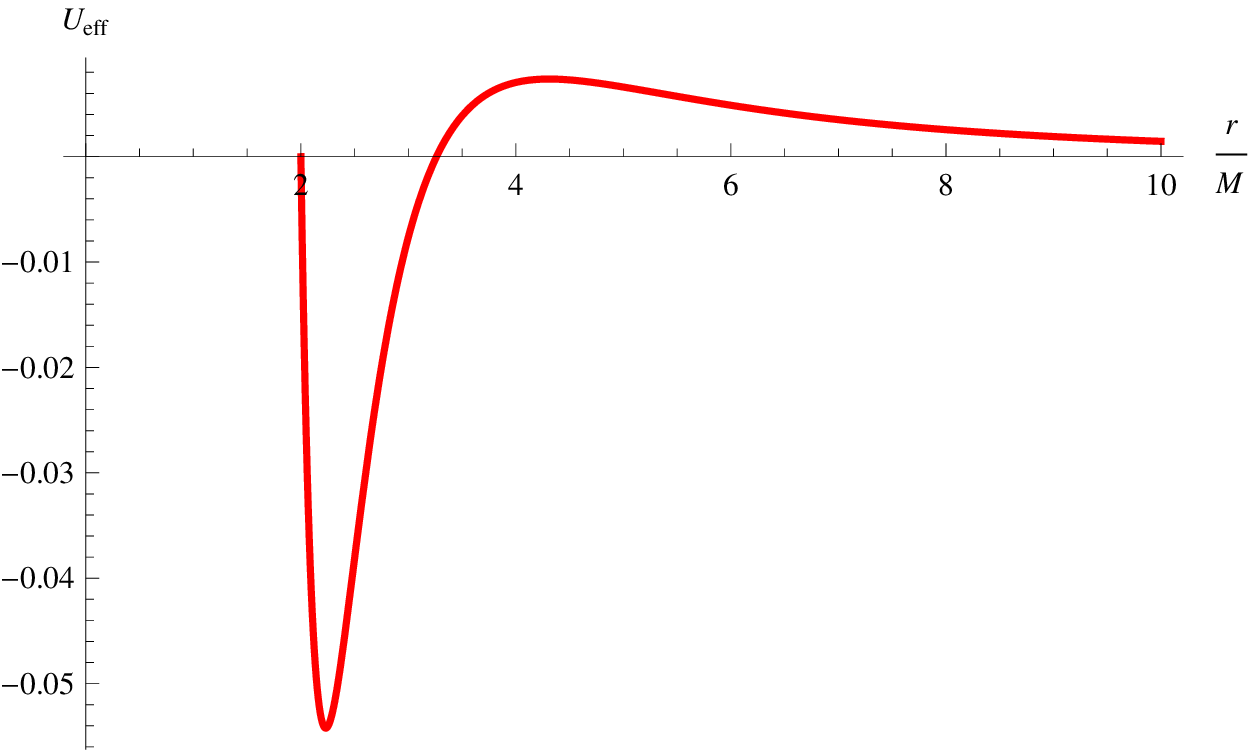}
\caption{
The same plots for $\beta=0.90$ and $\eta=0.99/[2(1-\beta)]$.
}
\label{figbeta4}
\end{center}
\end{figure}

\begin{figure}[h]
\unitlength=1.1mm
\begin{center}
  \includegraphics[height=4.5cm,angle=0]{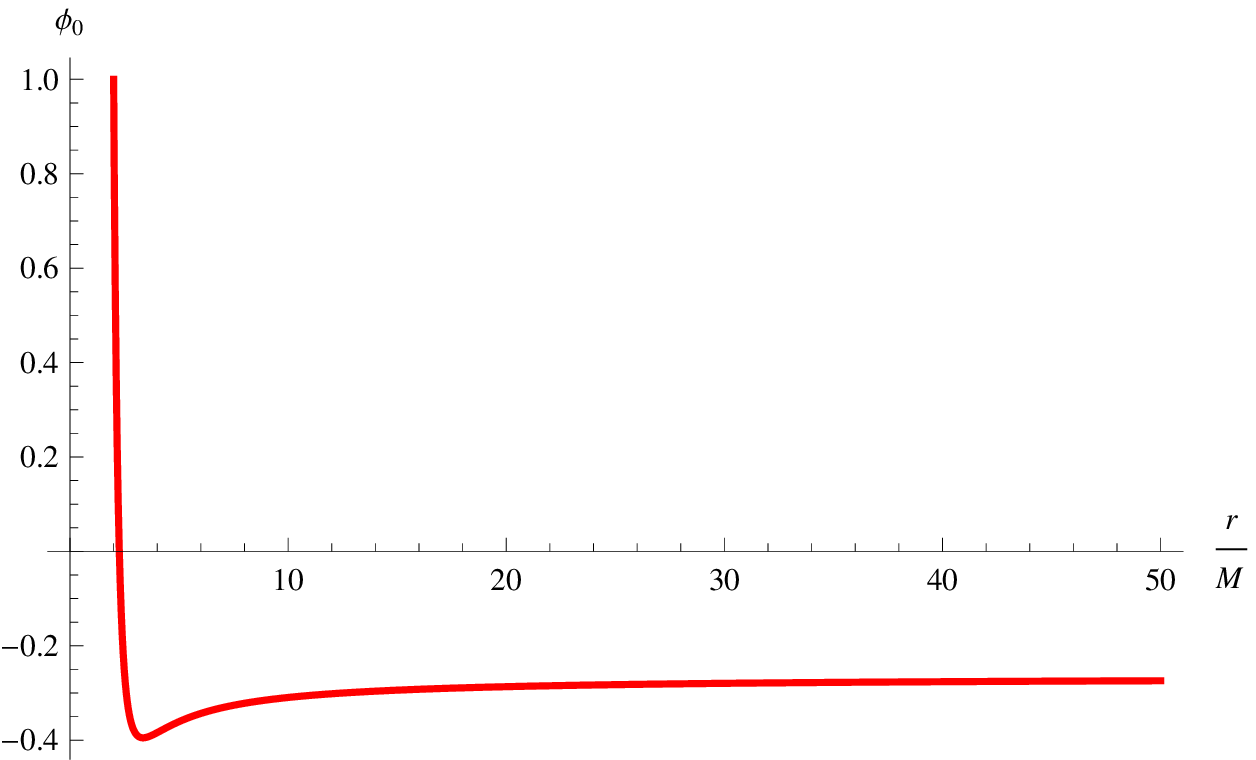}
  \includegraphics[height=4.5cm,angle=0]{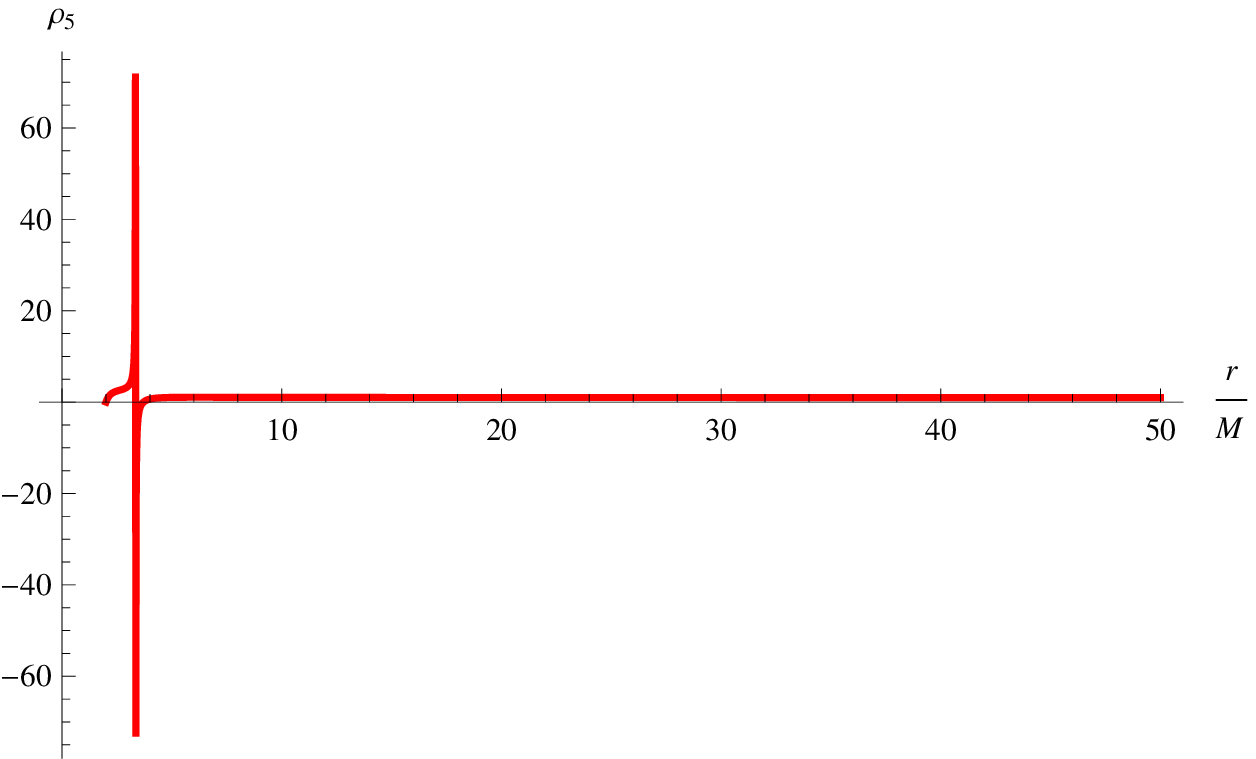}
  \includegraphics[height=4.5cm,angle=0]{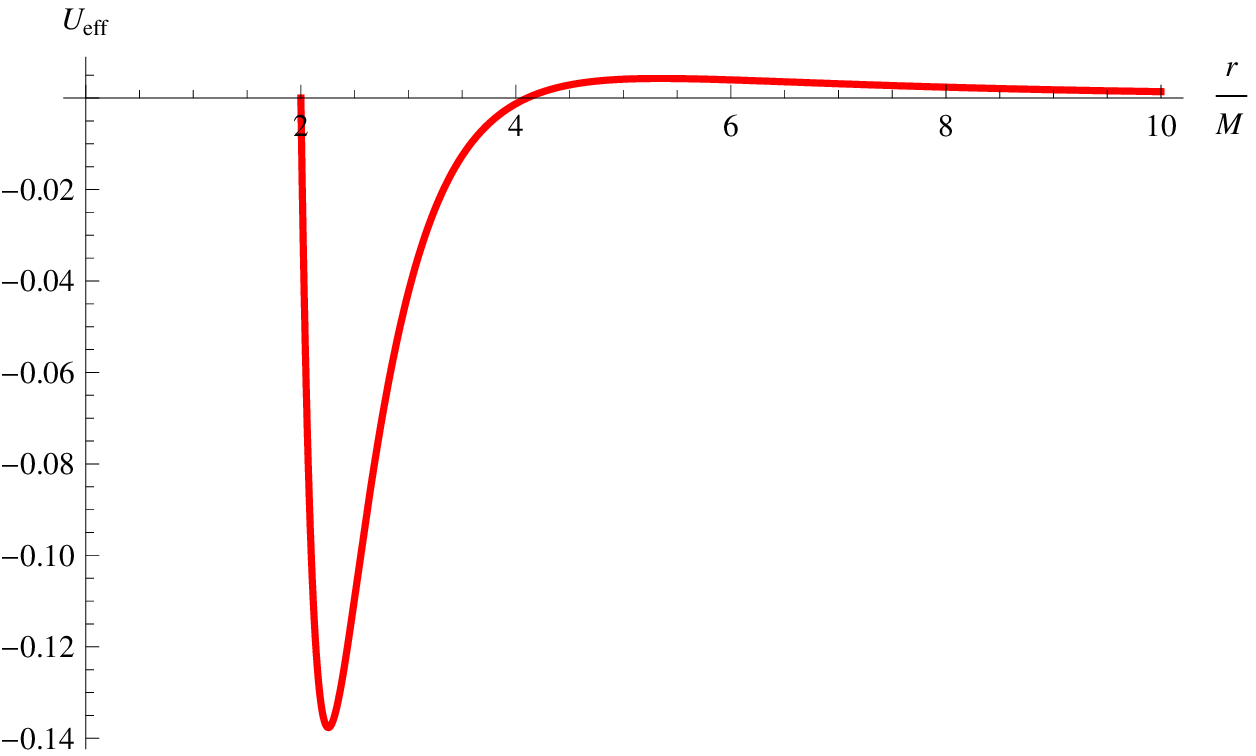}
\caption{
The same plots for $\beta=0.95$ and $\eta=0.99/[2(1-\beta)]$.
}
\label{figbeta5}
\end{center}
\end{figure}

The theory with $\alpha \neq 1$
can be rewritten
in terms of the theory with $\alpha=1$
by the redefinition of $\eta \to \eta/\alpha$.
Furthermore, the $M$ dependence can be eliminated
by the rescalings of $r\to r/M$ and $\eta\to \eta/M^2$.
Hence without loss of generality,
for the numerical analysis we may set $\alpha=1$ and $M=1$.
We also set $C_0=1$.
In Figs. \ref{figbeta2}-\ref{figbeta5},
the solution for $\phi_0$ is shown for $\beta=0.5, 0.75, 0.90, 0.95$ and $\eta=0.99/[2(1-\beta)]$
(for a given $\beta$),
respectively,
all of which satisfy the bound Eq. \eqref{hypg}.
For $\beta=0.95$,
$\rho_5$ blows up in the vicinity of the event horizon,
as $\phi_0'$ vanishes there before taking the limit of $C_0\to 0$. 
Thus,
in this region 
there is violation of hyperbolicity.
For all the other cases, $\beta=0.5, 0.75, 0.90$,
$\phi_0(r)$ is monotonically decreasing, 
and hence $\phi_0'(r)$ does not reach zero
and $\rho_5$ is positive definite everywhere outside the event horizon.
As $\beta$ increases,
the effective potential $U_{\rm eff}(r)$ defined in Eq. \eqref{ug4g5}
develops a negative region in the vicinity of the event horizon,
leading to a tachyonic instability.
In the next subsection,
we will analyze the stability of the model.

\subsection{Tachyonic instability}
\label{sec52}

In order to investigate the existence of the bound state with the pure imaginary frequency,
we will employ the $S$-deformation method.
Multiplying $\Psi^\ast$ on Eq. \eqref{sch2} and integrating from the event horizon to the infinity of the Schwarzschild spacetime,
we obtain 
\begin{eqnarray}
\label{integral}
\int_{-\infty}^\infty
dr_\ast
\Psi^\ast
\left[
-\frac{d^2}{dr_\ast^2}
+U_{\rm eff}
\right]
\Psi
=\omega^2
\int_{-\infty}^\infty
dr_\ast
\rho_5
 |\Psi|^2,
\end{eqnarray}
where we have assumed that the mode function $\Psi$
is square integrable.
By introducing an arbitrary function $S$,
Eq. \eqref{integral}
can be rewritten as 
\begin{eqnarray}
-
\left[
\Psi^\ast \frac{d}{dr_\ast}\Psi
+S |\Psi|^2
\right]^{\infty}_{-\infty}
+
\int_{-\infty}^\infty
dr_\ast
\left[
\left|
 \frac{d\Psi}{dr_\ast}
+S\Psi
\right|^2
+
\left(
U_{\rm eff}
+\frac{dS}{dr_\ast}
-S^2
\right)
|\Psi|^2
\right]
=\omega^2
\int_{-\infty}^\infty
dr_\ast
\rho_5
 |\Psi|^2.
\end{eqnarray}
Assuming the boundary conditions for which the first boundary terms vanish,
$\omega^2$ is bounded from below if 
there exists an $S$ satisfying 
\begin{eqnarray}
U_{\rm eff}
+\frac{dS}{dr_\ast}
-S^2
\geq 0.
\end{eqnarray}
A method to analyze the stability of BH was presented in Ref. \cite{Kimura:2018eiv},
which states that
if the first order differential equation,
\begin{eqnarray}
\label{s_eq}
U_{\rm eff}
+\frac{dS}{dr_\ast}
-S^2
= 0,
\end{eqnarray}
admits the regular solution for $S$,
there exists no eigenmode with $\omega^2<0$.
Introducing $\Psi_0$ by $S=-(d\Psi_0/dr_\ast)/\Psi_0$, 
Eq. \eqref{s_eq} reduces to 
\begin{eqnarray}
\left[
-\frac{d^2}{dr_\ast^2}
+U_{\rm eff}
\right]
\Psi_0
=0,
\end{eqnarray}
and hence $\Psi_0$ corresponds to the eigenmode function of the zero energy state $\omega=0$.
If Eq. \eqref{s_eq} admits only regular solutions for $S$,
$\Psi_0$ never crosses zero 
and 
hence the zero energy state $\omega=0$ has to be the lowest mode.
Thus, 
the existence of $S$ which never diverges
provides a direct proof for stability
against the given type of perturbations.
On the other hand, 
if $S$ diverges at some point,
it indicates the existence of nodes in the eigenmode function of the zero energy state
and hence the existence of the modes with $\omega^2<0$.
Since in this paper we will not explicitly 
investigate the eigenvalues of $\omega^2<0$ and the corresponding eigenmode functions,
the divergence of $S$ is not a direct evidence of the tachyonic instability. 
Nevertheless, 
in the case of the Einstein-scalar-GB theory,
the existence of the regular $S$
can be ensured for $\eta/(\alpha M^2)<2.903$,
which corresponds to the bifurcation point 
of hairy BH solutions with nodeless nontrivial profiles of the scalar field
from the Schwarzschild solution
with $\phi=0$
\cite{Blazquez-Salcedo:2018jnn,Minamitsuji:2018xde,Silva:2018qhn}.

\begin{figure}[h]
\unitlength=1.1mm
\begin{center}
  \includegraphics[height=6.5cm,angle=0]{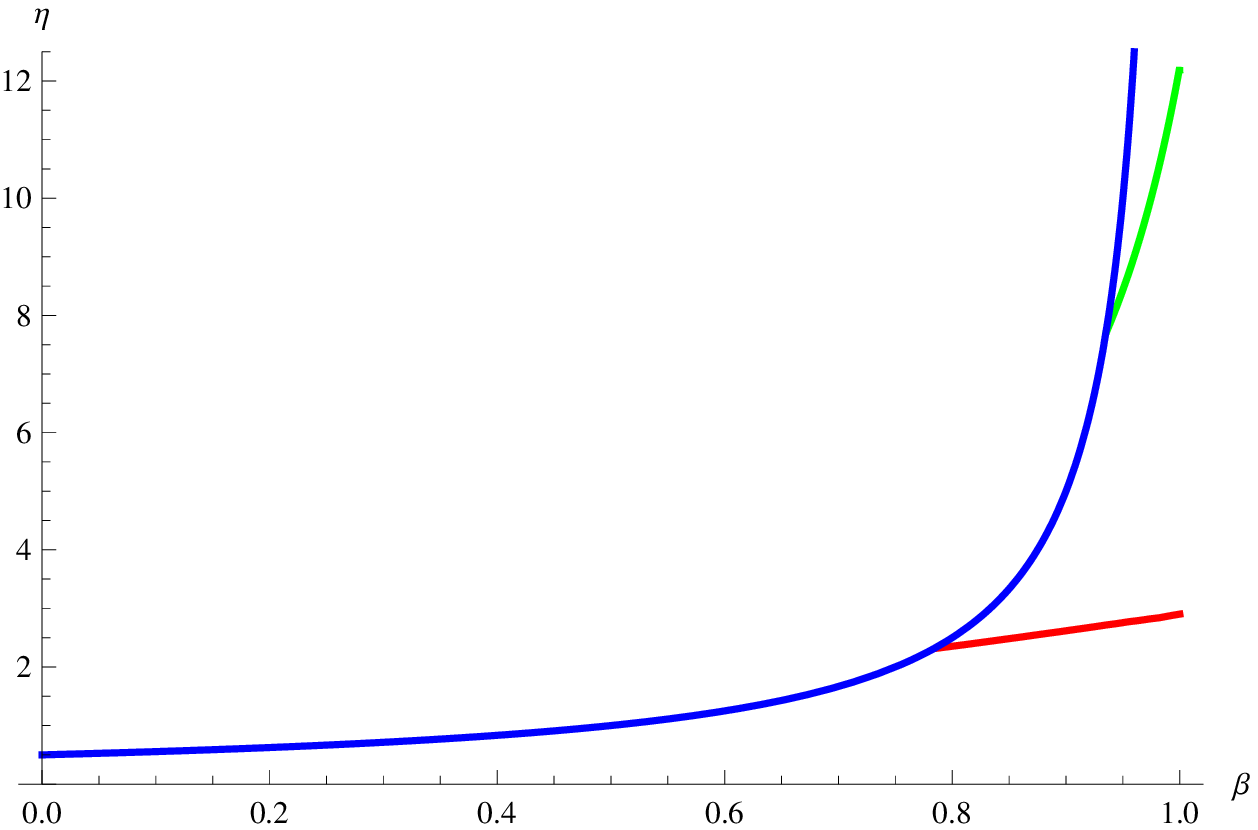}
\caption{
By setting $M=1$ and $\alpha=1$,
in the $(\beta,\eta)$-plane
the critical curve below which the regular solution $S$ for Eq. \eqref{s_eq} exists
is shown in red.
The blue curve corresponds to 
the values of $\eta$ given by Eq. \eqref{hypg},
above which hyperbolicity is broken in the vicinity of the event horizon.
The green curve represents the critical value of $\eta$,
above which
hyperbolicity is broken at a finite radius outside the event horizon where $\phi_0'(r)=0$,
even if it is satisfied in the vicinity of the event horizon (see Fig. \ref{figbeta5} as an example).
For the solutions located in the right region surrounded by these red, blue and green curves,
the radial perturbation about the Schwarzschild solution
satisfies the hyperbolicity,
while the stability of the Schwarzschild solution is not ensured,
since there is no regular $S$ satisfying Eq. \eqref{s_eq}
and $U_{\rm eff}$ would accommodate negative energy eigenmodes with $\omega^2<0$.
}
\label{figcrit}
\end{center}
\end{figure}

In Fig. \ref{figcrit},
by setting $M=1$ and $\alpha=1$,
in the $(\beta,\eta)$-plane
the critical curve below which the regular solution $S$ for Eq. \eqref{s_eq} exists
is shown in red.
The blue curve corresponds to 
the values of $\eta$ given by Eq. \eqref{hypg},
above which hyperbolicity is broken in the vicinity of the event horizon.
The green curve represents the critical value of $\eta$,
above which
hyperbolicity is broken at a finite radius outside the event horizon where $\phi_0'(r)=0$,
even if it is satisfied in the vicinity of the event horizon (see Fig. \ref{figbeta5} as an example).
For the solutions located in the right region surrounded by these red, blue and green curves,
the radial perturbation about the Schwarzschild solution
satisfies the hyperbolicity,
while the stability of the Schwarzschild solution is not ensured,
since there is no regular $S$ satisfying Eq. \eqref{s_eq}
and $U_{\rm eff}$ would accommodate negative energy eigenmodes with $\omega^2<0$.
We note
that the right end of Fig. \ref{figcrit} with $\beta=1$
corresponds to the case of the Einstein-scalar-GB theory,
and the intersection with the red curve is given by $2.903$,
which agrees with the value
obtained in Refs. \cite{Doneva:2017bvd,Silva:2017uqg,Blazquez-Salcedo:2018jnn,Minamitsuji:2018xde,Silva:2018qhn}.
Also,
we note that 
the right edge of the green curve
does not include the case of $\beta=1$,
for which always $\rho_5=1$ from Eq. \eqref{r5g4g5}.

Since the effective potential for the radial perturbation is very similar 
to that in the Einstein-scalar-GB theory of 
Refs. \cite{Doneva:2017bvd,Silva:2017uqg,Blazquez-Salcedo:2018jnn,Minamitsuji:2018xde,Silva:2018qhn},
we expect that
the end point of the tachyonic instability
is also an asymptotically flat scalarized BH
which is very similar to that in the Einstein-scalar-GB theory.
The numerical construction of scalarized BH solutions will be left for future work.

\section{Conclusions}
\label{sec6}

In this paper, 
we have investigated the possibility of spontaneous scalarization of 
static, spherically symmetric, and asymptotically flat BH solutions in the Horndeski theory.
Our studies
extended the previous analysis about the Einstein-scalar-GB theory
to the other classes of the Horndeski theory.

First, 
we have clarified the conditions that 
generalized galileon couplings in the Horndeski theory
could allow the constant scalar field solution on top of the Schwarzschild spacetime.
Without loss of generality, 
after some appropriate shift we could always set the constant scalar field to be $\phi=0$.  
For the coupling functions in the Horndeski theory which are regular at $X=0$,
where $X=-(1/2)g^{\mu\nu}\partial_\mu \phi \partial_\nu\phi$ is the ordinary kinetic term, 
all classes could possess the $\phi=0$ solution on top of the Schwarzschild spacetime,
but at the same time
their contribution to the radial perturbation automatically vanished.
On the other hand,
if the coupling functions are too singular at $X=0$,
no $\phi=0$ solution exists on top of the Schwarzschild spacetime.

In Sec. \ref{sec3},
in order for the $\phi=0$ solution to exist,
we have required that in all the components of the background equations of motion,
each contribution has to be regular 
in the simultaneous limits of $\phi\to 0$, $\phi_{,\tr}\to 0$, and $\phi_{,\tr\tr}\to 0$,
where $\tr$ is the proper length in the radial direction defined in Eq. \eqref{propr}.
However, 
since in general
the speed of the convergence of $\phi$, $\phi_{,\tr}$, and $\phi_{,\tr\tr}$ to $0$
is not known
unless a model is specified and the scalar field equation of motion is solved,
we have adopted the sufficient condition
that each contribution to the background equations of motion does not contain any inverse power of $\phi$, $\phi_{,\tr}$, and $\phi_{,\tr\tr}$.
We have also excluded the term which involves some power of $\ln \phi$ or $\ln \phi_{,\tr}$,
unless a positive power of $\phi$ or $\phi_{,\tr}$ is multiplied to the logarithmic term, respectively.
For each individual galileon coupling,
we have chosen the model with the minimal leading power of the galileon coupling function 
satisfying the above conditions.
The concrete models were given by Eqs. \eqref{g2_conc}, \eqref{g4_conc}, and \eqref{g5_conc}.

In Sec. \ref{sec4}, we have further investigated each model obtained in Sec. \ref{sec3}.
We have found 
that in model \eqref{g4_conc},
there was the $\phi=0$ solution on top of the Schwarzschild spacetime,
and the effective potential for the radial perturbation possesses a negative region in the intermediate length scales, leading to a tachyonic instability which does not affect the global Minkowski vacuum.
On the other hand,
for the models \eqref{g2_conc} and \eqref{g5_conc}, 
even if they allow for the $\phi=0$ solution on top of the Schwarzschild spacetime,
the radial perturbation 
was not suitable for spontaneous scalarization,
because of violation of hyperbolicity or no negative region in the effective potential.
Thus,
we have concluded 
that,
except for the model with the generalized quartic galileon coupling,
each individual galileon coupling could not realize a tachyonic instability of a Schwarzschild solution by itself.
We have also found that 
the behaviors of the model with the generalized quintic galileon coupling alone
are very similar to those in the case of the Einstein-scalar-GB model,
even if the effective potential could not possess the negative region
without violation of the hyperbolicity. The analysis  including the metric perturbations is left for future work.

In Sec. \ref{sec5},
we have investigated the model composed of generalized quartic and quintic couplings given by Eq. \eqref{g4g5},
which includes the Einstein-scalar-GB theory with the quadratic coupling as the special case.
We have shown
as one increases the relative contribution of the quartic coupling term in Eq. \eqref{g4g5},
the effective potential for the radial perturbation 
develops a negative region,
which could accommodate one or more states with pure imaginary frequencies.
In the two-dimensional parameter space,
we have clarified the region (1) where the hyperbolicity is preserved,
and using the $S$-deformation method,
the region (2) where the linear stability against the radial perturbation is ensured.
In the region inside the region (1) but outside the region (2),
the linear stability of the Schwarzschild solution against the radial perturbation is not ensured,
indicating the appearance of a tachyonic instability.
It implies that the theory in the region realizes spontaneous scalarization of a BH.
In the limit of the Einstein-scalar-GB theory,
the boundary of the region (2) coincides with
the value of the critical coupling constant
where the branch of scalarized hairy BHs is bifurcated from that of 
the Schwarzschild solution with the constant scalar field in the Einstein-scalar-GB theory.

One thing which we have neglected in our analysis is the coupling of the scalar field perturbation to the metric perturbations.
It would be necessary to clarify whether
the analysis including the metric perturbations could modify the results obtained in this paper or not. 
As we mentioned in Sec. \ref{sec22},
on the background of the Schwarzschild spacetime and the constant scalar field, i.e., $\phi_0'=0$,
the master equation for the radial perturbation agreed with the equation of the scalar field perturbation without the metric perturbations in the ordinary scalar-tensor and Einstein-scalar-GB theories \cite{Minamitsuji:2018xde}. In our analysis,
although we have finally taken the limit to the  $\phi_0=0$ solution, at the intermediate step to estimate the ratios $\phi_0''/\phi_0'$ and $\phi_0/\phi_0'$,
we have employed the general solution of the equation of $\phi_0(r)$. 
Thus, before the limit to the $\phi_0=0$ solution is taken,
the background scalar field $\phi_0(r)$ has a nontrivial profile, i.e., $\phi_0'(r)\neq 0$, and hence there
would be nontrivial couplings of the scalar field perturbation to the metric perturbations, if the metric perturbations are taken into consideration from the beginning. 
Although we expect that these couplings would vanish or be subleading in the limit to the $\phi_0=0$ solution,
we should explicitly confirm this by including the metric perturbations in our analysis, which would be left for the future studies.

Before closing this paper,  
we would like to mention the recent works
that studied the compatibility of
the conditions for spontaneous scalarization with cosmology
in the context of the Einstein-scalar-GB theory.
The authors of Ref. \cite{Anson:2019uto} argued 
that 
the scalar field $\phi$ in the Einstein-scalar-GB theory with the quadratic coupling 
exhibits a catastrophic instability during inflation
for the value of the coupling constant relevant for spontaneous scalarization of a BH,
by assuming that $\phi$ is produced quantum mechanically.
The authors of Ref. \cite{Franchini:2019npi} 
investigated
whether the scalar field exhibiting spontaneous scalarization of a BH 
is subdominant in the late-time cosmology,
compatible with the recent observational constraints from the measurements of GWs \cite{Monitor:2017mdv},
and 
argued 
that a mild tuning of initial conditions
is necessary.
The same issue may exist also for the Horndeski theory discussed in this paper.
Since the purpose of our study was the classification of scalar-tensor theories which are largely different from GR and the theories discussed in this paper were derived in the context of BH physics, however, such theories may not be relevant on the cosmological scales.
Their implications to cosmology in the early- and late-time universe would be left for future studies.

There will also be several extensions of the present work.
One of them is 
to construct the explicit hairy BH solutions, which may be the end point of the tachyonic instability.
It will also be interesting 
to extend the present analysis
to the more general scalar-tensor theories
such as GLPV \cite{Gleyzes:2014dya,Gleyzes:2014qga} and DHOST theories \cite{Langlois:2015cwa,BenAchour:2016fzp}.
We hope to come back to these issues
in future publications.

\acknowledgments{
We thank Masashi Kimura and Thomas Sotiriou for useful discussions.
M.M.\ was supported by the research grant under ``Norma Transit\'oria do DL 57/2016''.
T.~I. acknowledges financial support provided under the European Union's H2020 ERC Consolidator Grant ``Matter and strong- field gravity: New frontiers in Einstein's theory'' grant agreement no. MaGRaTh-646597, and under the H2020-MSCA-RISE-2015 Grant No. StronGrHEP-690904.  
}

\bibliography{ref-BH}

\begin{thebibliography}{42}%
\makeatletter
\providecommand \@ifxundefined [1]{%
 \@ifx{#1\undefined}
}%
\providecommand \@ifnum [1]{%
 \ifnum #1\expandafter \@firstoftwo
 \else \expandafter \@secondoftwo
 \fi
}%
\providecommand \@ifx [1]{%
 \ifx #1\expandafter \@firstoftwo
 \else \expandafter \@secondoftwo
 \fi
}%
\providecommand \natexlab [1]{#1}%
\providecommand \enquote  [1]{``#1''}%
\providecommand \bibnamefont  [1]{#1}%
\providecommand \bibfnamefont [1]{#1}%
\providecommand \citenamefont [1]{#1}%
\providecommand \href@noop [0]{\@secondoftwo}%
\providecommand \href [0]{\begingroup \@sanitize@url \@href}%
\providecommand \@href[1]{\@@startlink{#1}\@@href}%
\providecommand \@@href[1]{\endgroup#1\@@endlink}%
\providecommand \@sanitize@url [0]{\catcode `\\12\catcode `\$12\catcode
  `\&12\catcode `\#12\catcode `\^12\catcode `\_12\catcode `\%12\relax}%
\providecommand \@@startlink[1]{}%
\providecommand \@@endlink[0]{}%
\providecommand \url  [0]{\begingroup\@sanitize@url \@url }%
\providecommand \@url [1]{\endgroup\@href {#1}{\urlprefix }}%
\providecommand \urlprefix  [0]{URL }%
\providecommand \Eprint [0]{\href }%
\providecommand \doibase [0]{http://dx.doi.org/}%
\providecommand \selectlanguage [0]{\@gobble}%
\providecommand \bibinfo  [0]{\@secondoftwo}%
\providecommand \bibfield  [0]{\@secondoftwo}%
\providecommand \translation [1]{[#1]}%
\providecommand \BibitemOpen [0]{}%
\providecommand \bibitemStop [0]{}%
\providecommand \bibitemNoStop [0]{.\EOS\space}%
\providecommand \EOS [0]{\spacefactor3000\relax}%
\providecommand \BibitemShut  [1]{\csname bibitem#1\endcsname}%
\let\auto@bib@innerbib\@empty
\bibitem [{\citenamefont {Horndeski}(1974)}]{Horndeski:1974wa}%
  \BibitemOpen
  \bibfield  {author} {\bibinfo {author} {\bibfnamefont {G.~W.}\ \bibnamefont
  {Horndeski}},\ }\href {\doibase 10.1007/BF01807638} {\bibfield  {journal}
  {\bibinfo  {journal} {Int. J. Theor. Phys.}\ }\textbf {\bibinfo {volume}
  {10}},\ \bibinfo {pages} {363} (\bibinfo {year} {1974})}\BibitemShut
  {NoStop}%
\bibitem [{\citenamefont {Deffayet}\ \emph {et~al.}(2011)\citenamefont
  {Deffayet}, \citenamefont {Gao}, \citenamefont {Steer},\ and\ \citenamefont
  {Zahariade}}]{Deffayet:2011gz}%
  \BibitemOpen
  \bibfield  {author} {\bibinfo {author} {\bibfnamefont {C.}~\bibnamefont
  {Deffayet}}, \bibinfo {author} {\bibfnamefont {X.}~\bibnamefont {Gao}},
  \bibinfo {author} {\bibfnamefont {D.~A.}\ \bibnamefont {Steer}}, \ and\
  \bibinfo {author} {\bibfnamefont {G.}~\bibnamefont {Zahariade}},\ }\href
  {\doibase 10.1103/PhysRevD.84.064039} {\bibfield  {journal} {\bibinfo
  {journal} {Phys. Rev.}\ }\textbf {\bibinfo {volume} {D84}},\ \bibinfo {pages}
  {064039} (\bibinfo {year} {2011})},\ \Eprint {http://arxiv.org/abs/1103.3260}
  {arXiv:1103.3260 [hep-th]} \BibitemShut {NoStop}%
\bibitem [{\citenamefont {Kobayashi}\ \emph {et~al.}(2011)\citenamefont
  {Kobayashi}, \citenamefont {Yamaguchi},\ and\ \citenamefont
  {Yokoyama}}]{Kobayashi:2011nu}%
  \BibitemOpen
  \bibfield  {author} {\bibinfo {author} {\bibfnamefont {T.}~\bibnamefont
  {Kobayashi}}, \bibinfo {author} {\bibfnamefont {M.}~\bibnamefont
  {Yamaguchi}}, \ and\ \bibinfo {author} {\bibfnamefont {J.}~\bibnamefont
  {Yokoyama}},\ }\href {\doibase 10.1143/PTP.126.511} {\bibfield  {journal}
  {\bibinfo  {journal} {Prog. Theor. Phys.}\ }\textbf {\bibinfo {volume}
  {126}},\ \bibinfo {pages} {511} (\bibinfo {year} {2011})},\ \Eprint
  {http://arxiv.org/abs/1105.5723} {arXiv:1105.5723 [hep-th]} \BibitemShut
  {NoStop}%
\bibitem [{\citenamefont {Zumalac«¡rregui}\ and\ \citenamefont
  {Garc«¿a-Bellido}(2014)}]{Zumalacarregui:2013pma}%
  \BibitemOpen
  \bibfield  {author} {\bibinfo {author} {\bibfnamefont {M.}~\bibnamefont
  {Zumalac«¡rregui}}\ and\ \bibinfo {author} {\bibfnamefont
  {J.}~\bibnamefont {Garc«¿a-Bellido}},\ }\href {\doibase
  10.1103/PhysRevD.89.064046} {\bibfield  {journal} {\bibinfo  {journal} {Phys.
  Rev.}\ }\textbf {\bibinfo {volume} {D89}},\ \bibinfo {pages} {064046}
  (\bibinfo {year} {2014})},\ \Eprint {http://arxiv.org/abs/1308.4685}
  {arXiv:1308.4685 [gr-qc]} \BibitemShut {NoStop}%
\bibitem [{\citenamefont {Gleyzes}\ \emph
  {et~al.}(2015{\natexlab{a}})\citenamefont {Gleyzes}, \citenamefont
  {Langlois}, \citenamefont {Piazza},\ and\ \citenamefont
  {Vernizzi}}]{Gleyzes:2014dya}%
  \BibitemOpen
  \bibfield  {author} {\bibinfo {author} {\bibfnamefont {J.}~\bibnamefont
  {Gleyzes}}, \bibinfo {author} {\bibfnamefont {D.}~\bibnamefont {Langlois}},
  \bibinfo {author} {\bibfnamefont {F.}~\bibnamefont {Piazza}}, \ and\ \bibinfo
  {author} {\bibfnamefont {F.}~\bibnamefont {Vernizzi}},\ }\href {\doibase
  10.1103/PhysRevLett.114.211101} {\bibfield  {journal} {\bibinfo  {journal}
  {Phys. Rev. Lett.}\ }\textbf {\bibinfo {volume} {114}},\ \bibinfo {pages}
  {211101} (\bibinfo {year} {2015}{\natexlab{a}})},\ \Eprint
  {http://arxiv.org/abs/1404.6495} {arXiv:1404.6495 [hep-th]} \BibitemShut
  {NoStop}%
\bibitem [{\citenamefont {Langlois}\ and\ \citenamefont
  {Noui}(2016)}]{Langlois:2015cwa}%
  \BibitemOpen
  \bibfield  {author} {\bibinfo {author} {\bibfnamefont {D.}~\bibnamefont
  {Langlois}}\ and\ \bibinfo {author} {\bibfnamefont {K.}~\bibnamefont
  {Noui}},\ }\href {\doibase 10.1088/1475-7516/2016/02/034} {\bibfield
  {journal} {\bibinfo  {journal} {JCAP}\ }\textbf {\bibinfo {volume} {1602}},\
  \bibinfo {pages} {034} (\bibinfo {year} {2016})},\ \Eprint
  {http://arxiv.org/abs/1510.06930} {arXiv:1510.06930 [gr-qc]} \BibitemShut
  {NoStop}%
\bibitem [{\citenamefont {Crisostomi}\ \emph {et~al.}(2016)\citenamefont
  {Crisostomi}, \citenamefont {Koyama},\ and\ \citenamefont
  {Tasinato}}]{Crisostomi:2016czh}%
  \BibitemOpen
  \bibfield  {author} {\bibinfo {author} {\bibfnamefont {M.}~\bibnamefont
  {Crisostomi}}, \bibinfo {author} {\bibfnamefont {K.}~\bibnamefont {Koyama}},
  \ and\ \bibinfo {author} {\bibfnamefont {G.}~\bibnamefont {Tasinato}},\
  }\href {\doibase 10.1088/1475-7516/2016/04/044} {\bibfield  {journal}
  {\bibinfo  {journal} {JCAP}\ }\textbf {\bibinfo {volume} {1604}},\ \bibinfo
  {pages} {044} (\bibinfo {year} {2016})},\ \Eprint
  {http://arxiv.org/abs/1602.03119} {arXiv:1602.03119 [hep-th]} \BibitemShut
  {NoStop}%
\bibitem [{\citenamefont {Kimura}\ \emph {et~al.}(2017)\citenamefont {Kimura},
  \citenamefont {Naruko},\ and\ \citenamefont {Yoshida}}]{Kimura:2016rzw}%
  \BibitemOpen
  \bibfield  {author} {\bibinfo {author} {\bibfnamefont {R.}~\bibnamefont
  {Kimura}}, \bibinfo {author} {\bibfnamefont {A.}~\bibnamefont {Naruko}}, \
  and\ \bibinfo {author} {\bibfnamefont {D.}~\bibnamefont {Yoshida}},\ }\href
  {\doibase 10.1088/1475-7516/2017/01/002} {\bibfield  {journal} {\bibinfo
  {journal} {JCAP}\ }\textbf {\bibinfo {volume} {1701}},\ \bibinfo {pages}
  {002} (\bibinfo {year} {2017})},\ \Eprint {http://arxiv.org/abs/1608.07066}
  {arXiv:1608.07066 [gr-qc]} \BibitemShut {NoStop}%
\bibitem [{\citenamefont {Ben~Achour}\ \emph
  {et~al.}(2016{\natexlab{a}})\citenamefont {Ben~Achour}, \citenamefont
  {Crisostomi}, \citenamefont {Koyama}, \citenamefont {Langlois}, \citenamefont
  {Noui},\ and\ \citenamefont {Tasinato}}]{BenAchour:2016fzp}%
  \BibitemOpen
  \bibfield  {author} {\bibinfo {author} {\bibfnamefont {J.}~\bibnamefont
  {Ben~Achour}}, \bibinfo {author} {\bibfnamefont {M.}~\bibnamefont
  {Crisostomi}}, \bibinfo {author} {\bibfnamefont {K.}~\bibnamefont {Koyama}},
  \bibinfo {author} {\bibfnamefont {D.}~\bibnamefont {Langlois}}, \bibinfo
  {author} {\bibfnamefont {K.}~\bibnamefont {Noui}}, \ and\ \bibinfo {author}
  {\bibfnamefont {G.}~\bibnamefont {Tasinato}},\ }\href {\doibase
  10.1007/JHEP12(2016)100} {\bibfield  {journal} {\bibinfo  {journal} {JHEP}\
  }\textbf {\bibinfo {volume} {12}},\ \bibinfo {pages} {100} (\bibinfo {year}
  {2016}{\natexlab{a}})},\ \Eprint {http://arxiv.org/abs/1608.08135}
  {arXiv:1608.08135 [hep-th]} \BibitemShut {NoStop}%
\bibitem [{\citenamefont {Ben~Achour}\ \emph
  {et~al.}(2016{\natexlab{b}})\citenamefont {Ben~Achour}, \citenamefont
  {Langlois},\ and\ \citenamefont {Noui}}]{Achour:2016rkg}%
  \BibitemOpen
  \bibfield  {author} {\bibinfo {author} {\bibfnamefont {J.}~\bibnamefont
  {Ben~Achour}}, \bibinfo {author} {\bibfnamefont {D.}~\bibnamefont
  {Langlois}}, \ and\ \bibinfo {author} {\bibfnamefont {K.}~\bibnamefont
  {Noui}},\ }\href {\doibase 10.1103/PhysRevD.93.124005} {\bibfield  {journal}
  {\bibinfo  {journal} {Phys. Rev.}\ }\textbf {\bibinfo {volume} {D93}},\
  \bibinfo {pages} {124005} (\bibinfo {year} {2016}{\natexlab{b}})},\ \Eprint
  {http://arxiv.org/abs/1602.08398} {arXiv:1602.08398 [gr-qc]} \BibitemShut
  {NoStop}%
\bibitem [{\citenamefont {Koyama}(2016)}]{Koyama:2015vza}%
  \BibitemOpen
  \bibfield  {author} {\bibinfo {author} {\bibfnamefont {K.}~\bibnamefont
  {Koyama}},\ }\href {\doibase 10.1088/0034-4885/79/4/046902} {\bibfield
  {journal} {\bibinfo  {journal} {Rept. Prog. Phys.}\ }\textbf {\bibinfo
  {volume} {79}},\ \bibinfo {pages} {046902} (\bibinfo {year} {2016})},\
  \Eprint {http://arxiv.org/abs/1504.04623} {arXiv:1504.04623 [astro-ph.CO]}
  \BibitemShut {NoStop}%
\bibitem [{\citenamefont {Berti}\ \emph {et~al.}(2015)\citenamefont {Berti}
  \emph {et~al.}}]{Berti:2015itd}%
  \BibitemOpen
  \bibfield  {author} {\bibinfo {author} {\bibfnamefont {E.}~\bibnamefont
  {Berti}} \emph {et~al.},\ }\href {\doibase 10.1088/0264-9381/32/24/243001}
  {\bibfield  {journal} {\bibinfo  {journal} {Class. Quant. Grav.}\ }\textbf
  {\bibinfo {volume} {32}},\ \bibinfo {pages} {243001} (\bibinfo {year}
  {2015})},\ \Eprint {http://arxiv.org/abs/1501.07274} {arXiv:1501.07274
  [gr-qc]} \BibitemShut {NoStop}%
\bibitem [{\citenamefont {Damour}\ and\ \citenamefont
  {Esposito-Farese}(1993)}]{Damour:1993hw}%
  \BibitemOpen
  \bibfield  {author} {\bibinfo {author} {\bibfnamefont {T.}~\bibnamefont
  {Damour}}\ and\ \bibinfo {author} {\bibfnamefont {G.}~\bibnamefont
  {Esposito-Farese}},\ }\href {\doibase 10.1103/PhysRevLett.70.2220} {\bibfield
   {journal} {\bibinfo  {journal} {Phys. Rev. Lett.}\ }\textbf {\bibinfo
  {volume} {70}},\ \bibinfo {pages} {2220} (\bibinfo {year}
  {1993})}\BibitemShut {NoStop}%
\bibitem [{\citenamefont {Damour}\ and\ \citenamefont
  {Esposito-Farese}(1996)}]{Damour:1996ke}%
  \BibitemOpen
  \bibfield  {author} {\bibinfo {author} {\bibfnamefont {T.}~\bibnamefont
  {Damour}}\ and\ \bibinfo {author} {\bibfnamefont {G.}~\bibnamefont
  {Esposito-Farese}},\ }\href {\doibase 10.1103/PhysRevD.54.1474} {\bibfield
  {journal} {\bibinfo  {journal} {Phys. Rev.}\ }\textbf {\bibinfo {volume}
  {D54}},\ \bibinfo {pages} {1474} (\bibinfo {year} {1996})},\ \Eprint
  {http://arxiv.org/abs/gr-qc/9602056} {arXiv:gr-qc/9602056 [gr-qc]}
  \BibitemShut {NoStop}%
\bibitem [{\citenamefont {Harada}(1997)}]{Harada:1997mr}%
  \BibitemOpen
  \bibfield  {author} {\bibinfo {author} {\bibfnamefont {T.}~\bibnamefont
  {Harada}},\ }\href {\doibase 10.1143/PTP.98.359} {\bibfield  {journal}
  {\bibinfo  {journal} {Prog. Theor. Phys.}\ }\textbf {\bibinfo {volume}
  {98}},\ \bibinfo {pages} {359} (\bibinfo {year} {1997})},\ \Eprint
  {http://arxiv.org/abs/gr-qc/9706014} {arXiv:gr-qc/9706014 [gr-qc]}
  \BibitemShut {NoStop}%
\bibitem [{\citenamefont {Harada}(1998)}]{Harada:1998ge}%
  \BibitemOpen
  \bibfield  {author} {\bibinfo {author} {\bibfnamefont {T.}~\bibnamefont
  {Harada}},\ }\href {\doibase 10.1103/PhysRevD.57.4802} {\bibfield  {journal}
  {\bibinfo  {journal} {Phys. Rev.}\ }\textbf {\bibinfo {volume} {D57}},\
  \bibinfo {pages} {4802} (\bibinfo {year} {1998})},\ \Eprint
  {http://arxiv.org/abs/gr-qc/9801049} {arXiv:gr-qc/9801049 [gr-qc]}
  \BibitemShut {NoStop}%
\bibitem [{\citenamefont {Doneva}\ and\ \citenamefont
  {Yazadjiev}(2018)}]{Doneva:2017bvd}%
  \BibitemOpen
  \bibfield  {author} {\bibinfo {author} {\bibfnamefont {D.~D.}\ \bibnamefont
  {Doneva}}\ and\ \bibinfo {author} {\bibfnamefont {S.~S.}\ \bibnamefont
  {Yazadjiev}},\ }\href {\doibase 10.1103/PhysRevLett.120.131103} {\bibfield
  {journal} {\bibinfo  {journal} {Phys. Rev. Lett.}\ }\textbf {\bibinfo
  {volume} {120}},\ \bibinfo {pages} {131103} (\bibinfo {year} {2018})},\
  \Eprint {http://arxiv.org/abs/1711.01187} {arXiv:1711.01187 [gr-qc]}
  \BibitemShut {NoStop}%
\bibitem [{\citenamefont {Silva}\ \emph {et~al.}(2018)\citenamefont {Silva},
  \citenamefont {Sakstein}, \citenamefont {Gualtieri}, \citenamefont
  {Sotiriou},\ and\ \citenamefont {Berti}}]{Silva:2017uqg}%
  \BibitemOpen
  \bibfield  {author} {\bibinfo {author} {\bibfnamefont {H.~O.}\ \bibnamefont
  {Silva}}, \bibinfo {author} {\bibfnamefont {J.}~\bibnamefont {Sakstein}},
  \bibinfo {author} {\bibfnamefont {L.}~\bibnamefont {Gualtieri}}, \bibinfo
  {author} {\bibfnamefont {T.~P.}\ \bibnamefont {Sotiriou}}, \ and\ \bibinfo
  {author} {\bibfnamefont {E.}~\bibnamefont {Berti}},\ }\href {\doibase
  10.1103/PhysRevLett.120.131104} {\bibfield  {journal} {\bibinfo  {journal}
  {Phys. Rev. Lett.}\ }\textbf {\bibinfo {volume} {120}},\ \bibinfo {pages}
  {131104} (\bibinfo {year} {2018})},\ \Eprint
  {http://arxiv.org/abs/1711.02080} {arXiv:1711.02080 [gr-qc]} \BibitemShut
  {NoStop}%
\bibitem [{\citenamefont {Antoniou}\ \emph
  {et~al.}(2018{\natexlab{a}})\citenamefont {Antoniou}, \citenamefont
  {Bakopoulos},\ and\ \citenamefont {Kanti}}]{Antoniou:2017acq}%
  \BibitemOpen
  \bibfield  {author} {\bibinfo {author} {\bibfnamefont {G.}~\bibnamefont
  {Antoniou}}, \bibinfo {author} {\bibfnamefont {A.}~\bibnamefont
  {Bakopoulos}}, \ and\ \bibinfo {author} {\bibfnamefont {P.}~\bibnamefont
  {Kanti}},\ }\href {\doibase 10.1103/PhysRevLett.120.131102} {\bibfield
  {journal} {\bibinfo  {journal} {Phys. Rev. Lett.}\ }\textbf {\bibinfo
  {volume} {120}},\ \bibinfo {pages} {131102} (\bibinfo {year}
  {2018}{\natexlab{a}})},\ \Eprint {http://arxiv.org/abs/1711.03390}
  {arXiv:1711.03390 [hep-th]} \BibitemShut {NoStop}%
\bibitem [{\citenamefont {Antoniou}\ \emph
  {et~al.}(2018{\natexlab{b}})\citenamefont {Antoniou}, \citenamefont
  {Bakopoulos},\ and\ \citenamefont {Kanti}}]{Antoniou:2017hxj}%
  \BibitemOpen
  \bibfield  {author} {\bibinfo {author} {\bibfnamefont {G.}~\bibnamefont
  {Antoniou}}, \bibinfo {author} {\bibfnamefont {A.}~\bibnamefont
  {Bakopoulos}}, \ and\ \bibinfo {author} {\bibfnamefont {P.}~\bibnamefont
  {Kanti}},\ }\href {\doibase 10.1103/PhysRevD.97.084037} {\bibfield  {journal}
  {\bibinfo  {journal} {Phys. Rev.}\ }\textbf {\bibinfo {volume} {D97}},\
  \bibinfo {pages} {084037} (\bibinfo {year} {2018}{\natexlab{b}})},\ \Eprint
  {http://arxiv.org/abs/1711.07431} {arXiv:1711.07431 [hep-th]} \BibitemShut
  {NoStop}%
\bibitem [{\citenamefont {Bl«¡zquez-Salcedo}\ \emph
  {et~al.}(2018)\citenamefont {Bl«¡zquez-Salcedo}, \citenamefont {Doneva},
  \citenamefont {Kunz},\ and\ \citenamefont
  {Yazadjiev}}]{Blazquez-Salcedo:2018jnn}%
  \BibitemOpen
  \bibfield  {author} {\bibinfo {author} {\bibfnamefont {J.~L.}\ \bibnamefont
  {Bl«¡zquez-Salcedo}}, \bibinfo {author} {\bibfnamefont {D.~D.}\
  \bibnamefont {Doneva}}, \bibinfo {author} {\bibfnamefont {J.}~\bibnamefont
  {Kunz}}, \ and\ \bibinfo {author} {\bibfnamefont {S.~S.}\ \bibnamefont
  {Yazadjiev}},\ }\href {\doibase 10.1103/PhysRevD.98.084011} {\bibfield
  {journal} {\bibinfo  {journal} {Phys. Rev.}\ }\textbf {\bibinfo {volume}
  {D98}},\ \bibinfo {pages} {084011} (\bibinfo {year} {2018})},\ \Eprint
  {http://arxiv.org/abs/1805.05755} {arXiv:1805.05755 [gr-qc]} \BibitemShut
  {NoStop}%
\bibitem [{\citenamefont {Minamitsuji}\ and\ \citenamefont
  {Ikeda}(2019)}]{Minamitsuji:2018xde}%
  \BibitemOpen
  \bibfield  {author} {\bibinfo {author} {\bibfnamefont {M.}~\bibnamefont
  {Minamitsuji}}\ and\ \bibinfo {author} {\bibfnamefont {T.}~\bibnamefont
  {Ikeda}},\ }\href {\doibase 10.1103/PhysRevD.99.044017} {\bibfield  {journal}
  {\bibinfo  {journal} {Phys. Rev.}\ }\textbf {\bibinfo {volume} {D99}},\
  \bibinfo {pages} {044017} (\bibinfo {year} {2019})},\ \Eprint
  {http://arxiv.org/abs/1812.03551} {arXiv:1812.03551 [gr-qc]} \BibitemShut
  {NoStop}%
\bibitem [{\citenamefont {Silva}\ \emph {et~al.}(2019)\citenamefont {Silva},
  \citenamefont {Macedo}, \citenamefont {Sotiriou}, \citenamefont {Gualtieri},
  \citenamefont {Sakstein},\ and\ \citenamefont {Berti}}]{Silva:2018qhn}%
  \BibitemOpen
  \bibfield  {author} {\bibinfo {author} {\bibfnamefont {H.~O.}\ \bibnamefont
  {Silva}}, \bibinfo {author} {\bibfnamefont {C.~F.~B.}\ \bibnamefont
  {Macedo}}, \bibinfo {author} {\bibfnamefont {T.~P.}\ \bibnamefont
  {Sotiriou}}, \bibinfo {author} {\bibfnamefont {L.}~\bibnamefont {Gualtieri}},
  \bibinfo {author} {\bibfnamefont {J.}~\bibnamefont {Sakstein}}, \ and\
  \bibinfo {author} {\bibfnamefont {E.}~\bibnamefont {Berti}},\ }\href
  {\doibase 10.1103/PhysRevD.99.064011} {\bibfield  {journal} {\bibinfo
  {journal} {Phys. Rev.}\ }\textbf {\bibinfo {volume} {D99}},\ \bibinfo {pages}
  {064011} (\bibinfo {year} {2019})},\ \Eprint
  {http://arxiv.org/abs/1812.05590} {arXiv:1812.05590 [gr-qc]} \BibitemShut
  {NoStop}%
\bibitem [{\citenamefont {Herdeiro}\ \emph {et~al.}(2018)\citenamefont
  {Herdeiro}, \citenamefont {Radu}, \citenamefont {Sanchis-Gual},\ and\
  \citenamefont {Font}}]{Herdeiro:2018wub}%
  \BibitemOpen
  \bibfield  {author} {\bibinfo {author} {\bibfnamefont {C.~A.~R.}\
  \bibnamefont {Herdeiro}}, \bibinfo {author} {\bibfnamefont {E.}~\bibnamefont
  {Radu}}, \bibinfo {author} {\bibfnamefont {N.}~\bibnamefont {Sanchis-Gual}},
  \ and\ \bibinfo {author} {\bibfnamefont {J.~A.}\ \bibnamefont {Font}},\
  }\href {\doibase 10.1103/PhysRevLett.121.101102} {\bibfield  {journal}
  {\bibinfo  {journal} {Phys. Rev. Lett.}\ }\textbf {\bibinfo {volume} {121}},\
  \bibinfo {pages} {101102} (\bibinfo {year} {2018})},\ \Eprint
  {http://arxiv.org/abs/1806.05190} {arXiv:1806.05190 [gr-qc]} \BibitemShut
  {NoStop}%
\bibitem [{\citenamefont {Doneva}\ \emph {et~al.}(2010)\citenamefont {Doneva},
  \citenamefont {Yazadjiev}, \citenamefont {Kokkotas},\ and\ \citenamefont
  {Stefanov}}]{Doneva:2010ke}%
  \BibitemOpen
  \bibfield  {author} {\bibinfo {author} {\bibfnamefont {D.~D.}\ \bibnamefont
  {Doneva}}, \bibinfo {author} {\bibfnamefont {S.~S.}\ \bibnamefont
  {Yazadjiev}}, \bibinfo {author} {\bibfnamefont {K.~D.}\ \bibnamefont
  {Kokkotas}}, \ and\ \bibinfo {author} {\bibfnamefont {I.~Z.}\ \bibnamefont
  {Stefanov}},\ }\href {\doibase 10.1103/PhysRevD.82.064030} {\bibfield
  {journal} {\bibinfo  {journal} {Phys. Rev.}\ }\textbf {\bibinfo {volume}
  {D82}},\ \bibinfo {pages} {064030} (\bibinfo {year} {2010})},\ \Eprint
  {http://arxiv.org/abs/1007.1767} {arXiv:1007.1767 [gr-qc]} \BibitemShut
  {NoStop}%
\bibitem [{\citenamefont {Stefanov}\ \emph {et~al.}(2008)\citenamefont
  {Stefanov}, \citenamefont {Yazadjiev},\ and\ \citenamefont
  {Todorov}}]{Stefanov:2007eq}%
  \BibitemOpen
  \bibfield  {author} {\bibinfo {author} {\bibfnamefont {I.~Z.}\ \bibnamefont
  {Stefanov}}, \bibinfo {author} {\bibfnamefont {S.~S.}\ \bibnamefont
  {Yazadjiev}}, \ and\ \bibinfo {author} {\bibfnamefont {M.~D.}\ \bibnamefont
  {Todorov}},\ }\href {\doibase 10.1142/S0217732308028351} {\bibfield
  {journal} {\bibinfo  {journal} {Mod. Phys. Lett.}\ }\textbf {\bibinfo
  {volume} {A23}},\ \bibinfo {pages} {2915} (\bibinfo {year} {2008})},\ \Eprint
  {http://arxiv.org/abs/0708.4141} {arXiv:0708.4141 [gr-qc]} \BibitemShut
  {NoStop}%
\bibitem [{\citenamefont {Motohashi}\ \emph {et~al.}(2016)\citenamefont
  {Motohashi}, \citenamefont {Suyama},\ and\ \citenamefont
  {Takahashi}}]{Motohashi:2016prk}%
  \BibitemOpen
  \bibfield  {author} {\bibinfo {author} {\bibfnamefont {H.}~\bibnamefont
  {Motohashi}}, \bibinfo {author} {\bibfnamefont {T.}~\bibnamefont {Suyama}}, \
  and\ \bibinfo {author} {\bibfnamefont {K.}~\bibnamefont {Takahashi}},\ }\href
  {\doibase 10.1103/PhysRevD.94.124021} {\bibfield  {journal} {\bibinfo
  {journal} {Phys. Rev.}\ }\textbf {\bibinfo {volume} {D94}},\ \bibinfo {pages}
  {124021} (\bibinfo {year} {2016})},\ \Eprint
  {http://arxiv.org/abs/1608.00071} {arXiv:1608.00071 [gr-qc]} \BibitemShut
  {NoStop}%
\bibitem [{\citenamefont {Kobayashi}\ \emph {et~al.}(2012)\citenamefont
  {Kobayashi}, \citenamefont {Motohashi},\ and\ \citenamefont
  {Suyama}}]{Kobayashi:2012kh}%
  \BibitemOpen
  \bibfield  {author} {\bibinfo {author} {\bibfnamefont {T.}~\bibnamefont
  {Kobayashi}}, \bibinfo {author} {\bibfnamefont {H.}~\bibnamefont
  {Motohashi}}, \ and\ \bibinfo {author} {\bibfnamefont {T.}~\bibnamefont
  {Suyama}},\ }\href {\doibase 10.1103/PhysRevD.96.109903,
  10.1103/PhysRevD.85.084025} {\bibfield  {journal} {\bibinfo  {journal} {Phys.
  Rev.}\ }\textbf {\bibinfo {volume} {D85}},\ \bibinfo {pages} {084025}
  (\bibinfo {year} {2012})},\ \bibinfo {note} {[Erratum: Phys.
  Rev.D96,no.10,109903(2017)]},\ \Eprint {http://arxiv.org/abs/1202.4893}
  {arXiv:1202.4893 [gr-qc]} \BibitemShut {NoStop}%
\bibitem [{\citenamefont {Kobayashi}\ \emph {et~al.}(2014)\citenamefont
  {Kobayashi}, \citenamefont {Motohashi},\ and\ \citenamefont
  {Suyama}}]{Kobayashi:2014wsa}%
  \BibitemOpen
  \bibfield  {author} {\bibinfo {author} {\bibfnamefont {T.}~\bibnamefont
  {Kobayashi}}, \bibinfo {author} {\bibfnamefont {H.}~\bibnamefont
  {Motohashi}}, \ and\ \bibinfo {author} {\bibfnamefont {T.}~\bibnamefont
  {Suyama}},\ }\href {\doibase 10.1103/PhysRevD.89.084042} {\bibfield
  {journal} {\bibinfo  {journal} {Phys. Rev.}\ }\textbf {\bibinfo {volume}
  {D89}},\ \bibinfo {pages} {084042} (\bibinfo {year} {2014})},\ \Eprint
  {http://arxiv.org/abs/1402.6740} {arXiv:1402.6740 [gr-qc]} \BibitemShut
  {NoStop}%
\bibitem [{\citenamefont {Kase}\ \emph {et~al.}(2014)\citenamefont {Kase},
  \citenamefont {Gergely},\ and\ \citenamefont {Tsujikawa}}]{Kase:2014baa}%
  \BibitemOpen
  \bibfield  {author} {\bibinfo {author} {\bibfnamefont {R.}~\bibnamefont
  {Kase}}, \bibinfo {author} {\bibfnamefont {L.~A.}\ \bibnamefont {Gergely}}, \
  and\ \bibinfo {author} {\bibfnamefont {S.}~\bibnamefont {Tsujikawa}},\ }\href
  {\doibase 10.1103/PhysRevD.90.124019} {\bibfield  {journal} {\bibinfo
  {journal} {Phys. Rev.}\ }\textbf {\bibinfo {volume} {D90}},\ \bibinfo {pages}
  {124019} (\bibinfo {year} {2014})},\ \Eprint {http://arxiv.org/abs/1406.2402}
  {arXiv:1406.2402 [hep-th]} \BibitemShut {NoStop}%
\bibitem [{\citenamefont {Franciolini}\ \emph {et~al.}(2019)\citenamefont
  {Franciolini}, \citenamefont {Hui}, \citenamefont {Penco}, \citenamefont
  {Santoni},\ and\ \citenamefont {Trincherini}}]{Franciolini:2018uyq}%
  \BibitemOpen
  \bibfield  {author} {\bibinfo {author} {\bibfnamefont {G.}~\bibnamefont
  {Franciolini}}, \bibinfo {author} {\bibfnamefont {L.}~\bibnamefont {Hui}},
  \bibinfo {author} {\bibfnamefont {R.}~\bibnamefont {Penco}}, \bibinfo
  {author} {\bibfnamefont {L.}~\bibnamefont {Santoni}}, \ and\ \bibinfo
  {author} {\bibfnamefont {E.}~\bibnamefont {Trincherini}},\ }\href {\doibase
  10.1007/JHEP02(2019)127} {\bibfield  {journal} {\bibinfo  {journal} {JHEP}\
  }\textbf {\bibinfo {volume} {02}},\ \bibinfo {pages} {127} (\bibinfo {year}
  {2019})},\ \Eprint {http://arxiv.org/abs/1810.07706} {arXiv:1810.07706
  [hep-th]} \BibitemShut {NoStop}%
\bibitem [{\citenamefont {Hui}\ and\ \citenamefont
  {Nicolis}(2013)}]{Hui:2012qt}%
  \BibitemOpen
  \bibfield  {author} {\bibinfo {author} {\bibfnamefont {L.}~\bibnamefont
  {Hui}}\ and\ \bibinfo {author} {\bibfnamefont {A.}~\bibnamefont {Nicolis}},\
  }\href {\doibase 10.1103/PhysRevLett.110.241104} {\bibfield  {journal}
  {\bibinfo  {journal} {Phys. Rev. Lett.}\ }\textbf {\bibinfo {volume} {110}},\
  \bibinfo {pages} {241104} (\bibinfo {year} {2013})},\ \Eprint
  {http://arxiv.org/abs/1202.1296} {arXiv:1202.1296 [hep-th]} \BibitemShut
  {NoStop}%
\bibitem [{\citenamefont {Babichev}\ and\ \citenamefont
  {Charmousis}(2014)}]{Babichev:2013cya}%
  \BibitemOpen
  \bibfield  {author} {\bibinfo {author} {\bibfnamefont {E.}~\bibnamefont
  {Babichev}}\ and\ \bibinfo {author} {\bibfnamefont {C.}~\bibnamefont
  {Charmousis}},\ }\href {\doibase 10.1007/JHEP08(2014)106} {\bibfield
  {journal} {\bibinfo  {journal} {JHEP}\ }\textbf {\bibinfo {volume} {08}},\
  \bibinfo {pages} {106} (\bibinfo {year} {2014})},\ \Eprint
  {http://arxiv.org/abs/1312.3204} {arXiv:1312.3204 [gr-qc]} \BibitemShut
  {NoStop}%
\bibitem [{\citenamefont {Tattersall}\ and\ \citenamefont
  {Ferreira}(2018)}]{Tattersall:2018nve}%
  \BibitemOpen
  \bibfield  {author} {\bibinfo {author} {\bibfnamefont {O.~J.}\ \bibnamefont
  {Tattersall}}\ and\ \bibinfo {author} {\bibfnamefont {P.~G.}\ \bibnamefont
  {Ferreira}},\ }\href {\doibase 10.1103/PhysRevD.97.104047} {\bibfield
  {journal} {\bibinfo  {journal} {Phys. Rev.}\ }\textbf {\bibinfo {volume}
  {D97}},\ \bibinfo {pages} {104047} (\bibinfo {year} {2018})},\ \Eprint
  {http://arxiv.org/abs/1804.08950} {arXiv:1804.08950 [gr-qc]} \BibitemShut
  {NoStop}%
\bibitem [{\citenamefont {Saravani}\ and\ \citenamefont
  {Sotiriou}(2019)}]{Saravani:2019xwx}%
  \BibitemOpen
  \bibfield  {author} {\bibinfo {author} {\bibfnamefont {M.}~\bibnamefont
  {Saravani}}\ and\ \bibinfo {author} {\bibfnamefont {T.~P.}\ \bibnamefont
  {Sotiriou}},\ }\href@noop {} {\  (\bibinfo {year} {2019})},\ \Eprint
  {http://arxiv.org/abs/1903.02055} {arXiv:1903.02055 [gr-qc]} \BibitemShut
  {NoStop}%
\bibitem [{\citenamefont {Babichev}\ \emph {et~al.}(2017)\citenamefont
  {Babichev}, \citenamefont {Charmousis},\ and\ \citenamefont
  {Lehebel}}]{Babichev:2017guv}%
  \BibitemOpen
  \bibfield  {author} {\bibinfo {author} {\bibfnamefont {E.}~\bibnamefont
  {Babichev}}, \bibinfo {author} {\bibfnamefont {C.}~\bibnamefont
  {Charmousis}}, \ and\ \bibinfo {author} {\bibfnamefont {A.}~\bibnamefont
  {Lehebel}},\ }\href {\doibase 10.1088/1475-7516/2017/04/027} {\bibfield
  {journal} {\bibinfo  {journal} {JCAP}\ }\textbf {\bibinfo {volume} {1704}},\
  \bibinfo {pages} {027} (\bibinfo {year} {2017})},\ \Eprint
  {http://arxiv.org/abs/1702.01938} {arXiv:1702.01938 [gr-qc]} \BibitemShut
  {NoStop}%
\bibitem [{\citenamefont {Sotiriou}\ and\ \citenamefont
  {Zhou}(2014)}]{Sotiriou:2014pfa}%
  \BibitemOpen
  \bibfield  {author} {\bibinfo {author} {\bibfnamefont {T.~P.}\ \bibnamefont
  {Sotiriou}}\ and\ \bibinfo {author} {\bibfnamefont {S.-Y.}\ \bibnamefont
  {Zhou}},\ }\href {\doibase 10.1103/PhysRevD.90.124063} {\bibfield  {journal}
  {\bibinfo  {journal} {Phys. Rev.}\ }\textbf {\bibinfo {volume} {D90}},\
  \bibinfo {pages} {124063} (\bibinfo {year} {2014})},\ \Eprint
  {http://arxiv.org/abs/1408.1698} {arXiv:1408.1698 [gr-qc]} \BibitemShut
  {NoStop}%
\bibitem [{\citenamefont {Kimura}\ and\ \citenamefont
  {Tanaka}(2018)}]{Kimura:2018eiv}%
  \BibitemOpen
  \bibfield  {author} {\bibinfo {author} {\bibfnamefont {M.}~\bibnamefont
  {Kimura}}\ and\ \bibinfo {author} {\bibfnamefont {T.}~\bibnamefont
  {Tanaka}},\ }\href {\doibase 10.1088/1361-6382/aadc13} {\bibfield  {journal}
  {\bibinfo  {journal} {Class. Quant. Grav.}\ }\textbf {\bibinfo {volume}
  {35}},\ \bibinfo {pages} {195008} (\bibinfo {year} {2018})},\ \Eprint
  {http://arxiv.org/abs/1805.08625} {arXiv:1805.08625 [gr-qc]} \BibitemShut
  {NoStop}%
\bibitem [{\citenamefont {Anson}\ \emph {et~al.}(2019)\citenamefont {Anson},
  \citenamefont {Babichev}, \citenamefont {Charmousis},\ and\ \citenamefont
  {Ramazanov}}]{Anson:2019uto}%
  \BibitemOpen
  \bibfield  {author} {\bibinfo {author} {\bibfnamefont {T.}~\bibnamefont
  {Anson}}, \bibinfo {author} {\bibfnamefont {E.}~\bibnamefont {Babichev}},
  \bibinfo {author} {\bibfnamefont {C.}~\bibnamefont {Charmousis}}, \ and\
  \bibinfo {author} {\bibfnamefont {S.}~\bibnamefont {Ramazanov}},\ }\href@noop
  {} {\  (\bibinfo {year} {2019})},\ \Eprint {http://arxiv.org/abs/1903.02399}
  {arXiv:1903.02399 [gr-qc]} \BibitemShut {NoStop}%
\bibitem [{\citenamefont {Franchini}\ and\ \citenamefont
  {Sotiriou}(2019)}]{Franchini:2019npi}%
  \BibitemOpen
  \bibfield  {author} {\bibinfo {author} {\bibfnamefont {N.}~\bibnamefont
  {Franchini}}\ and\ \bibinfo {author} {\bibfnamefont {T.~P.}\ \bibnamefont
  {Sotiriou}},\ }\href@noop {} {\  (\bibinfo {year} {2019})},\ \Eprint
  {http://arxiv.org/abs/1903.05427} {arXiv:1903.05427 [gr-qc]} \BibitemShut
  {NoStop}%
\bibitem [{\citenamefont {Abbott}\ \emph {et~al.}(2017)\citenamefont {Abbott}
  \emph {et~al.}}]{Monitor:2017mdv}%
  \BibitemOpen
  \bibfield  {author} {\bibinfo {author} {\bibfnamefont {B.~P.}\ \bibnamefont
  {Abbott}} \emph {et~al.} (\bibinfo {collaboration} {Virgo, Fermi-GBM,
  INTEGRAL, LIGO Scientific}),\ }\href {\doibase 10.3847/2041-8213/aa920c}
  {\bibfield  {journal} {\bibinfo  {journal} {Astrophys. J.}\ }\textbf
  {\bibinfo {volume} {848}},\ \bibinfo {pages} {L13} (\bibinfo {year}
  {2017})},\ \Eprint {http://arxiv.org/abs/1710.05834} {arXiv:1710.05834
  [astro-ph.HE]} \BibitemShut {NoStop}%
\bibitem [{\citenamefont {Gleyzes}\ \emph
  {et~al.}(2015{\natexlab{b}})\citenamefont {Gleyzes}, \citenamefont
  {Langlois}, \citenamefont {Piazza},\ and\ \citenamefont
  {Vernizzi}}]{Gleyzes:2014qga}%
  \BibitemOpen
  \bibfield  {author} {\bibinfo {author} {\bibfnamefont {J.}~\bibnamefont
  {Gleyzes}}, \bibinfo {author} {\bibfnamefont {D.}~\bibnamefont {Langlois}},
  \bibinfo {author} {\bibfnamefont {F.}~\bibnamefont {Piazza}}, \ and\ \bibinfo
  {author} {\bibfnamefont {F.}~\bibnamefont {Vernizzi}},\ }\href {\doibase
  10.1088/1475-7516/2015/02/018} {\bibfield  {journal} {\bibinfo  {journal}
  {JCAP}\ }\textbf {\bibinfo {volume} {1502}},\ \bibinfo {pages} {018}
  (\bibinfo {year} {2015}{\natexlab{b}})},\ \Eprint
  {http://arxiv.org/abs/1408.1952} {arXiv:1408.1952 [astro-ph.CO]} \BibitemShut
  {NoStop}%
\end{thebibliography}%
\end{document}